\documentclass[amsmath,amssymb, aps, pra,reprint,longbibliography]{revtex4-1}

\usepackage{graphicx}
\usepackage[retainorgcmds]{IEEEtrantools}
\usepackage{color}

\def\bd{\begin{displaymath}}
\def\be{\begin{equation}}
\def\ed{\end{displaymath}}
\def\ee{\end{equation}}
\def\bsub{\begin{subequations}}
\def\esub{\end{subequations}}
\def\sz{\,\sigma_z }
\def\sx{\,\sigma_x}

\newcommand{\Eq}[1]{Eq.~(\ref{#1})}
\newcommand{\Fig}[1]{Fig.~\ref{#1}}
\newcommand{\abs}[1]{\left| #1 \right|}

\makeatletter
\newcommand*\dashline{\rotatebox[origin=c]{90}{$\dabar@\dabar@\dabar@$}}
\makeatother
\def\dw{\mspace{-4mu} \dashline \mspace{-4mu}}

\begin{document}

\title{\bf Abelian and non-Abelian statistics in the coherent state representation}

\author{John Flavin}
\author{Alexander Seidel}
\affiliation{
Department of Physics and Center for Materials Innovation,
Washington University, St. Louis, MO 63136, USA}


\begin{abstract}
We further develop an approach
to identify the braiding statistics associated to a given
fractional quantum Hall state through adiabatic
transport of quasiparticles. 
This approach
is based on the notion of adiabatic continuity
between quantum Hall states on the torus and simple product 
states---or ``patterns''---in the thin torus limit, together with
a suitable coherent state ansatz for localized quasiholes
that respects the modular invariance of the torus.
We give a refined and unified account of the application of this
method to the Laughlin and Moore-Read states, which may
serve as a pedagogical introduction to the nuts and bolts of this technique.
Our main result is that the approach is also 
applicable---without further assumptions---to
more complicated non-Abelian states.
We demonstrate this in great detail for the level $k=3$ 
Read-Rezayi state at filling factor $\nu=3/2$.
These results may serve as an independent check of
other techniques, where the statistics are inferred from
conformal block monodromies.
Our approach has the benefit of giving rise to intuitive
pictures representing the transformation of topological sectors during braiding,
and allows for a self-consistent derivation of non-Abelian statistics without
heavy mathematical machinery.

\end{abstract}


\maketitle

\section{Introduction}
The discovery of the fractional quantum Hall (FQH) effect \cite{TSG} has demonstrated that under the right conditions, an interacting electron system may enter a state with topological quantum order \cite{wenniu}. Laughlin's seminal treatment \cite{laughlin} of this new state of matter predicted that FQH systems should display a unique and rich phenomenology beyond the quantized Hall conductance that had led to its discovery. This includes the presence of robust gapless chiral excitations at the edge, as well as fractionally charged bulk excitations, which furthermore obey fractional statistics. These characteristics allow one to distinguish a great wealth of different classes of FQH states including, possibly, ones where the statistics 
of the quasiparticle-type excitations are non-Abelian \cite{mooreread, wenblok}. The latter might facilitate a particularly robust route to fault tolerant quantum computing \cite{kitaev, freedman}. 
Experimental control of the quasiparticle excitations and their possible utilization for schemes of quantum computing remain among the foremost challenges of the field. On the theoretical side, things are largely under control due to a  remarkable correspondence between FQH trial wave functions and conformal blocks in certain rational conformal field theories \cite{mooreread}. This formal correspondence has given rise to powerful field theoretic mappings that allow one, among other things, to infer the quasiparticle braiding statistics of the state in question \cite{nayakwilczek}. This inference, however, remains 
without microscopic justification in most cases. Such a justification requires showing that the procedure followed agrees with the result of adiabatic transport of quasiparticles (as defined by trial wave functions), which ultimately defines the statistics. This remains challenging in many cases of interest. For Laughlin quasiparticles, this program has been carried out early on by Arovas, Schrieffer, and Wilczek \cite{ASW}. The non-Abelian case has proven to be a profound technical challenge. For the Moore-Read, or ``Pfaffian'', state, a proof has recently been put forth \cite{bonderson_gurarie_nayak} following a series of insightful papers \cite{gurarie_nayak97, read09,read_arxiv_08} further developing the plasma mapping \cite{laughlin} (for $p+ip$ wave paired superfluids, a proof was given earlier in Ref. \onlinecite{read09}).

Prior to that, a number of non-rigorous techniques had been developed to independently confirm the conformal field theory (CFT) result for the braiding statistics of the Pfaffian state.
 These techniques have the additional merit of recasting the non-Abelian statistics of the Pfaffian state into a different language that makes no use of conformal field theory or related modular tensor categories. Such alternative languages might be particularly desirable in the Pfaffian case, which is believed to be relevant to the experimentally observed plateau at filling factor $\nu=5/2$ \cite{fivehalves}. The first such approach is based on the interpretation \cite{read_green,ivanov,oppen,stonechung,oshikawa} of the Pfaffian state as a $p+ip$ wave Bardeen-Cooper-Schrieffer (BCS) state of composite fermions \cite{jain}. The second approach employs a strategy that has been successfully  applied to interacting many body-systems since Landau's concept of a Fermi liquid, but only recently to electrons in the fractional quantum Hall regime. This strategy is to view the complicated interacting many-body state of interest as the adiabatic descendant of a simple, non-interacting state. As demonstrated in a series of recent works \cite{seidel1, seidel2, seidel_lee, seidelyang, seidel_pfaffian, seidel_sduality, seidelyang2, karlhede1,karlhede2,karlhede3,karlhede4,karlhede5,karlhede09}, an adiabatic continuity with the desired features is given for a large class of FQH states by taking the thin torus or cylinder limit, which, as a formal limit, had been considered earlier in Ref. \onlinecite{haldane_rezayi}. This approach gives rise to a language of simple strings of integers, or patterns, that are associated to various incompressible FQH states and their quasiparticle-type excitations. The same patterns also play a central role in the recently discussed connection between FQH states and Jack polynomials \cite{haldanebernevig,haldanebernevig3, Bernevig2008}, and are intimately related to the ``patterns of zeros'' describing these states \cite{wenwang, wenwang2, barkeshli1, barkeshli2}. The adiabatic continuity between such ``thin torus patterns'' and FQH states on general tori has been utilized in Refs. \onlinecite{seidel_lee, seidel_pfaffian} to derive the statistics of various Abelian FQH states and of the Moore-Read state, respectively.

For the Pfaffian state, there are thus a number of alternatives to the standard CFT method of obtaining the statistics, including rigorous results. However, the generalization of these alternative methods to more complicated non-Abelian states has thus far been limited. It has been argued in Ref. \onlinecite{seidel_pfaffian} that the method using thin torus patterns and adiabatic continuity should in principle be generalizable to other non-Abelian states. At the same time, it is by no means obvious that the approach chosen there will always give rise to a sufficiently constraining set of equations to determine the statistics. 
That this is the case might naively be expected from the fact that the set of inequivalent solutions obtained in Ref. \onlinecite{seidel_pfaffian} is identical to those obtained \cite{bonderson_thesis, kitaev06}  from an assumed knowledge of the underlying CFT fusion rules, 
together with Moore-Seiberg polynomial equations \cite{moore_seiberg}.
Indeed, in both cases one obtains eight distinct solutions that are all related by overall Abelian phase factors. Furthermore, it is known that the thin torus patterns efficiently encode information about fusion rules \cite{Ardonne2008, ardonne_domain}. One might thus conjecture that both methods generally produce the same results. We will show below that this is not the case. To see why this need not be surprising, we note up front 
a number of important differences between these two methods at both the conceptual and technical levels. A key step in relating fusion rules to braid matrices is to impose the validity of the ``hexagon'' and ``pentagon'' equations as they appear naturally in rational CFTs \cite{moore_seiberg}. 
These equations will not be explicitly enforced in our approach.
Indeed,
our framework requires
no {\em a priori} assumption 
that the result of adiabatic transport of quasiparticles along braiding paths is purely topological in nature.
 Rather, this fact emerges naturally---together with the proper {\em non}-topological (Aharonov-Bohm) contributions---as a result of adiabatic transport. The assumptions underlying these two methods are thus quite different. It is hence not immediately
 clear whether the ``thin torus approach'' can be generalized to more complicated non-Abelian states. The main purpose of this paper is to give an affirmative answer to this question, and to flesh out this scheme of attack in greater generality, by applying it to the level $k=3$ Read-Rezayi state \cite{readrezayi}.

Even more generally, 
an efficient route from thin torus patterns to the statistics of the underlying state might be 
to use the information encoded in the patterns to relate them to some rational CFT.
As a general disclaimer, this is not what we will attempt to do in this work. Rather, we will use adiabatic continuity and related assumptions to work out the statistics of a given quantum Hall state within a consistent framework that is {\em independent} of the field-theoretic assumptions that are traditionally employed in this field. The fact that we obtain results that are consistent with the CFT framework can thus be regarded as a non-rigorous, but independent, confirmation of the latter. We note that the connection between CFT and the patterns of zeros of a quantum Hall state, together with its implications about braiding statistics, 
have been studied in detail by Lu et al \cite{Lu}.

We believe that an added benefit to the method developed here lies in the fact that an intuitive language is provided to describe non-Abelian statistics, and this language does not require the reader to have much background in mathematical physics. To proceed, we thus give a refined and more detailed account of simpler cases already studied by this method. The basic ideas and underlying assumptions are fleshed out in Sec. \ref{laughlin}, where the simplest Abelian state, the $\nu=1/2$ Laughlin state, is studied. The generalization to the non-Abelian case is studied in Sec. \ref{pfaff}, where the Pfaffian state is considered. This section gives a rather more detailed and somewhat improved account 
of results first presented in
Ref. \onlinecite{seidel_pfaffian}. In Sec. \ref{rr}, we then show that exactly the same set of assumptions that sufficed to treat the Pfaffian case can also be used to determine the statistics of the $\nu=3/2$ (level 3) Read-Rezayi state essentially uniquely (up to an Abelian phase and complex conjugation). Our results concerning the statistics of this state, as represented though thin torus patterns, are summarized in Sec. \ref{discussion}. The reader not interested in technical details, but rather more in this representation, is recommended to glimpse over Secs. \ref{outline},  \ref{ttl}, \ref{cstates}, and \ref{braid}, to gather the nuts and bolts of the basic language used in this work, and then skip ahead to Sec. \ref{discussion} (for the Read-Rezayi state) or Secs. \ref{pf_braid},
\ref{pf_braidgroup} 
(for the Pfaffian state). In the remainder of the present section, we will review some basic formalism (\ref{LLstructure}), and then proceed to outline our general scheme of attack in \ref{outline}. Some technical details are relegated to two appendices.
\subsection{\label{LLstructure}Physics of LLL} 
We work on a torus, identified as a rectangular 2D domain of dimensions $L_x$ and $L_y$ subject to (magnetic) periodic  boundary conditions. We take the magnetic vector potential to be in Landau gauge, $\mathbf{A}=(0,x)$. The magnetic length is set equal to $1$ such that $L_x L_y = 2\pi L$, where $L$ equals the number of magnetic flux quanta through the surface of the torus, which also equals the number of orbitals in the lowest Landau level (LLL). An infinite cylinder is obtained in the limit $L_x\rightarrow\infty$, with $L_y$ kept finite. We first construct a basis of the LLL on such a cylinder. It is given by  $\varphi^c_n(z) = \xi^n \exp(-\frac{1}{2} x^2 - \frac{1}{2} \kappa^2 n^2 )$, where $\kappa = 2\pi / L_y$, $z=x+i y$ is the particle's complex coordinate, and $\xi=\exp(\kappa z)$. From the LLL  states $\varphi_n^c$ on the infinite cylinder one can construct LLL states $\varphi_n$ that satisfy proper periodic magnetic boundary conditions  (cf. Ref. \onlinecite{haldanePRL85}) on a torus with finite $L_x=\kappa L$. Fixing some unimportant overall phases, these boundary conditions read
\begin{equation}
   \begin{split}\label{mbcs}
	\varphi_n (z+L_x)&=e^{i\kappa y}\varphi_n(z)\\
	\varphi_n(z+iL_y)&=\varphi_n(z)\;,
   \end{split}
\end{equation}
for the present gauge, and the orbitals $\varphi_n(z)$ satisfying these conditions are then simply obtained by ``repeating" the LLL orbitals of the cylinder along the $x$ direction:
\be\label{varphi}
 	\varphi_n(z) = \sum_j \varphi^c_{n+j L}(z)\;.
\ee
\begin{figure}  
	\includegraphics[width=.75\columnwidth]{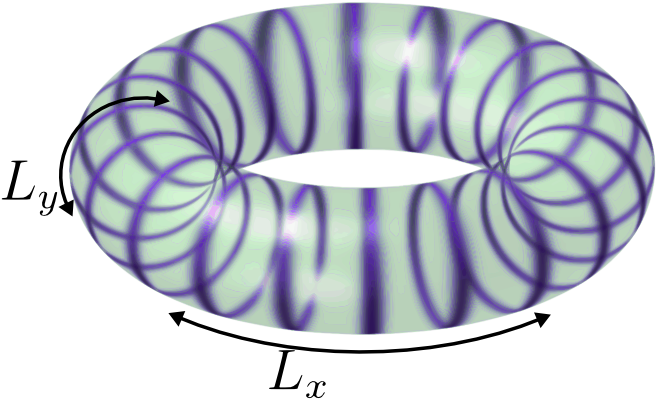} 
	\caption{Landau level basis on the torus of dimensions $L_x\times L_y$. 
	          The orbitals  $\varphi_n(z)$ form a 1D 
		periodic ``lattice'' in the $x$ direction. Each orbital $\varphi_n(z)$ localizes a 
		particle at $x=\kappa n$ while being delocalized in the $y$ direction,
		leading to a ``ring shape'' geometry.
		Consecutive orbitals are separated by a distance $\kappa$. 
		}
	\label{rings}
 \end{figure}
For both the cylinder and the torus (with sufficiently large $L_x$),  the $n$-th LLL orbital has the ``ring shape'' geometry shown in \Fig{rings}. The orbital $\varphi_n(z)$ localizes a particle in the $x$ direction around $x=\kappa n$ to within one magnetic length, such that consecutive orbitals are separated by a distance $\kappa$. At the same time, each  orbital is completely delocalized in $y$. We can view the orbitals $\varphi_n$ as forming a 1D periodic ``lattice'' along the $x$ direction, with each orbital representing a lattice site. Note that we have $\varphi_{n+L}(z)=\varphi_n(z)$, and in this sense the ``orbital lattice''  satisfies ordinary periodic boundary conditions in $n$. A ``thin torus limit" \cite{haldane_rezayi94, seidel1, seidel2, seidel_lee, seidelyang, seidel_pfaffian, seidel_sduality, seidelyang2, karlhede1,karlhede2,karlhede3,karlhede4,karlhede5,karlhede09} can be defined as $\kappa\gg 1$. In this limit, the orbitals in the basis \eqref{varphi} are well separated and have negligible overlap.

\begin{figure}  
	\includegraphics[width=\columnwidth]{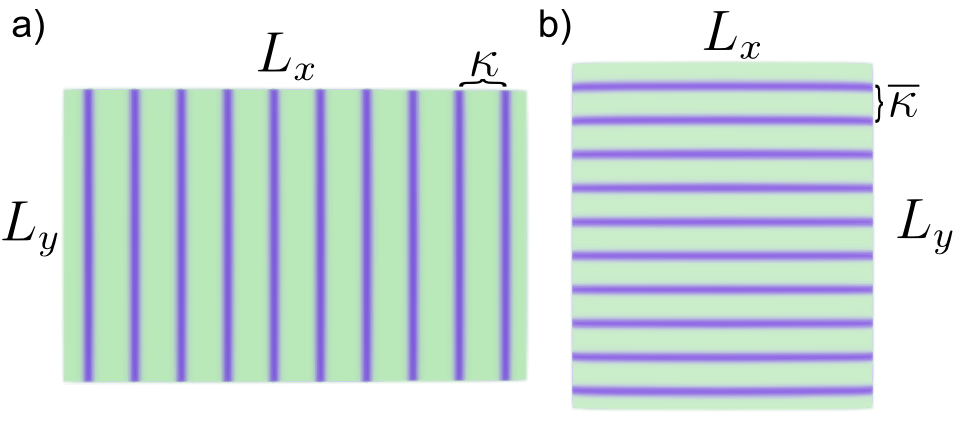} 
	\caption{The two Landau level bases $\varphi_n(z)$ and 
	          $\overline\varphi_n(z)$ on the torus, shown here as a rectangular strip. 
		a) The ``original'' basis, in which orbitals $\varphi_n(z)$ form a lattice in $x$ 
		with lattice spacing $\kappa=2\pi/L_y$ and encircle the torus in $y$. 
		b) The dual basis, in which orbitals $\overline\varphi_n(z)$ form a lattice in $y$ 
		with lattice spacing $\overline\kappa=2\pi/L_x$ and encircle the torus in $x$. } 
	\label{orig_dual}
 \end{figure}

It is clear that the choice of LLL orbital basis made above  treats the $x$ direction on the torus differently from the $y$ direction. However, nothing prevents us from exchanging the roles of $x$ and $y$. A ``dual'' basis of states $\overline\varphi_n$ localized at $y=\overline\kappa n$ (for $\overline\kappa=2 \pi / L_x$), encircling the torus in the $x$ direction (\Fig{orig_dual}), can be obtained by formally ``rotating" the $\varphi_n$ basis, followed by a gauge transformation, via
\be\label{dualvarphi} 
	\overline\varphi_n(z) = \exp(i x y) \varphi_n(-i z)|_{\kappa \rightarrow \overline\kappa }\;.
\ee
Alternatively, it can be shown (via Poisson resummation) that the  $\overline\varphi_n$ basis thus defined is related to the original basis \eqref{varphi} through a discrete Fourier transform, i.e
\be \label{dualvarphi2}
	\overline\varphi_n(z) 
	= \frac1{\sqrt{L}} \sum_{n'} \exp(-i \frac{2 \pi}L n n') \varphi_{n'}(z)\,.
\ee

In the presence of the magnetic field, the single-particle Hamiltonian commutes with two magnetic translation operators, whose form in the chosen gauge is given by
\be 
   \begin{split}
	t_x &= e^{-\kappa (\partial_x-iy)} \\ 
	t_y &= e^{- \overline\kappa\partial_y }\;.
   \end{split}
\ee 
The orbital bases ${\varphi}_n$ and $\overline{\varphi}_n$ have simple transformation properties under the action of these two non-commuting translation operators. One easily verifies that
\begin{subequations} \label{t}
   \begin{align} \label{t_x} 
	t_x \varphi_n(z) & = \varphi_{n+1}(z) &
	t_x \overline\varphi_n(z) & = e^{\frac{2 \pi i}L n} \overline\varphi_n(z) \\ 
   \label{t_y} 
	t_y \varphi_n(z) & = e^{-\frac{2 \pi i}{L} n}\varphi_n(z)  &
	t_y \overline\varphi_n(z) & = \overline\varphi_{n+1}(z)\;.
   \end{align}
\end{subequations}
All orbitals are thus invariant under the action of the operators $t_x^{\phantom{x}L}$ and $t_y^{\phantom{y}L}$, which represent magnetic translations by $L_x$ and $L_y$ in the respective direction. This is equivalent to the observation that both the $\varphi_n$ as well as the $\overline\varphi_n$ orbitals satisfy the {\em same} periodic magnetic boundary conditions \eqref{mbcs} appropriate to the gauge $\mathbf{A}=(0,x)$.

We finally mention some other important symmetries of the problem under consideration. Inversion symmetry acts on wave functions via $I \psi(z)=\psi(-z)$, and on the basis states defined above via
\be \label{inversion}
	I \varphi_n(z) =\varphi_{-n}(z)\;,
	\quad  I \overline\varphi_n(z)=\overline\varphi_{-n}(z)\,.
\ee
Similarly, while there is neither time reversal symmetry nor mirror symmetry in the presence of the constant magnetic field, the combined symmetry does exist. We denote by $\tau$ the antilinear operator that acts on wave functions via $\tau\psi(z)=\psi(-z^\ast)^\ast$, and on basis states via
\be \label{mirror1}
	\tau \varphi_n(z)=\varphi_{-n}(z)\;,
	\quad \tau \overline\varphi_n(z)=\overline\varphi_{n}(z)\,,
\ee
where the second equation follows from the first with \Eq{dualvarphi2}. The reflectional part of $\tau$ is obviously a reflection about the $y$ axis. We can similarly define an antilinear operator $\bar\tau$ that performs a reflection about the $x$ axis in conjunction with time reversal, and which acts on basis states via
\be \label{mirror2}
	\bar \tau \varphi_n(z)=\varphi_n(z)\;,
	\quad \bar\tau \overline\varphi_n(z)=\overline\varphi_{-n}(z)\,.
\ee
%
\subsection{\label{outline}Outline of the method}
Our method of inferring the statistics of a quantum Hall state can be broken down into a few elementary steps, which can in principle be applied to any quantum Hall state that can be assigned well-defined ground-state patterns through a thin torus limit \cite{seidel1, seidel2, seidel_lee, seidelyang, seidel_pfaffian, seidel_sduality,  karlhede2, karlhede3, karlhede4, karlhede5,karlhede09}. Here we give a brief summary of  the individual steps and the underlying ideas. A detailed development of these ideas will be given in the subsequent chapters, with applications to Laughlin, Moore-Read, and $k=3$ Read-Rezayi states. 

\emph{1. Identify integer patterns characterizing the state}

It has recently become appreciated that a large class of trial wave functions can be characterized by simple sequences of integers. These patterns may be identified through various interrelated approaches. We we focus here on the approach based on the thin torus limit  and adiabatic continuity \cite{seidel1,seidel2, seidel_lee, seidelyang, seidel_pfaffian, seidel_sduality,seidelyang2, karlhede1,karlhede2,karlhede3,karlhede4, karlhede5,karlhede09}. This is not done for mere convenience, as it turns out that adiabatic continuity will play an essential role in the following. However, the patterns themselves can also be identified using various other methods, such as the Jack polynomial construction \cite{haldanebernevig, haldanebernevig3, Bernevig2008} and through ``patterns of zeros" \cite{wenwang, wenwang2, barkeshli1, barkeshli2}.

To be more specific, we will assume that the following program can be successfully carried out for the quantum Hall state in question. We assume that a Hamiltonian has been identified whose ground state lies within the desired phase. This Hamiltonian is assumed be local and to induce no Landau level mixing. The Hamiltonian can then be deformed into a limit that describes a thin torus, where either $L_y\ll 1$ or $L_x\ll 1$. In this limit the ground state will approach a trivial product state of LLL orbitals, either in the $\varphi_n$ basis (for $L_y\ll 1$) or in the $\overline \varphi_n$ basis (for $L_x\ll 1$). These limiting states can be simply labeled by the pattern of occupancy numbers of successive LLL orbitals. Examples include $100100100\dotsc$ for the $\nu=1/3$ Laughlin state, or  $202020\dotsc$ and $111111\dotsc$ for the degenerate ground states of the Moore-Read state at $\nu=1$. In these examples it has been demonstrated numerically \cite{seidel1,seidel2} that the deformation of the Hamiltonian into this limit can be done adiabatically, i.e., the gap of the incompressible fluid is maintained along the way. We believe that these observations can be extended, at the least, to all classes of trial wave functions for which local parent Hamiltonians, whose spectrum remains gapped in the 2D limit of an infinite plane, may be identified.

\emph{2. Use adiabatic continuity to organize the space of elementary quasiparticle-type excitations}

It will further be assumed that adiabatic continuity, as described above, does not only hold in the sector spanned by the incompressible ground states, but also in the sectors obtained by adding quasiparticles or quasiholes to the system. Specifically, in known cases of special trial wave functions for which parent Hamiltonians can be identified, the set of states obtained from the incompressible  ground states by adding $n\geq 0$ quasiholes may usually be characterized as the set of zero-energy states (or ``zero-modes") of this Hamiltonian \cite{readrezayiXX}. All examples discussed in the following will be of this kind. Adiabatic evolution from the 2D torus (by this we will always mean the regime $L_x,L_y\gg1$) into the thin torus limit will then take states with $n$ quasiholes into states with $n$ domain walls between different integer ground-state patterns \cite{seidel1, seidel2, seidel_lee, seidelyang, seidel_pfaffian, seidel_sduality, seidelyang2, karlhede1,karlhede2,karlhede3,karlhede4,karlhede5,karlhede09}. For example, a state with $n=2$ quasiholes in a $\nu=1$ Pfaffian may evolve into a thin torus state corresponding to the following occupancy pattern:
\be \label{samplepattern}
	2020202020202020111111111111111110202020202020202\;.
\ee
A formal thin torus limit of the known zero-mode wave functions reveals the types of domain walls that represent quasihole states. The assumption of adiabatic continuity implies that at any aspect ratio of the torus, the zero-mode states will be in one-to-one correspondence with thin torus states of the form Eq. \eqref{samplepattern}. One can thus define a complete basis of zero modes via adiabatic continuation of the thin torus basis. The simple patterns that characterize the thin torus limit of a given basis state may still be used as ``state labels'' away from the thin torus limit. These labels carry important information about the transformation properties of basis states under magnetic translations. They also provide information about the change of topological sector for certain topologically nontrivial rearrangements of quasiparticles on the torus. The fact that these ``thin torus labels" remain meaningful,  i.e., can be used to organize the space of zero-mode states {\em away from this torus limit}, will allow us to obtain information about the braiding statistics of the state, 
even though braiding statistics are well defined only on a (nonthin) 2D torus.

\emph{3. Form coherent states describing localized quasiholes}

Individually, the adiabatically continued domain-wall states defining the quasihole basis described above do not
correspond to states of well localized quasiholes. This is so because these states have a well defined momentum about the ``quantization axis", by which we mean the $x$ axis when states are defined in the $L_y\rightarrow 0$ limit (where $\varphi_n$ orbitals are used to define product states), and the $y$ axis when states are defined in the $L_x\rightarrow 0$ limit (where $\overline\varphi_n$ orbitals are used instead). Localized quasiholes exhibiting nontrivial braiding statistics are given by coherent state superpositions formed by 
 states in the basis defined above.  
 The general form of these coherent states is highly constrained by symmetries, non-commutative geometry (i.e., $[x,y]\propto i$ within a Landau level), and other consistency requirements that will be discussed as we go along. The validity of this general form may also be checked rather directly in the case of Laughlin states, see Sec. \ref{laughlin}.

\emph{4. Determine transition functions describing a change of basis between dual coherent state descriptions}

A serious limitation of the coherent state ansatz mentioned in the preceding step is that its validity is restricted to quasiholes that are well separated along the quantization axis (as defined above).  However, there are two quantization axes at our disposal. These correspond to taking the opposite (mutually dual) thin torus limits, $L_y\rightarrow 0$ and $L_x\rightarrow 0$, respectively, giving rise to different ways of organizing the zero-mode space into adiabatically continued domain-wall states. These two ways of organizing the zero modes are related by a modular $S$ transformation of the torus, which essentially exchanges the roles of $x$ and $y$. We will say that the  corresponding different coherent state descriptions of local quasiholes are related by $S$ duality. $S$ duality allows us to write down a coherent state description for basically any local configuration of quasiholes. However, we will need to translate back and forth between mutually dual coherent state expressions along a braiding path. This change of basis is performed by matrix-valued transition functions, whose elements are sufficiently constrained by symmetry, topological considerations, and locality requirements to be discussed below.

\emph{5. Adiabatically move the quasiholes along a braiding path}

The coherent state ansatz together with the transition functions allows the calculations of adiabatic transport of quasiholes along a given exchange path. In all cases studied, this confirms that the result of braiding is purely topological up to an Aharonov-Bohm phase, even though this is by no means a basic assumption made in our approach.

In the following, we will demonstrate the utility of this method for increasingly complex quantum Hall states. We start by discussing the simplest fractional quantum Hall state, the $\nu=1/2$ (bosonic) Laughlin state.
\section{\label{laughlin}The Laughlin state} 
\subsection{\label{ttl}Thin torus limits}

Laughlin's $\nu=1/m$ wave functions \cite{laughlin} are the most elementary examples of a rich class of quantum Hall trial wave functions. These wave functions are generally characterized by a set of analytic requirements, 
the most basic of which enforces
that the wave function is entirely contained in the lowest Landau level (LLL). Laughlin's original construction of incompressible quantum liquids in a 2D planar geometry has been generalized by Haldane to states living on a sphere \cite{haldane_hierarchy} enclosing monopole charges
and to states on a torus \cite{haldanePRL85}. The torus construction has also revealed that the $\nu=1/m$ Laughlin state is $m$-fold degenerate on the torus, while it is nondegenerate on the sphere. 
The nontrivial torus degeneracy was later understood to be the hallmark of topological order \cite{wenniu}, 
and to be a necessary condition for  the presence of anyonic excitations \cite{Einarsson}.
Here we focus on the torus. Let $|\psi^c\rangle$, where $c=0\dotsc m-1$, denote the $m$ incompressible Laughlin-type ground-state wave functions  at filling factor $\nu=1/m$ on the torus. We may expand the states $|\psi^c\rangle$ in the basis of the LLL Fock space that is derived from the singleparticle basis $\varphi_n$: 
\be \label{Cmn}
	|\psi^c \rangle=\sum_{\{m_n\}} C_{\{m_n\}}|m_1, m_2\dotsc m_L\rangle\,.
\ee
Here, $m_n$ denotes the number of particles in the state $\varphi_n$, and we consider a system with a fixed number $L=L_xL_y/2\pi$ of flux quanta or LLL orbitals. For the time being, we will use $L_y$ to parameterize the aspect ratio of the torus. The coefficients $C_{\{m_n\}}$ depend on the $y$ perimeter $L_y$ of the torus. In the thin torus limit $L_y\rightarrow 0$, the states \eqref{Cmn} evolve into states dominated by a {\em single} pattern of occupancy numbers   ${\{m_n\}}$. E.g,  the state with $c=0$ evolves into the Fock state $|100\dotsc100\dots\rangle$ (where dots indicate that $1$'s are separated by $m-1$ zeros), and states with $c>0$ are obtained by repeated application of the translation operator $T_x$. $T_x$ is the many-particle version of the single particle translation operator $t_x$ discussed above, and acts on a thin torus pattern such as $100\dotsc100\dots$ as a right shift. For any value of the perimeter $L_y$, the Laughlin states $|\psi^c\rangle$ are ground states of a ``pseudopotential'' Hamiltonian \cite{haldane_hierarchy, trugman}, whose action within the LLL explicitly depends on $L_y$. 
The evolution of the states $|\psi^c\rangle$ with $L_y$ can be understood as the adiabatic evolution of the ground states of the pseudopotential Hamiltonian $H(L_y)$  as the parameter $L_y$ is slowly changed. 
This has been studied in some detail for $m=3$ in Ref. \onlinecite{seidel1}, where is was shown numerically that the gap above the ground states never closes as a function of $L_y$.

The thin torus states discussed here are formally identical to the Tau-Thouless states proposed in Ref. \onlinecite{TT}. When considered in the ``2D-limit'' $L_x=L_y=\infty$, these states do not have long range charge density wave (CDW) order. In contrast, the thin torus states considered here can be characterized as 1D CDW states breaking the translational symmetry of the system. This is so since in the thin torus limit, the LLL orbitals $\varphi_n$ are well separated by a distance $\kappa=2\pi/L_y$ (\Fig{orig_dual}), and the symmetry breaking pattern of occupancy numbers becomes visible as a CDW modulation. The findings of Ref. \onlinecite{seidel1} imply that the Laughlin states retain the CDW order of the thin torus limit on any torus with at least one of the dimensions $L_x$, $L_y$ finite. Related rigorous results have been discussed in Ref. \onlinecite{lieb}. However, as long as both $L_x$ and $L_y$ are large compared to the magnetic length, the CDW order is exponentially small. The physics of the incompressible fluid is thus quickly approached as $L_x$, $L_y$ become large, and in particular the notion of braiding statistics can be made arbitrarily well defined on a large but finite torus. This, together with the fact that the states on such a torus are adiabatically connected to simple product states sharing all their essential quantum numbers, is the foundation of the method discussed here.

For simplicity we will now focus on the case $m=2$, the bosonic $\nu=1/2$ Laughlin state with ground-state patterns $101010\dotsc$ and $010101\dotsc$, respectively. The general case was worked out in Ref. \onlinecite{seidel_lee}. However, here we will discuss an improved variant of the method, which was used in Ref. \onlinecite{seidel_pfaffian}
to derive the statistics of the Pfaffian state. The two degenerate $m=2$  Laughlin states on the torus are the  unique zero-energy eigenstates of the $\hat V_0$ Haldane pseudopotential at filling factor $\nu=1/2$. As in other cases where parent Hamiltonians for incompressible trial states are known, further zero-energy states exist at smaller filling factors: The excitations associated with elementary quasihole-type excitations are in one-to-one correspondence with the zero modes
of the parent Hamiltonian at filling factor $\nu<1/2$. This is again true at any value of the perimeter $L_y$, and in particular the number of zero modes for any fixed number of constituent particles (electrons) $N$ does not depend on $L_y$. We will extend the assumption of adiabatic continuity to the entire zero-mode sector. The thin torus limit of a Laughlin state with $n$ quasiholes can easily be worked out directly from the $L_y\rightarrow 0$ limit of the Hamiltonian \cite{seidel1}, or from the same limit of the wave function on the torus or cylinder \cite{haldane_rezayi94}. A state with a single Laughlin quasihole evolves into a thin torus state that has a single domain wall between the two ground-state patterns. We can distinguish domain-wall states in two ``topological sectors", according to the two possible phases of the charge density wave to the left and to the right of the domain wall, i.e., $1010\dw0101...$ or $01010\dw010...$ . The 1D domain walls can be ascribed a fractional charge by means of the usual ``Su-Schrieffer'' counting argument \cite{suschrieffer}. This charge (here $1/2$) generally agrees \cite{seidel1, karlhede2} with the charge of Laughlin quasiholes, as it should by adiabatic continuity.

We introduce notation $\left| a,c \right)$ for LLL product states with a domain wall at position $a$ in topological sector $c$:
\begin{subequations} \label{laugh_barestates0}
   \begin{align}
	\label{laugh_bare_01} 
	\left| a,0 \right) &=  
	\left| \dots 1010101010\dw01010101010\dots \right) \\
	\label{laugh_bare_02} 
	\left| a,1 \right) &= 
	\left| \dots 01010101010\dw0101010101\dots \right)
   \end{align}
\end{subequations}
The curved ket indicates that these are ``bare" product states to be distinguished from states that have undergone adiabatic evolution, which we will discuss below. The number $a$ is a half-odd integer labeling the  domain-wall position relative to the LLL orbitals, such that $a\pm 1/2$ are the orbital indices of the LLL orbitals adjacent to the domain wall. The two possible values of the topological sector label $c$ distinguish the sequence of ground-state patterns in the two states of Eq. \eqref{laugh_barestates}. It is worth noting that in principle, the topological sector is already determined by the value of $\exp(i\pi{a})$ and so the notation of Eq. \eqref{laugh_barestates0} may seem slightly redundant. We find it advantageous, though, to include the topological sector information explicitly into the sector label, especially with regard to more general cases discussed later.

\begin{figure} 
	\includegraphics[width=\columnwidth]{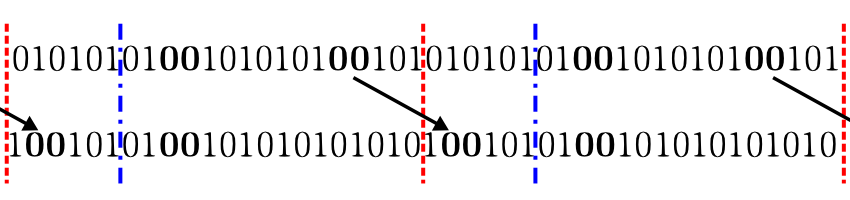}
	\caption{Top: A possible arrangement of two domain walls is shown in a 
		``repeated zone scheme'' with the domain-wall positions marked by the bold $\mathbf{00}$ strings. 
		The dotted line (red) marks the boundaries of the $0$ frame and the dot-dashed line (blue) 
		marks a shifted frame, the $6$ frame. Bottom: The second domain wall moves to a new 
		position. When viewed from the $0$ frame, this domain wall moves across the frame 
		boundary where it becomes the first domain wall in a different topological sector. 
		Viewed from the $6$ frame, the domain wall does 
		not move across the boundary and the topological sector does not change.} 
	\label{frames}
 \end{figure}

The above observations immediately generalize to states with two quasiholes, whose thin torus limits are given by product states corresponding to patterns with two domain walls. These states are labeled $\left| a_1,a_2,c \right)$, with occupation number patterns for the values of $c=0,1$ given by:
\begin{subequations} \label{laugh_barestates}
   \begin{align}
	\label{laugh_bare_1} 
	\left| a_1,a_2,0 \right) &=  
	\left| \dots 1010\dw010101010\dw0101010\dots \right) \\
	\label{laugh_bare_2} 
	\left| a_1,a_2,1 \right) &=  
	\left| \dots 01010\dw010101010\dw010101\dots \right)
   \end{align}
\end{subequations}
We will always take $a_1$ to be less than $a_2$, such that $a_1$ and $a_2$ refer to the first and second domain wall, respectively. It is clear from Eq. \eqref{laugh_barestates} that the two domain-wall positions are also subject to the constraint
\be \label{constraint}
	a_2-a_1=1 \mod 2\;.
\ee
Again, the label $c$ explicitly distinguishes the two possible sequences of ground-state patterns, even though in principle this information is also contained in the values of  $\exp(i\pi{a_1})$ or $\exp(i\pi{a_2})$. The labels $a_1$, $a_2$, and $c$ describing a given two--domain-wall state are unique when the condition
\be \label{frame0}
	0< a_1<a_2<L
\ee
is imposed. Whenever the domain-wall positions satisfy \eqref{frame0}, we will say that they are given ``in the default frame". However, since we are working on the torus and LLL orbitals satisfy the periodic boundary condition $\varphi_n\equiv\varphi_{n+L}$, it is desirable to admit domain-wall positions that refer to more general reference frames also. We thus define the states $\left| a_1,a_2,c \right)$ for all $a_1$, $a_2$ satisfying
\be \label{frame}
	a_1<a_2<a_1+L\;,
\ee
together with the following identification:
\be \label{identify}
	\left| a_1,a_2,c \right) \equiv \left| a_2-L,a_1,c' \right)\;,
\ee
where $c'=1+c \text{ mod }m$ (here m=2). We will say that the domain-wall positions $a_1$, $a_2$ lie in an $f$ frame if
\be\label{fframe}
	f< a_1<a_2<f+L\;.
\ee
The standard frame is the $0$ frame. If necessary, repeated application of \Eq{identify} allows one to transform domain-wall positions between different frames, where the roles of the first and second domain wall may be exchanged; whenever this happens, the topological sector label $c$ changes also, as stated in \Eq{identify}. 
This fact follows from \Eq{constraint}, since the value of $c$ is determined by the value of the position of, say, the first domain wall modulo 2, as discussed above. Note that $L=2N+2$ is even for states with two domain walls. The topological sector label is therefore frame dependent. This is of a piece with the fact that the topological sector changes when one quasihole is transported around one of the ``holes'' of the torus, as we will discuss in detail below (see \Fig{frames}).
The transformation properties of topological sectors under the exchange of two quasiholes along nontrivial loops (going once around the torus) are thus encoded in the thin torus patterns. This is a key ingredient of the method presented here, and sector transformation rules analogous to \Eq{identify} will be of much importance especially in the non-Abelian states to be discussed below.
\subsection{Delocalized quasihole states\label{deloc}}
\begin{figure} 
	\includegraphics[width=.8\columnwidth]{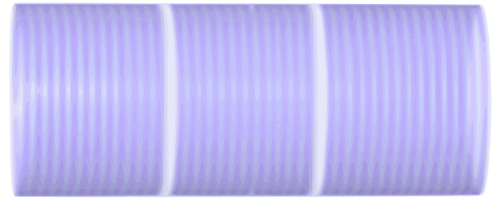}
	\caption{A depiction of a two-quasihole dressed domain-wall state. These 
		states are adiabatically evolved from the ``bare'' thin torus domain-wall states but live on the 
		full-sized torus. The quasiholes described by this state are localized at some 
		position in the $x$ direction but delocalized in the $y$ direction.} 
	\label{delocalizedholes}
 \end{figure}

The notion of braiding is not well defined in the thin torus limit. 
In order for a well-defined statistics to emerge 
from an adiabatic exchange of quasiholes, 
throughout the exchange the quasiholes must be spatially localized in both $x$ and $y$, 
and at the same time must be kept away from each other at distances large compared to their individual spatial extent. 
Both are not simultaneously possible in the thin torus limit. Hence, in order to ``braid'' quasiholes through adiabatic transport, we will need to work with states that live not on a thin torus but on a full-sized torus with $L_x,L_y$ both large. Formally, the assumption of adiabatic continuity means the following. There exists a family of unitary operators $\hat S(L_y,L_y')$ that describe the adiabatic evolution of the eigenstates (in particular the zero modes) of the pseudopotential Hamiltonian at perimeter $L_y'$ into those at $L_y$. In particular, we define $\hat S(L_y)\equiv \hat S(L_y,0)$, the unitary operator that evolves thin torus states, Eqs.  \eqref{laugh_barestates0}, \eqref{laugh_barestates}, into states at finite $L_y$. We hence define the ``dressed'' or adiabatically evolved domain-wall states as the descendants of thin torus states via the operator $\hat S(L_y)$.
In particular, for states with a single domain wall, we write 
\be\label{dressed}
	\left| a,c,L_y \right\rangle=\hat S(L_y) \left| a, c \right)\;,
\ee
where we will suppress the label $L_y$ whenever no confusion can arise, using the regular ket to denote dressed states as opposed to bare domain-wall states. For sufficiently large $L_y$ ({\em and} $L_x=2\pi L/L_y$), the states in \Eq{dressed} describe a quasihole immersed into a (here: $\nu=1/2$) Laughlin liquid. The quasihole is localized in $x$ around $x=\kappa a$. However, it is entirely delocalized in the $y$ direction. To see this, we consider the operator $T_y$ which is the many-body analogue of the single-particle translation operator $t_y$ discussed above. The bare domain-wall states are $T_y$ eigenstates by construction, with eigenvalues that are easily calculated from the pattern of occupation numbers. Since the pseudopotential Hamiltonian commutes with the magnetic translation operators for any value of $L_y$, so does the adiabatic evolution operator $\hat S(L_y)$. It follows that the dressed domain-wall states 
transform under magnetic translations in the same manner as the bare ones do.
The states in \Eq{dressed} are thus still $T_y$ eigenstates, with eigenvalues identical to those of their bare counterparts. It is clear that in such a state, the quasihole must be completely delocalized in the $y$ direction (see \Fig{delocalizedholes}). 
Again, these observations can be extended to states with two quasiholes,
\be\label{dressed2}
   \begin{split}
	\left| a_1, a_2,c,L_y \right\rangle&=\hat S(L_y) \left| a_1,a_2, c \right)\;.
   \end{split} 
\ee
Here, two Laughlin quasiholes in the topological sector $c$ are localized in $x$ around $x_1=\kappa a_1$ and $x_2=\kappa a_2$, respectively, and are both delocalized in $y$.   Note that the $x$ separation between the two quasiholes depends on $L_y$ via $\Delta x= \kappa\Delta a=2\pi(a_2-a_1)/L_y$. The two delocalized quasiholes in the state $|a_1,a_2,c,L_y\rangle$ will be uncorrelated as long as $\Delta x$ is much larger than a magnetic length (set equal to $1$). There are certainly no such correlations in the thin torus limit, and even at finite $L_y$ both the correlation length of the incompressible fluid and the range of the interaction remain on the order of a magnetic length. As we increase $L_y$, the adiabatic evolution will therefore not induce any correlations between the two quasiholes as long as $\Delta x\gg 1$ remains satisfied. In this case, the local properties of each of the quasiholes will be the same as those of the single quasihole
described by \Eq{dressed}.

We emphasize once more that the adiabatically continued domain-wall states in Eqs. \eqref{dressed} and \eqref{dressed2} are neither simple product states, nor are they any longer ``thin torus states'' in any sense. 
Rather, the assumption of adiabatic continuity  
allows one to organize the zero-mode subspace 
into a basis labeled by 1D patterns for any value of $L_y$. These patterns carry information about the properties under magnetic translations not only of the thin torus states, but also of their adiabatically descended counterparts at finite $L_y$. Finally, it will be of some significance that, since the adiabatic evolution operator $\hat S(L_y)$ is unitary, the dressed states of Eqs. \eqref{dressed} and \eqref{dressed2} are orthonormal, since the thin torus product states certainly are.
\subsection{\label{cstates}Coherent states}
Individually, the members of the basis of zero-mode states defined above 
describe delocalized Laughlin quasiholes. In order to analyze the braiding statistics of these quasiholes, we need to form states where quasiholes are localized in both $x$ and $y$.
Laughlin has constructed analytic wave functions for such states \cite{laughlin},
which are 
also zero-energy eigenstates of the pseudopotential Hamiltonian. It must therefore be possible to write these localized quasihole states as superpositions, or coherent states, in the zero-mode basis defined in the preceding section. 

We consider the single-quasihole case first. According to the above, it must be possible to write
\be \label{1hole}
	|\psi_c(h)\rangle=\sum_a C(h,a) |a,c\rangle
\ee
for a state with a quasihole localized at complex coordinate $h=h_x+ih_y$. Here, we anticipate that to localize a quasihole, it is sufficient to include states of a single topological sector into the superposition, such that the localized quasihole state still carries a well-defined sector label. The left-hand side of \Eq{1hole} is assumed to be a Laughlin single-hole state. Interestingly, as long as we assume that a zero-mode basis $|a,c\rangle$ with the properties claimed in the preceding section exists, the coefficients $C(h,a)$ of this expansion are fully determined. To this end, we note that 
\be \label{ortho}
	(a',c|a,c\rangle=\mbox{const}\,\times\,\delta_{a,a'}\;.
\ee
The vanishing of \Eq{ortho} for $a\neq a'$ follows since for different domain-wall positions the bare state $|a',c)$ and the dressed state $|a,c\rangle$ have different $T_y$ eigenvalues, as is easily seen by writing out the corresponding domain-wall patterns and calculating the action of $T_y$. On the other hand, the constant in \Eq{ortho} does not depend on $a$, since states with different domain-wall position $a$ are related by repeated application of $T_x$. From Eqs. \eqref{1hole} and \eqref{ortho}, it follows that
\be \label{Cha}
	C(h,a)\propto (a,c|\psi_c(h)\rangle\,.
\ee
We also expect only those states $|a,c\rangle$ to have any appreciable weight in the coherent state \eqref{1hole} whose domain-wall position $x=\kappa a$ is close to the $x$ position $h_x$ of the quasihole. 
We will assume that the coefficients  $C(h,a)$ in this region are not affected by a change from periodic to open boundary conditions, 
as long as the torus is cut  into a cylinder by a cut along $y$ that is far away in $x$ from the quasihole. 
In particular, it is clear from the discussion in Sec. \ref{LLstructure} that such a cut would affect the local structure of the $\varphi_n$ LLL basis (in terms of which the states $|a,c\rangle$ have been defined) only by negligible amounts (for large $L_x$). For cylindrical topology, however, it is possible to evaluate the right-hand side of \Eq{Cha} explicitly. For definiteness, we explicitly write out the wave function for the Laughlin state $|\psi_c(h)\rangle$ on a cylinder of perimeter $L_y$:
\be \label{laughwf}
	\psi_c(h;z_1\dotsc z_N)=\prod_i(\xi_i-\eta)\,\prod_{i<j} (\xi_i-\xi_j)^{2}\,\times\,e^{-\frac 12 \sum_i x_i^2}\,.
\ee
Here, $\xi_i=\exp(\kappa z_i)$ and $\eta=\exp(\kappa h)$. 
Evaluating \Eq{Cha} amounts to evaluating the coefficients of ``dominance patterns'' in the polynomial
of \Eq{laughwf}. This can be done using ``squeezed lattice'' methods discussed in Refs. \cite{seidelyang,seidelyang2}.  This shows that the above wave function does indeed lie in a definite topological sector, as defined by the thin cylinder limit.\footnote{The other sector can be reached by multiplication with $\prod_i \xi_i$.}  One finds:\footnote{An $h_y$-dependent phase has been dropped for simplicity. As a result, note that the original Laughlin state \Eq{laughwf} is single valued in $h_y$, whereas \Eq{1hole2} is not.}
\be \label{1hole2}
	|\psi_c(h)\rangle={\cal N} \sum_a \phi(h,\kappa a) |a,c\rangle\,,
\ee
where
\be \label{phi}
	\phi(h,x) = \exp \left[ \frac{1}{2} i (h_y + \pi/\kappa) x - \frac{1}{4} (h_x - x)^2 \right] \,,
\ee
and $\cal N$ is a normalization constant independent of $h$. The general form of the coherent state wave function \Eq{phi} could have been guessed based on the following observations. As a function of $x$, $\phi(h,x)$ can be interpreted as a ``minimum uncertainty" coherent state of a particle confined to one spatial dimension. This is consistent with the fact that, after projection into a single Landau level, the $x$ and $y$ components of the position operator do not commute, but satisfy a position-momentum--type commutation relation $[x,y]\propto i$. $y$ position can thus be regarded as $x$ momentum, and vice versa. It is thus natural that the $y$ position of the quasihole enters as a momentum-like phase twist in Eqs. \eqref{1hole2}, \eqref{phi}. On the other hand, as a function of $h$, $\phi(h,x)$ looks like a lowest Landau level orbital of a charge $1/2$ degree of freedom in the same magnetic field that is felt by the underlying electrons. These heuristic considerations will later allow us to generalize the coherent state form \Eq{1hole2} to more complicated cases, where a direct derivation of the kind outlined here is not straightforward. 

The next logical step is to generalize the expression \eqref{1hole2} to states with two localized quasiholes. This is not difficult, as long as the two quasiholes at complex positions $h_1$ and $h_2$ are well separated along the $x$ axis, i.e., $h_{2,x}-h_{1,x}\gg 1$. In this case, we can argue that the presence of the one quasihole does not influence the other, and the natural generalization of the coherent state \Eq{1hole2} takes on the following form:
\be \label{2holes} 
	\left| \psi_c(h_1,h_2) \right> = 
	{\cal N}^2 \sideset{}{'}\sum_{a_1<a_2} \phi(h_1,\kappa a_1) \phi(h_2, \kappa a_2) \left| a_1,a_2,c \right> \;.
\ee
%
The function $\phi(h,x)$ is just as defined in \Eq{phi}. 
The prime in the above sum denotes the restriction of the domain-wall positions to values corresponding to the topological sector $c$. These are different for $a_1$ and $a_2$, as a result of \Eq{constraint}. To be precise, we can define the topological sector $c$ for two quasiholes via the following constraint on the domain-wall positions:
\be \label{constraint2}
	a_1=2n_1-1/2+c\,,\qquad a_2=2n_2+1/2+c
\ee
with integers $n_2\geq n_1$. By default, the sum in \Eq{2holes} is further restricted to domain-wall positions within the default frame, \Eq{frame0}. The restriction to a different frame according to \eqref{fframe} will be indicated by a subscript $f$, $\left| \psi_c(h_1,h_2) \right>_f$.

For as long as the condition $h_{2,x}-h_{1,x}\gg 1$ holds, \Eq{2holes} can be inferred from \Eq{1hole2} in a more formal way, using assumptions about the action of local operators on the adiabatically continued domain-wall basis. Locality arguments of this kind will play an important role in the following, and we will devote the next section to the development these arguments.
\subsection{\label{locality}Locality} 
It is useful to formalize  the assumptions that enter the factorized two-quasihole ansatz, \Eq{2holes}. This naturally leads to general assumptions about the matrix elements of local operators within the zero-mode basis of adiabatically continued domain-wall states defined above, which will be of further relevance in much of the following. 
Let  $\hat \rho(\vec r)$ be a local operator, localized at some position $\vec r = (r_x,r_y)$. We will later 
consider $\hat\rho(\vec r)$ to be the operator for the local charge density at $\vec r$, but for now we wish to
consider a generic (not necessarily single-particle) local operator. 
The action of this operator within the LLL Fock space depends on
the aspect ratio of the torus.
We first consider the action of $\hat \rho(\vec r )$ on a bare domain-wall state $|a_1,a_2,a_3,\dots,c)$
(which for finite $L_y$ is not an eigenstate of the pseudopotential Hamiltonian). 
Quite obviously, the operator $\hat \rho(\vec r )$ can only generate  matrix elements between this state
and some other domain-wall state $|b_1,b_2,b_3,\dots,c)$ if the associated pattern of orbital occupancy numbers differs only locally between these two states, for orbitals whose location lies within a magnetic length of $r_x$. We will usually be interested in cases where the domain-wall positions $\kappa a_1, \kappa a_2, \kappa a_3,\dots$ are all separated by much more than a magnetic length. In this case, for the matrix element between these two states to be finite, it is clear that either $a_i=b_i$ for all $i$, or there is a single $j$ such that  $a_j\neq b_j$, with both $\kappa a_j$ and $\kappa b_j$  in the vicinity of $r_x$.
Otherwise the patterns associated with the two states would differ even 
in orbitals that are far removed from $r_x$ along the $x$ axis,
and their matrix element would be exponentially small. In particular, matrix elements between states in different topological sectors are not possible (in the thermodynamic limit). Although at large $L_y$, the dressed domain-wall states $|a_1,a_2,a_3\dots,c\rangle$ are quite different from their bare counterparts, they still describe topological defects inserted into the torus at $x$ positions $\kappa a_i$. We will assume here and in the following that if the associated patterns of two dressed domain-wall states differ by many microscopic degrees of freedom, then this is also true for dressed states themselves. 
In particular,
 if the patterns of two states differ
 in orbitals whose separation along the $x$ axis is large compared to one magnetic length,
 we assume that their matrix element for any local operator will be negligible. For states with well separated domain walls, the observation made above for bare states then extends to their dressed counterparts. I.e., non-zero matrix elements are of the form
\be\label{rhomatrix}
	\langle \dotsc a_i\dotsc|\hat \rho(\vec r)|\dotsc b_i\dotsc \rangle= \rho(a_i,b_i)\;,
\ee
where the ellipses represent other domain-wall positions,
 which must remain fixed but otherwise do not affect the value of the matrix element, and again $\kappa a_j\approx \kappa b_j \approx r_x$ to within a magnetic length. With these assumptions, we can easily show that \Eq{2holes} describes two localized quasiholes, assuming that \Eq{1hole2} describes a single localized quasihole. Let now
$\hat\rho(\vec r)$ be the local density operator. We consider the expectation value $\langle\psi_c(h_1,h_2)|\hat\rho(\vec r)|\psi_c(h_1,h_2)\rangle$ for $|h_{2x}-r_x|\gg 1$, and show that this expectation value reduces exactly to that of $\langle\psi_c(h_1)|\hat\rho(\vec r)|\psi_c(h_1)\rangle$, which we know to describe a single quasihole at position $h_1$. Using \Eq{rhomatrix}, we have
\begin{widetext}
\be
   \begin{split}
	\langle\psi_c(h_1,h_2)|\hat\rho(\vec r)|\psi_c(h_1,h_2)\rangle 
	&= {\cal N}^4 \sideset{}{'}\sum_{a_1,a_2}\sideset{}{'}\sum_{b_1,b_2}
	\phi(h_1,\kappa a_1)^\ast \phi(h_2, \kappa a_2)^\ast 
	\phi(h_1,\kappa b_1) \phi(h_2, \kappa b_2)\left\langle a_1,a_2,c\right| \hat\rho(\vec r) \left| b_1,b_2,c \right> \\
	&\simeq{\cal N}^4\sideset{}{'}\sum_{a_1,a_2,b_1}\phi(h_1,\kappa a_1)^\ast \phi(h_2, \kappa a_2)^\ast 
	\phi(h_1,\kappa b_1) \phi(h_2, \kappa a_2)\left\langle a_1,a_2,c\right| \hat\rho(\vec r) \left| b_1,a_2,c \right> \\
	&={\cal N}^4\sideset{}{'}\sum_{a_1,a_2,b_1}\phi(h_1,\kappa a_1)^\ast 
	\phi(h_1,\kappa b_1)\phi(h_2, \kappa a_2)^\ast  \phi(h_2, \kappa a_2)\,\rho(a_1,b_1)\\
	&\simeq{\cal N}^2\sideset{}{'}\sum_{a_1,b_1}\phi(h_1,\kappa a_1)^\ast  \phi(h_1,\kappa b_1)\, \rho(a_1,b_1)\;.
   \end{split}
\ee
\end{widetext}
In the above, the primes on the sums enforce all the necessary constraints such that the bras and kets correspond to domain-wall patterns in the topological sector $c$, cf. \Eq{constraint2}. In the second line, we have used that the matrix elements are diagonal in the second domain-wall position for $|h_{2x}-r_x|\gg 1$. Furthermore, for $h_{2x}-h_{1x}\gg 1$ the constraint $a_1,b_2<a_2$ which the domain-wall positions obey becomes irrelevant due to the Gaussian nature of the $\phi$ functions, and the sum over $a_2$ in the third line simply yields the normalization of the single-quasihole state, \Eq{1hole2}. The last line is, however, identical to $\langle\psi_c(h_1)|\hat\rho(\vec r)|\psi_c(h_1)\rangle$. In words, this shows that when $\vec r$ is far away along the $x$ axis from the second quasihole, the expectation value of $\hat\rho(\vec r)$ reduces to that of a state with a single quasihole at $h_1$. Similar arguments show that if $\vec r$ is far away along the $x$ axis from the first  quasihole, $\langle \hat\rho(\vec r)\rangle$ reduces to that of a state with a single quasihole at $h_2$. Together, this shows that for $h_{2x}-h_{1x}\gg 1$, the state \eqref{2holes} describes two quasiholes localized at $h_1$ and $h_2$.
\subsection{\label{duality}Dual description} 

The coherent state expression \eqref{2holes} is in principle suited to calculate the Berry connection governing 
adiabatic transport \cite{berry,simon,wilczek}. However, as the arguments in the preceding section have made clear,
 \Eq{2holes} can be expected to be accurate only in the limit of quasiholes that are well separated along the 
$x$ axis. As can be seen in \Fig{exchange}, the $x$ separation of the quasiholes must vanish at some point for any
exchange path, even though the absolute distances between the quasiholes remain large throughout.
As a result, \Eq{2holes} is by itself not sufficient to fully calculate the result of adiabatic transport.

The resolution to this problem lies in making use of the modular $S$ invariance of the torus. 
Though we have so far only used the thin torus limit $L_y\rightarrow 0$, the physics must be invariant under
an exchange of $x$ and $y$. In doing so, we may now define a zero-mode basis 
by working from the limit $L_x\rightarrow 0$. In this limit, the  zero modes of the pseudopotential Hamiltonian are domain-wall states that are occupation number eigenstates in the $\overline\varphi_n$ basis. The corresponding ground-state and domain-wall patterns are the same as those appearing in the $L_y\rightarrow 0$ limit, except that the associated charge density waves extend along the $y$ direction of the torus. We denote the bare domain-wall states in the $\overline\varphi_n$ basis with an overline, e.g. $\overline{\left| a_1,a_2, c \right)}$ for a two--domain-wall state. We now proceed in a manner that is completely analogous to the definition of the ``original'' zero-mode basis on a general torus, \Eq{dressed2}. To this end, we define a unitary operator ${\overline{S}}(L_x)$ that describes the adiabatic evolution of states from the ``narrow $x$ limit'' to a finite value of $L_x$. We then define the general zero-mode basis for two-quasihole states via
\be\label{dualdressed}
	\overline{\left| a_1,a_2, c, L_x \right>} ={\overline{S}}(L_x) \overline{\left| a_1,a_2, c \right)}\;,
\ee
where again, we will drop the label $L_x$ on the left-hand side whenever no confusion is possible. The states in \Eq{dualdressed} describe quasiholes that are localized in $y$ but delocalized around the torus along $x$. Similar definitions are made for states with $n$ quasiholes. We can form localized quasihole states in a manner completely analogous to \Eq{2holes}. So long as \Eq{2holes} describes two localized quasiholes at positions $h_1$ and $h_2$ for any aspect ratio of the torus, invariance of the physics under exchange of $x$ and $y$ implies that the following expression will do the same in terms of the dual zero-mode basis \Eq{dualdressed}:
\be \label{2holesdual} 
	\overline{\left| \psi_c (h_1,h_2) \right>} 
	= {\cal\bar N}^2 \sum_{a_1<a_2} \overline{\phi}(h_1,\overline{\kappa} a_1) 
	\overline{\phi}(h_2, \overline{\kappa} a_2) \overline{\left| a_1,a_2,c \right>} 
\ee
where 
\be
	\overline{\phi}(h,y) = \phi(-i h, y)|_{\kappa \rightarrow \overline{\kappa} } 
	= \exp \left[ - \frac{i}{2}(h_x+\pi/\overline{\kappa})y - \frac{1}{4}(h_y - y)^2 \right] \;, 
\ee
and \Eq{2holesdual} is now applicable to the case $ h_{2y}-h_{1y}\gg 1$.
We thus have at least one valid coherent state expression for any configuration of the two quasiholes along the exchange path shown in \Fig{exchange}. At some points along the path, however, we will be forced to translate back and forth between the two coherent state expressions \eqref{2holes} and \eqref{2holesdual}. This task is nontrivial. To see this, it is important to note that the topological sector label $c$ has different meanings in the original zero-mode basis \Eq{dressed2} and the dual zero-mode basis \Eq{dualdressed}: in the former, it means that the state evolves into a well defined charge density wave product state in the limit $L_y\rightarrow 0$, characterized by a certain sequence of ground-state patterns separated by domain walls; in the latter, it means the same in the opposite thin torus limit, $L_x\rightarrow 0$. It will turn out that a state that carries a definite sector label $c$ in the original basis, \Eq{dressed2}, is a superposition of states carrying {\em different} topological sector labels in the dual basis \Eq{dualdressed}, and vice versa. The same is true for the coherent state expressions Eqs. \eqref{2holes} and \eqref{2holesdual}. While the relation between the sets of states ${\left| \psi_{c }(h_1,h_2) \right>}$ and  $\overline{\left| \psi_{c }(h_1,h_2) \right>}$ is thus not diagonal in the topological sector label $c$, for given quasihole coordinates $h_1$, $h_2$ both sets span the same subspace, namely the space associated with having quasiholes localized at $h_1$, $h_2$. The relation between the states ${\left| \psi_{c }(h_1,h_2) \right>}$ and  $\overline{\left| \psi_{c }(h_1,h_2) \right>}$ is thus diagonal in the quasihole positions, and we may write
\be \label{udef} 
	\left| \psi_c(h_1,h_2) \right> = 
	\sum_{c'} u^\sigma_{cc'}(h_1,h_2)\overline{\left| \psi_{c'}(h_1,h_2) \right>} \;.
\ee
Note that above, we had defined $|\psi_c(h_1,h_2) \rangle $ only for $h_{2x}>h_{1x}$, and $\overline{\left| \psi_{c'}(h_1,h_2) \right>} $ only for $h_{2y}>h_{1y}$. While we will stick to these restrictions most of the time, we will generally let $|\psi_c(h_2,h_1) \rangle\equiv|\psi_c(h_1,h_2) \rangle$ and $\overline{\left| \psi_{c'}(h_2,h_1) \right>}\equiv\overline{\left| \psi_{c'}(h_1,h_2) \right>} $ for convenience. This allows us to write relations such as \Eq{udef} without distinguishing different cases. The transition functions $u^\sigma_{cc'}(h_1,h_2)$ are then meaningful in regions where both $|h_{1x}-h_{2x}|\gg 1$ {\em and} $|h_{1y}-h_{2y}|\gg 1$, since it is only in these regions where we have defined both  $|\psi_c(h_1,h_2) \rangle $ and $\overline{| \psi_{c'}(h_1,h_2) \rangle} $ through coherent state expressions. The final technical obstacle is to sufficiently determine these transition functions from symmetries and topological considerations.

\begin{figure} 
	\includegraphics[width=0.75\columnwidth]{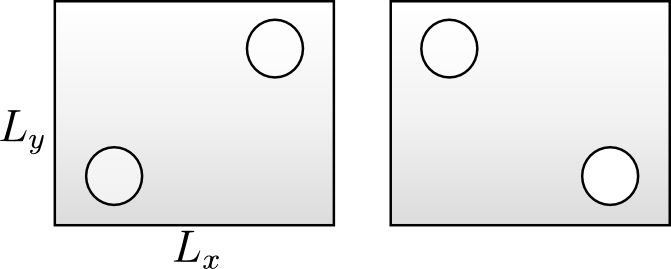}
	\caption{The two possible configurations of two quasiholes, which are distinguished by the 
		value of $\sigma=\text{sgn}(h_{1x}-h_{2x})\text{sgn}(h_{1y}-h_{2y})$. 
		Left: $\sigma=+$. Right: $\sigma=-$.} 
	\label{config}
 \end{figure}

To this end, we begin by distinguishing two regions of the 2-hole configuration space. Let $\sigma=\text{sgn}(h_{1x}-h_{2x})\text{sgn}(h_{1y}-h_{2y})$. $\sigma=\pm 1$ then refers to first and second quasihole configuration in \Fig{config}, respectively. We will first be interested in the ``local'' dependence of the transition functions on coordinates within each of these regions. Later we will use the fact that these regions are actually connected by  ``global'' trajectories where one quasihole is taken around one of the holes of the torus (\Fig{globalpaths}). For now we will not allow these global moves. Within each of these regions, we now show that the local dependence of the $u$ functions on coordinates is as follows,
\be \label{laugh_def_xi} 
	u^\sigma_{cc'}(h_1,h_2) 
	= \xi^\sigma_{cc'} \,e^{\frac i2 (h_{1x}h_{1y}+h_{2x}h_{2y}) } 
	\equiv \xi^\sigma_{cc'}u(h_1,h_2)\,,
\ee
where the parameters $\xi^\sigma_{cc'}$ are complex constants and $u(h_1,h_2)$ is the phase function $e^{\frac{i}{2}(h_{1x}h_{1y}+h_{2x}h_{2y})}$.

The $h_1,h_2$ dependence of $u^{\sigma}_{cc'}$ can be locally determined from the Berry connections. Using the coherent state expressions in Eqs. \eqref{2holes} and \eqref{2holesdual} on the full-sized torus ($\kappa, \overline{\kappa} \ll 1$), the Berry connections can be calculated
to be
\begin{align} \label{berrycons}
   \begin{split}
	i \left< \psi_c(h_1,h_2) \right| \boldsymbol{\nabla}_{h_{1,2}} \left| \psi_c(h_1,h_2) \right> 
	&= -\frac{1}{2} ( 0 , h_{1x,2x} ) \\
	i \overline{\left< \psi_c(h_1,h_2) \right|} \boldsymbol{\nabla}_{h_{1,2}} \overline{\left| \psi_c(h_1,h_2) \right>} 
	&= \frac{1}{2} ( h_{1y,2y} , 0 ). 
   \end{split}
\end{align}
An essential ingredient in the above is the fact that the zero-mode basis states 
we have defined are orthonormal, as explained at the end of Sec. \ref{deloc}.
This is where the assumption of adiabatic continuity is crucial in our approach.
Obtaining \Eq{berrycons} is then straightforward, since in the limit
$\kappa, \overline{\kappa} \ll 1$, the remaining sums can be replaced by
Gaussian integrals.

Let us consider an adiabatic process where one quasihole is fixed at $h_1$ and the other is dragged from $h_2$ to $h_2'$ (which are both in the same region $\sigma$).
This process is described by a unitary operator,
which acts separately on each term on both sides
 of  \Eq{udef}, yielding
\begin{widetext}
\be \label{uloc}
	\exp{ \left(i \int_{h_2}^{h_2'} \mathrm{d}\boldsymbol{h}\cdot 
	\left[ -\frac{1}{2}\left( 0, h_x \right)\right]\right) } \left| \psi_c(h_1,h_2') \right> 
	= \sum_{c'} u^{\sigma}_{cc'}(h_1,h_2) \, \exp{ \left(i \int_{h_2}^{h_2'} \mathrm{d}\boldsymbol{h}\cdot 
	\left[ \frac{1}{2}\left( h_y, 0 \right)\right]\right) } \, \overline{\left| \psi_{c'}(h_1,h_2') \right>} \;.
\ee
\end{widetext}
The above equation may be compared to
 \Eq{udef} evaluated at $(h_1,h_2')$ instead of $(h_1,h_2)$. 
 This yields a relationship between the $u$ functions at these two locations,
\be \label{uloc2}
   \begin{split}
	u^{\sigma}_{cc'}(h_1,h_2') &= u^{\sigma}_{cc'}(h_1,h_2) \, 
	\exp{\left( i \int_{h_2}^{h_2'} \mathrm{d} \boldsymbol{h} \cdot 
	\left[\frac{1}{2} \boldsymbol{\nabla}_h h_xh_y \right] \right)} \\
	&= u^{\sigma}_{cc'}(h_1,h_2) \, \exp{ \left(\frac{1}{2}i (h_{2x}'h_{2y}' - h_{2x}h_{2y}) \right)}
   \end{split}
\ee
where we used the fact that $\left( h_y, 0 \right) = -\left( 0, h_x \right) + \boldsymbol{\nabla}_h h_xh_y$. In order to satisfy \Eq{uloc2}, the dependence of $u$ on $h_2$ must be proportional to $e^{\frac{1}{2}ih_{2x}h_{2y} }$. Using a similar argument in which the quasihole at $h_2$ remains fixed while the quasihole at $h_1$ is moved, we find that the dependence of $u$ on $h_1$ is proportional to $e^{\frac{1}{2}ih_{1x}h_{1y} }$. Therefore the general form of the $u$ functions is given by \Eq{laugh_def_xi}.
\subsection{\label{symmetries}Symmetries and further simplifications} 
With the above considerations, the transition functions $u^\sigma_{cc'}$ have been reduced to parameters $\xi^\sigma_{cc'}$, of which there are eight at $\nu=1/2$. We will now establish further relations between these parameters using symmetries and adiabatic transport along the ``global'' trajectories mentioned above.

First, we derive relations arising from properties under magnetic translations. The magnetic  many-body translation operators $T_x$, $T_y$ introduced above have the following effect on the dressed domain-wall states:
\be \label{Tx}
   \begin{split}
	T_x \left| a_1,a_2,c \right> & = \left| a_1+1,a_2+1,1-c \right> \\
	T_x \overline{\left| a_1,a_2,c \right>} & = e^{\frac{2\pi i}{L}\sum_j n_j}\overline{\left| a_1,a_2,c \right>} \,
   \end{split} 
\ee
\be \label{Ty}
   \begin{split} 
	T_y \left| a_1,a_2,c \right> & =  e^{-\frac{2\pi i}{L}\sum_j n_j}\left| a_1,a_2,c \right> \\
	T_y \overline{\left| a_1,a_2,c \right>} &=  \overline{\left| a_1+1,a_2+1,1-c \right>}
   \end{split}
\ee
where $c=0,1$, and $n_j$ is the orbital index of the orbital occupied by the $j$-th particle in the thin torus pattern associated with the state. For the bare product states associated with these patterns, the above identities are direct consequences of Eqs. \eqref{t} for the single particle translation operators. 
However, the properties under magnetic translations remain the same for the dressed states, as explained in Sec. \ref{deloc}.
Note that the basis states $\left| a_1,a_2,c \right>$ are eigenstates of $T_y$ whereas $T_x$ changes the topological sector label, and vice versa for the basis states $\overline{\left| a_1,a_2,c \right>}$.

Equations \eqref{Tx} and \eqref{Ty} allow us to work out the properties of the coherent states under magnetic translations. The fact that both sides of \Eq{udef} must transform the same way under these translations poses severe constraints on the coefficients $\xi^\sigma_{cc'}$. 
Observing that for given domain-wall positions,
\be
\sum_j n_j = \frac 1 2 L ( \frac 1 2 L + c ) - \frac 1 2 ( a_1+a_2 ),
\ee
it is a simple thing to verify the
following properties of the coherent states under magnetic translations:
\be \label{Tx_coherent}
   \begin{split}
	T_x \left| \psi_c (h_1,h_2) \right> &= 
	e^{-\frac{1}{2}i \kappa(h_{1y}+h_{2y})+i\pi} \left| \psi_{1-c} (h_1+\kappa,h_2+\kappa) \right>  \\
	T_x \overline{\left| \psi_c (h_1,h_2) \right>} &= 
	e^{i \pi \lambda + i \pi c} \overline{\left| \psi_c (h_1+\kappa,h_2+\kappa) \right>}
   \end{split} 
\ee
\be \label{Ty_coherent}
   \begin{split}
	T_y \left| \psi_c (h_1,h_2) \right> &= 
	e^{ i \pi \lambda+i \pi c} \left| \psi_c (h_1+i\overline{\kappa},h_2+i\overline{\kappa}) \right> \\
	T_y \overline{\left| \psi_c (h_1,h_2) \right>} &= 
	e^{ \frac{1}{2}i \overline{\kappa}(h_{1x}+h_{2x})+i\pi} 
	\overline{\left| \psi_{1-c} (h_1+i\overline{\kappa},h_2+i\overline{\kappa}) \right>}
   \end{split} 
\ee
where we define $\lambda = \nu L$ (which in the present case, evaluates to the {\em integer} $L/2=N+1$).

We can use these translational properties to constrain the eight $\xi_{cc'}^\sigma$s. We recast \Eq{udef} in matrix form,
\be \label{laugh_Xidef}
	\left( \begin{array}{c}
		\left| \psi_0 (h_1,h_2) \right\rangle \\
		\left| \psi_1 (h_1,h_2) \right\rangle
	\end{array} \right) =
	u(h_1,h_2) \Xi^\sigma
	\left( \begin{array}{c}
		\overline{\left| \psi_0 (h_1,h_2) \right\rangle} \\
		\overline{\left| \psi_1 (h_1,h_2) \right\rangle}
	\end{array} \right) ,
\ee
where we have used \Eq{laugh_def_xi}, and  $\Xi^\sigma$ is the matrix with elements $\xi^\sigma_{cc'}$. Let us apply $T_y$ to \Eq{laugh_Xidef}.
\begin{IEEEeqnarray}{rCl} 
	\IEEEeqnarraymulticol{3}{l}{
		e^{i \pi \lambda} \sz
		\left( \begin{array}{c}
			\left| \psi_0 (h_1',h_2') \right\rangle \\
			\left| \psi_1 (h_1',h_2') \right\rangle 
		\end{array} \right)}\nonumber\\
	\quad &=&
		u(h_1',h_2') \Xi^\sigma
		(e^{i \pi}) \sx
		\left( \begin{array}{c}
			\overline{\left| \psi_0 (h_1',h_2') \right\rangle} \\
			\overline{\left| \psi_1 (h_1',h_2') \right\rangle}
		\end{array} \right) \label{laugh_XiTy}
\end{IEEEeqnarray}
The positions $h_j'=h_j+i\overline\kappa$ for $j=1,2$, and the $u(h_1,h_2)$ function has been shifted by absorbing the spatially dependent phase in \Eq{Ty_coherent}. If we compare \Eq{laugh_XiTy} to \Eq{laugh_Xidef} evaluated at the shifted positions $(h_1',h_2')$, 
we find that the two equations are consistent, provided that the $\Xi^\sigma$ matrix satisfies the following constraint:
%
\be \label{laugh_Xicon1}
	\Xi^\sigma = e^{i\pi\lambda+i\pi}
	\sz \Xi^\sigma \sx .
\ee
We can derive another constraint using the same logic after translating \Eq{laugh_Xidef} with $T_x$:
\be \label{laugh_Xicon2}
	\Xi^\sigma = e^{i\pi\lambda+i\pi}
	\sx \Xi^\sigma \sz.
\ee
These two sets of equations constrain the $\Xi^\sigma$ matrix to be of the following form,
%
%
\be \label{Xi_mat}
	\Xi^\sigma=
	\frac{\xi^\sigma}{\sqrt2}
	\left(  \begin{array}{cc}
		1 &  e^{i \pi \lambda+i\pi} \\
		e^{i \pi \lambda+i\pi} & -1
	\end{array} \right),
\ee
where $\xi^\sigma$ is a pure phase, and the overall normalization factor $1/\sqrt{2}$ has been determined from
the requirement that $\Xi^\sigma$ is a unitary matrix.
Thus, after using translations we have only two unknowns remaining, the overall phases $\xi^+$ and $\xi^-$. 
It is only the relative phase between the two that will have physical significance.

\begin{figure} 
	\includegraphics[width=.75\columnwidth]{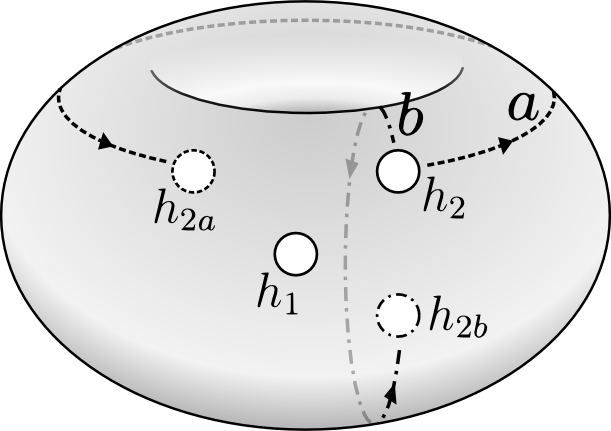}
	\caption{Different configurations $\sigma$ can be connected by dragging one quasihole 
		along a global path. Initially, the two quasiholes at $h_1$ and $h_2$ are in 
		configuration $\sigma=+$. Keeping the quasihole at $h_1$ fixed, the quasihole 
		at $h_2$ can be moved along one of two paths: path $a$, in which the quasihole at $h_2$ 
		moves around the torus in the $x$ direction to $h_{2a}$, 
		or path $b$, in which the quasihole at $h_2$ moves around the torus in the $y$ direction to 
		$h_{2b}$. Both paths can be used to change the configuration $\sigma$ while keeping 
		 quasiholes well separated in both $x$ and $y$. At the same time, the topological sector 
		  also changes. } 
	\label{globalpaths}
 \end{figure}
In order to fix this relative phase, we will now drag one of the quasiholes in a two quasihole state along a ``global path'', i.e., a path where the quasihole disappears on one end of the standard frame (see Sec. \ref{laughlin} and \Fig{frames}) and reappears at the other. 
The merit of such a path is that it connects the $\sigma=+$ and $\sigma=-$ configuration while 
maintaining both conditions $|h_{1x}-h_{2x}|\gg 1$, $|h_{1y}-h_{2y}|\gg 1$.
Let us consider the coherent state $|\psi_c(h_1, h_2)\rangle$, \Eq{2holes}, with two quasiholes in the topological sector $c$ in the $\sigma\!=\!+$ configuration. We will drag the second quasihole along path ``$a$'' as shown in \Fig{globalpaths}. We will do so by continuously changing the position of this quasihole from a value $h_2^i$ with $h^i_{2x}$ well within the boundaries $0$ and $L_x$ to a value $h^f_2$ with $L_x<h^f_{2x}<h_{1x}+L_x$. The default frame introduced in Sec. \ref{ttl} is not suited to describe this process continuously.  We thus choose an $f$ frame as described  in Secs. \ref{ttl} and \ref{cstates}, and consider the state $|\psi_c(h_1, h_2)\rangle_f$, i.e., the coherent state \eqref{2holes} with the sum restricted to the $f$ frame. For this we choose a parameter $f$ such that $\kappa f<h_{1x}<h^i_{2x}<h^f_{2x}<\kappa(f+L)=\kappa f+L_x$.
Note that as long as the $x$ position $h_{2x}$ of the second quasihole is well between $h_{1x}$ and $L_x$, one has $|\psi_c(h_1, h_2)\rangle_f\doteq|\psi_c(h_1, h_2)\rangle$, where $\doteq$ denotes equality up to exponentially small terms. In this case the weight of both Gaussians in the coherent state is well contained within both frames, and so $|\psi_c(h_1, h_2)\rangle_f$ and $|\psi_c(h_1, h_2)\rangle$ may be used interchangeably. However, as soon as $h_{2x}$ approaches $L_x$, we must work with $|\psi_c(h_1, h_2)\rangle_f$. In this regime, we will see that the coherent state $|\psi_c(h_1, h_2)\rangle_f$ is identical up to a phase to the (default frame) state $|\psi_{c'}(h_2-L_x, h_1)\rangle$. That is, the second quasihole reappears on the left end of the standard frame, thus becoming the new `first' quasihole (Figs. \ref{frames} and \ref{globalpaths}). However, in the default frame the final state will be in a different topological sector with $c'=1-c$. At the same time, the quasiholes are now in the $\sigma=-$ configuration. This allows us to obtain one more relation between the transition functions $u^\sigma_{cc'}$ and their defining parameters $\xi^\sigma_{cc'}$.

We first establish the precise relationship between $|\psi_c(h_1, h_2)\rangle_f$ and $|\psi_{c'}(h_2-L_x, h_1)\rangle$, where $h_{2x}$ exceeds $L_x$ by more than a magnetic length. One finds:
\begin{align} \label{psi_trafo}
	|\psi_c(h_1, h_2)\rangle_f
	&= e^{\frac{1}{2}i h_{2y}L_x+i\pi \eta +i \pi } |\psi_{1-c}(h_2-L_x, h_1)\rangle_{f-L} \nonumber\\
	&\doteq e^{\frac{1}{2}i h_{2y}L_x+i\pi \eta +i \pi } |\psi_{1-c}(h_2-L_x, h_1)\rangle
\end{align}
where in the first identity we have passed to the ${f-L}$ frame by straightforwardly plugging the identification \eqref{identify} into the coherent state \eqref{2holes}. The second identity follows from the fact that for $h_{2x}$ well exceeding $L_x$, the states $|\psi_{c'}(h_2-L_x, h_1)\rangle_{f-L}$ and $|\psi_{c'}(h_2-L_x, h_1)\rangle$ are again identical up to exponentially small terms, as discussed above.

Next we look at the comparatively trivial issue of how the dual state $\overline{|\psi_{c'}(h_1, h_2)\rangle}$ transforms along the same path, where $h_2$ is again taken from $h_2^i$ to $h_2^f$. Since the motion is chiefly along the $x$ direction, there is no need for a change of the frame for the $\overline{|a_1,a_2,c' \rangle}$ basis states. By inspection of \Eq{2holesdual}, it is easy to see that we have
\be \label{psibar_trafo}
      \overline{|\psi_{c'}(h_1, h_2)\rangle}=  e^{-i \pi (\frac{1}{2}+c')} \overline{|\psi_{c'}(h_1, h_2-L_x)\rangle} \,.
\ee
While the states $\overline{|\psi_{c'}(h_1, h_2)\rangle}$ are not single valued under a shift of quasihole positions by $L_x$, path $a$ in \Fig{globalpaths} can be described continuously without leaving the default frame. Since we have established that both  $|\psi_c(h_1, h_2)\rangle_f$ and $\overline{|\psi_{c'}(h_1, h_2)\rangle}$ describe states with quasiholes in the same position for $h_1$ fixed and $h_2$ along the path $a$ in \Fig{globalpaths}, a relation of the form 
\be \label{udef2} 
	\left| \psi_c(h_1,h_2) \right>_f = \sum_{c'} u^{+}_{cc'}(h_1,h_2)\overline{\left| \psi_{c'}(h_1,h_2) \right>} \;.
\ee
must again hold for (some neighborhood of) this path. It is clear that the coefficient functions $u^{+}_{cc'}(h_1,h_2)$ appearing in there must be the analytic continuation (for $h_{2x}>L_x$) of those already defined, since 1) the arguments leading to the functional dependence \Eq{laugh_def_xi} can be extended to the regime $h_{2x}>L_x$ and 2) for $h_{2x}<L_x$ the functions in \Eq{udef2} must be identical to those in \Eq{udef}. At the same time, for $h_{2x}>L_x$  we have by definition
\be \label{udef3} 
	\left| \psi_c(h_2-L_x,h_1) \right>= \sum_{c'} u^{-}_{cc'}(h_1,h_2)\overline{\left| \psi_{c'}(h_1,h_2-L_x) \right>} \;.
\ee
After plugging Eqs. \eqref{psi_trafo} and \eqref{psibar_trafo} into \Eq{udef2}, and further Eqs. \eqref{laugh_def_xi} and \eqref{Xi_mat} into both Eqs. \eqref{udef2} and \eqref{udef3}, comparing coefficients leads to the following additional relation between the $\xi$ parameters:
\be
	\xi^- = \xi^+ \, e^{-i \frac{\pi}{2}}
\ee 
All $\xi$ parameters are thus defined up to some overall phase $\xi$. We have
\begin{align} \label{laugh_xi_result}
	\Xi^{+} &=
	\frac{\xi}{\sqrt{2}}
	\left(  \begin{array}{cc}
		1 &  e^{i \pi \lambda+i\pi} \\
		e^{i \pi \lambda+i\pi} & -1
	\end{array} \right) \nonumber \\
	\Xi^{-} &=
	\frac{\xi}{\sqrt{2}} e^{-i\frac{\pi}{2}}
	\left(  \begin{array}{cc}
		1 &  e^{i \pi \lambda+i\pi} \\
		e^{i \pi \lambda+i\pi} & -1
	\end{array} \right) \,.
\end{align}

We note that processes similar to our moves along global paths play a fundamental role
in all studies of anyonic statistics on the torus (see, e.g., Ref. \onlinecite{Einarsson}).
Unlike in the present case, it is usually assumed from the beginning that these anyons
are entities carrying a representation of the braid group. Typically, complete
monodromies are considered, where the particle moves back into its original position
after following a path associated with one of the generators of the fundamental group of the torus.
In the present case, it is of some importance that these global moves end before the
quasihole crosses over back into a configuration labeled by the initial $\sigma$ value,
thus changing the value of $\sigma$.

\subsection{\label{braid}Braiding} 
\begin{figure} 
	\includegraphics[width=.75\columnwidth]{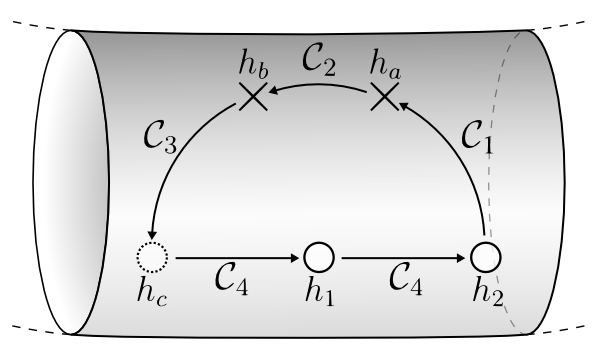}
	\caption{Exchange path for two quasiholes. First, the quasihole at $h_2$ is dragged 
		along path $\mathcal C_1$ to $h_a$. There the coherent state representation is 
		changed from the original basis to the dual basis using \Eq{udef}. The quasihole at $h_a$ 
		is then dragged along $\mathcal C_2$ to $h_b$, and the representation is changed 
		back to the original basis. The quasihole at $h_b$ is moved along $\mathcal C_3$ to 
		$h_c$. At this point both quasiholes are moved to their final positions: the quasihole at 
		$h_1$ goes to $h_2$ and the quasihole at $h_c$ goes to $h_1$.} 
	\label{exchange}
 \end{figure}
With the transition functions \Eq{udef} now fully defined via Eqs. \eqref{laugh_def_xi} and \eqref{laugh_xi_result}, the result of adiabatic transport along an exchange path as shown in \Fig{exchange} can be calculated without difficulty. We assume that in the beginning, the quasiholes are arranged at positions $h_1$ and $h_2$ as shown, with $h_{2x}-h_{1x}\gg 1$. The quasihole initially at $h_2$ is then dragged into the position $h_c$ directly opposite the other quasihole, via path segments ${\mathcal C}_1$, ${\mathcal C}_2$, ${\mathcal C}_3$ which are separated by points $h_a$, $h_b$. Finally, the quasihole at $h_1$ is moved into position $h_2$, and the other quasihole is moved from $h_c$ into $h_1$, completing the exchange. When the one quasihole  reaches the point $h_a$, we pass from the coherent state expression \eqref{2holes} to the dual expression \eqref{2holesdual} via the transition functions, and use the dual coherent state expression to calculate the adiabatic transport along the path segment ${\mathcal C}_2$. At the point $h_b$, the state is again re-expressed in terms of the original coherent state expression \eqref{2holes}, which may be used to describe the completion of the exchange.

Let the initial state be $|\psi_c(h_1, h_2)\rangle$, the state that lies in the topological sector $c$ as defined by the $L_y\rightarrow 0 $ limit. Adiabatic transport along the path ${\mathcal C}_1$ will change the coherent state according to
\be \label{laugh_braid_c1} 
	\left| \psi_c(h_1,h_2) \right> \rightarrow e^{i \gamma_1} \left| \psi_c(h_1,h_a) \right> 
\ee
where, using \Eq{berrycons},
\begin{align} 
	\gamma_1 &= i \int_{\mathcal C_1} d\boldsymbol{h}_2' \cdot 
	\left< \psi_c(h_1,h_2') \right| \boldsymbol{\nabla}_{h_{2}'} \left| \psi_c(h_1,h_2') \right>\nonumber \\
	&= \int_{\mathcal C_1} d\boldsymbol{h} \cdot \left[-\frac{1}{2}( 0, h_x ) \right]\,.\label{gam1}
\end{align}
At $h_a$ we reexpress the state in the dual basis, using Eqs. \eqref{udef} and \eqref{laugh_def_xi}:
\be \label{laugh_braid_trans1} 
	e^{i \gamma_1}\left| \psi_c(h_1,h_a) \right> =  
	e^{i \gamma_1} u(h_1,h_a) \sum_{c'} \xi_{cc'}^{+}\,\overline{\left| \psi_{c'}(h_1,h_a) \right>}\,.
\ee
We proceed by moving the same quasihole along the path segment ${\mathcal C}_2$. This process is easily described in terms of the dual basis states $\overline{\left|\psi_{c'}(h_1,h_2') \right>}$, which appear on the right-hand side of \Eq{laugh_braid_trans1}. In this basis the adiabatic process is simply described by the acquisition of a phase $e^{i\gamma_2}$, where, using again \Eq{berrycons},
\begin{align} 
	\gamma_2 &= i \int_{{\cal C}_2} d\boldsymbol{h}_2' \cdot
	\overline{\left< \psi_{c'}(h_1,h_2') \right|} \boldsymbol{\nabla}_{h_{2}'} 
	\overline{\left| \psi_{c'}(h_1,h_2') \right>}\nonumber \\
	&= \int_{{\cal C}_2} d\boldsymbol{h}\cdot \left[\frac{1}{2} ( h_{y}, 0 ) \right] \;,
\end{align}
which does not depend on the ``dual'' sector label $c'$. At the endpoint $h_b$ of  ${\mathcal C}_2$ we have thus transitioned into the state
\be \label{laugh_braid_c2} 
	e^{i \gamma_1+i\gamma_2} u(h_1,h_a) \sum_{c'} \xi^+_{cc'}\,\overline{\left| \psi_{c'}(h_1,h_b) \right>}\,.
\ee
The key observation is that this state is still in the topological sector $c$ as defined in the original coherent state basis, i.e., is of the form $|\psi_c(h_b, h_1)\rangle$ times a phase. To see this, note that the quasiholes are now in the $\sigma=-1$ configuration, and we have from \Eq{laugh_xi_result}
\be \label{laugh_statistics}
	\xi^+_{cc'}= e^{i\frac{\pi}{2}}\xi^-_{cc'}\,.
\ee
The state \eqref{laugh_braid_c2} can thus be rewritten as
\be \label{laugh_braid_c22} 
   \begin{split}
	&e^{i\pi/2} e^{i \gamma_1+i\gamma_2} u(h_1,h_a)u(h_b,h_1)^{-1} 
	\sum_{c'} u(h_b,h_1)\xi^-_{cc'}\,\overline{\left| \psi_{c'}(h_1,h_b) \right>}\\
	&=e^{i\pi/2} e^{i \gamma_1+i\gamma_2} u(h_1,h_a)u(h_b,h_1) ^{-1}{\left| \psi_{c}(h_b,h_1) \right>}
   \end{split}
\ee
The rest of the exchange path is trivially described using the coherent states $|\psi_c(h_1',h_2')\rangle$. The phase $\gamma_3$ associated with the path segment ${\mathcal C}_3$ is again given by an integral over a Berry connection of the form \Eq{gam1}. The final move along the ``baseline'' ${\mathcal C}_4$ is carried out by moving both quasiholes, one from $h_c$ into $h_1$, and the other from $h_1$ into $h_2$. The components of the Berry connection associated with each complex coordinate are, however, both of the same form, \Eq{berrycons}. For the remaining phases we thus get
 \be 
	\gamma_{3,4}= \int_{{\cal C}_{3}+{\cal C}_4} d\boldsymbol{h} \cdot \left[-\frac{1}{2}( 0 , h_{x} ) \right] \,.
\ee
The entire exchange process thus results in the following transformation of the state:
\be \label{laugh_braid_end} 
	\left| \psi_c(h_1,h_2) \right> \rightarrow 
	e^{i\pi/2}e^{i \sum_{i=1}^4 \gamma_i} u(h_1,h_a) u(h_b,h_1)^{-1}
	\left| \psi_c(h_1,h_2) \right>
\ee
As apparent from \Eq{laugh_def_xi}, the $u$ factors in the above equation equal $i(h_{ax}h_{ay}-h_{bx}h_{by})/2=-\frac{i}{2}\int_{{\mathcal C}_2} d\boldsymbol{h} \cdot (h_y, h_x)$. When combined with the expression for $\gamma_2$, all contour integrals can be combined into a single integral equal to the Aharonov-Bohm phase $\Phi_{AB}=\int_{\mathcal C} d\boldsymbol{h}\cdot \left[-\frac{1}{2}(0,h_x)\right]$, corresponding to a charge $-1/2$ particle moving in a unit magnetic field. We thus recover the well-known result \cite{ASW} that the exchange of two Laughlin quasiparticles results in the acquisition of a phase, which is equal to the sum of the Aharonov-Bohm phase and a purely topological, statistical part:
\be \label{laugh_braid_simplified} 
	\left| \psi_c(h_1,h_2) \right> \rightarrow 
	e^{i \Phi_{AB}} e^{i \frac{\pi}{2} } \left| \psi_c(h_1,h_2) \right>\,.
\ee

We emphasize once more that we did not assume \emph{a priori} that any aspect of this phase is topological. Rather, this result followed naturally from the coherent state ansatz Eqs. \eqref{2holes}, \eqref{2holesdual}, and the constraints we have derived. Note that one can read the statistical phase of $\pi/2$ directly off \Eq{laugh_statistics}, which relates the transition functions for different quasihole configurations. While we have focused on the simplest case of $\nu=1/2$ for clarity, the case $\nu=1/m$ can be treated by the same method through straightforward generalization \cite{seidel_lee}$^,$\footnote{Some care must be given
to fermion negative signs at odd denominator filling factors, in equations such as \eqref{Tx}, \eqref{Ty}, and \eqref{identify}. See Ref. \onlinecite{seidel_lee}.}.
\section{\label{pfaff}The Moore-Read state} 
\subsection{Generalized coherent state ansatz\label{gencoherent}}
An appealing aspect of the method developed above, thus far for Laughlin states, is that the Berry connections \Eq{berrycons} are trivial, i.e., essentially 
contributing
only to the AB-phase. In contrast, all aspects relating to the statistics are manifest in the transition functions (cf. \Eq{laugh_statistics}), which need to be evaluated only at two isolated points. This fact might suggest that the same method may be amenable to discuss non-Abelian states in relatively simple terms as well, if suitably generalized. That this is so has been shown in Ref. \onlinecite{seidel_pfaffian} for the special case of the Moore-Read (Pfaffian) state. In the following, we will review this method, emphasizing aspects that need nontrivial generalization when compared to the Laughlin case. We will later show that the same method may then, 
with little or no {further} modification, be applied to more complicated non-Abelian states also.

 The $\nu = 1$ (bosonic) Moore-Read, in planar geometry, is the state described by the following wave function:
\be
	\psi_\text{Pf}(z_1,\dots,z_N)=
	\text{Pfaff} \! \left[ \frac 1{z_i-z_j} \right] \prod_{i<j}(z_i-z_j)^2 e^{-\sum_i \abs{z_i}^2/4}
\ee
The torus degeneracy of this state is 3, and torus wave functions for the three ground states have been worked out in Ref. \onlinecite{greiterwenwilczek}. A program similar  to the one described for Laughlin states can now be implemented. A study \cite{seidel_pfaffian} of the special Hamiltonian \cite{greiterwenwilczek} associated with the Pfaffian state has demonstrated that again, the three ground states are adiabatically connected to a thin torus limit, in which the ground-state patterns $111111...$, $020202...$, and $202020...$ emerge.
 
The elementary quasihole-type excitations, which are again zero modes of the special Hamiltonian, turn out to evolve into charge 1/2 domain walls between $1111\dotsc$ and $2020\dotsc$ ground-state patterns. Periodic boundary conditions on the torus then require such domain walls to occur in even numbers. This observation is the thin torus statement of the well-known fact that the elementary Pfaffian quasiholes may only be created in pairs \cite{mooreread}. For the minimum number of two quasiholes, one thus has four topological sectors corresponding to the sequences of thin torus ground-state patterns shown in Table \ref{pf_sectors}.

As in the Laughlin case, we denote these two--domain-wall states $\left| a_1,a_2,c \right)$,  and their adiabatically continued counterparts by $|a_1,a_2,c\rangle$. We assume that a coherent ansatz of a form similar to \Eq{2holes} and its dual version \Eq{2holesdual} also describe localized quasiholes in this non-Abelian state. In particular, we assume a Gaussian form for the coherent state form factors $\phi(h, x)$ in the expression
\be \label{pf_psi_def} 
	\left| \psi_c(h_1,h_2) \right> = 
	{\cal N} \sum_{a_1<a_2} \phi(h_1,\kappa a_1) \phi(h_2, \kappa a_2) \left| a_1,a_2,c \right> 
\ee
for quasiholes well separated along the $x$ axis. A Gaussian form for $\phi(h,x)$ is essentially dictated by the fact that $x$ and $y$ are conjugate variables, as argued in Sec. \ref{cstates}. Unlike in the case of Laughlin quasiholes, however, we cannot extract all the parameters entering this expression from the analytic wave functions. Instead, we will have to rely more on symmetries and other consistency requirements to do this. We will thus initially assume $\phi(h,x)$ to be of the following generic form:
\be \label{phigeneral}
	\phi(h,x) = \exp \left[ i \beta (h_y + \delta/\kappa) x - \gamma (h_x - x)^2 \right] .
\ee
Unlike in the case of the Laughlin state, we cannot derive \Eq{phigeneral} analytically from the Pfaffian 2-hole wave functions \cite{mooreread, greiterwenwilczek}. We observe, however, that these wave functions are holomorphic in the quasihole positions $h_1$, $h_2$. We thus require the same for the coherent state \eqref{pf_psi_def}, except for an overall normalization factor that depends on the quasihole positions only (and in particular does not depend on the parameters $a_1$, $a_2$ in \Eq{pf_psi_def}). Equation \eqref{phigeneral} is certainly the simplest expression that satisfies all these requirements, and is consistent with the fact that $x$ and $y$ are conjugate variables, the latter implying that $y$ position enters as $x$ momentum. The discussion of Sec. \ref{cstates} then makes it natural to expect that, as a function of $h$, \Eq{phigeneral} should have the form of a LLL orbital for a charge $1/2$ degree of freedom in a unit magnetic field (for some choice of vector potential, and where boundary conditions in $h_y$ may be twisted). This implies $\beta=1/2$, $\gamma=1/4$, as for the $\nu=1/2$ Laughlin state.
We will show shortly that  $\beta=1/2$ also follows more rigorously from duality requirements. The parameter $\gamma$ merely controls the shape of the quasiholes. Its precise value will not be needed in the following.

Naively, it appears that the parameter $\delta$ can be formally absorbed into a shift of the coordinate origin. This is, however, not quite right. We will again require that there is a formally equivalent way to write two-hole states in the dual basis, defined as before via adiabatic evolution of domain-wall states:
\be \label{pf_psibar_def} 
	\overline{\left| \psi_{c }(h_i,h_j) \right>} = 
	{\cal N}' \sum_{a_1<a_2} \overline{\phi}(h_i,\overline{\kappa} a_1) \overline{\phi}(h_j, \overline{\kappa} a_2) 
	\overline{\left| a_1,a_2,c \right>} 
\ee
where 
\be \begin{split} \label{phibargeneral}
	\overline{\phi}(h,y) &= \phi(-i h, y)|_{\kappa \rightarrow \overline{\kappa} }  \\
	&= (\textrm{const}) \exp \left[ - i\beta(h_x+\delta/\overline{\kappa})y - \gamma (h_y - y)^2 \right] \; .
\end{split} \ee
It is clear that the formal equivalence between \Eq{pf_psi_def} and \Eq{pf_psibar_def} does not survive arbitrary shifts of the origin for the quasihole coordinates $h_1$, $h_2$. It is also clear that the coherent state expressions \eqref{pf_psi_def}-\eqref{phibargeneral} assume definite relations between the orbital indices in the LLL bases $\varphi_n$ and $\bar\varphi_n$, respectively, which define the properties of these orbitals under magnetic translations, \Eq{t}, and
determine the positions of these orbitals in space. 
\footnote{Note that a coordinate shift in particular changes both the magnetic vector potential and the quasiperiodic boundary condition in $x$ on wave functions. The constant $\Delta$ in $A=(0,x+\Delta)$ determines the locations of the LLL orbitals $\varphi_n$. An additional phase twist in the magnetic  boundary condition in $x$ does the same for the orbitals $\overline\varphi_n$.  In this sense, fixing $\Delta$ and the magnetic boundary conditions leads to a preferred set of coordinate systems on the torus, which up to scaling ($\kappa\rightarrow\bar\kappa$) is symmetric with respect to the LLL bases $\phi_n$ and $\bar\phi_n$. Here, the index $n$ is always defined via properties under magnetic translations, \Eq{t}.}
The choice of coordinate system, and its relation to the orbital indices, is also encoded in the definition of the symmetry operators $I$, $\tau$, $\bar\tau$, Eqs. \eqref{inversion}-\eqref{mirror2}, together with their geometric interpretation given above. We may use this to severely constrain the possible values of $\delta$. Indeed, these symmetries fix $\delta$ to be a multiple of $\pi$. Since the same conclusion will also emerge from duality arguments below, we will not pause here to show this in detail \cite{delta_note}.
The final result for the braid matrix will depend on $\delta$ only via $e^{2i\delta}$, which is fully determined
and equals unity.

There is one more parameter entering the generalized coherent state ansatz that is not yet explicit in Eqs. \eqref{pf_psi_def} and \eqref{pf_psibar_def}. This parameter enters when generalizing \Eq{constraint2}, which fixes the relation between the domain-wall positions $a_{1,2}$ entering the coherent states and 
an adjacent LLL orbital with index, e.g., $2n_{1,2}$.
 In the case of Laughlin states, a single domain wall has inversion symmetry, and this symmetry clearly demands that the position $a$ of this domain wall is defined as shown in \Eq{laugh_barestates}, i.e., as the position halfway in between the adjacent ground-state patterns. More precisely, it must be the distance $h_x-a$ between this domain-wall position and the $x$ position of a quasihole that suppresses the amplitude in the coherent states \eqref{1hole} or \eqref{2holes}. 
 There is no similar symmetry argument for the Pfaffian domain-wall patterns. Here, quasiholes must always come in pairs, as mentioned above.
Consider a 2-hole coherent state, \Eq{pf_psi_def}, in the topological sector $c=1$, Table \ref{pf_sectors}. It is clear that the domain-wall position $a_1$ entering the coherent state must be of the form $a_1=2n_1-s$, where $2n_1$ is the position of the first $0$ of the string, and $s$ is a shift parameter that defines the position of the domain wall relative to this leading $0$. For suitably chosen quasihole positions, an inversion symmetry leaving the coherent state invariant will map one quasihole onto the other. This does not fix the parameter $s$, but merely implies that the second domain wall must be assigned the position $a_2=2n_2+s$, where $2n_2$ is the position of the last $0$. In the topological sector $c$, we can thus write
\be \label{ai}
	a_i=2n_i+f_i(c)\,,
\ee
where $f_1(1)=-s$, $f_2(1)=+s$ as discussed above, and the values for $f_i(c)$ for $c>1$ can be related to those for $c=1$ by magnetic translations in $x$ as shown in Table \ref{pf_sectors}. Here, we have defined $\eta=0$ for even particle number $N$, $\eta=1$ for $N$ odd. Note that the even- or oddness of the particle number $N$ is just determined by the length of the $1111\dots$ string in the patterns of Table \ref{pf_sectors}.

Equations \eqref{pf_psi_def}-\eqref{phibargeneral}, together with the shifts in the domain-wall positions given by \Eq{ai} and Table \ref{pf_sectors}, define the generalized coherent state ansatz. We will now show that this ansatz can be used to make precise statements about the statistics of the Pfaffian, and other non-Abelian states.

\begin{table}[!tbp] 
   \begin{tabular}{c c c | c || c | c } 
	& $c$ && Thin torus pattern & $f_1(c)$ & $f_2(c)$  \\
	\hline
	& 1 && $1111111\underline{0}2020202\underline{0}1111111$ & $-s$ & $s$ \\
	& 2 && $1111111\underline{1}0202020\underline{2}0111111$ & $-s+1$ & $s+1$ \\
	& 3 && $0202020\underline{1}1111111\underline{0}2020202$ & $s-1$ & $-s+\eta$ \\
	& 4 && $2020202\underline{0}1111111\underline{1}0202020$ & $s$ & $-s+\eta+1$ \\
   \end{tabular} 
   \caption{Thin torus patterns for a two--domain-wall Moore-Read state, and the offset functions 
   	of those domain walls. The latter are defined in terms of the shift parameter $s$,
	and relate domain-wall positions $a_i$ to orbital positions $2n_i$ (underlined) 
	via $a_i=2n_i + f_i(c)$. Some offset functions depend on the particle number parity $\eta$,
	 with $\eta = 0$ ($\eta=1$) when $N$ is even (odd).}  
   \label{pf_sectors} 
 \end{table}
%

\subsection{\label{cons_pfaff1}The transition matrix: Constraints from translational symmetry} 

With the generalized coherent state ansatz in place, we continue by carrying out steps similar to those described in Secs.
\ref{duality} and \ref{symmetries} for Laughlin states. 
Equation \eqref{udef}, the general relation between the coherent state in the two mutually dual bases, can be carried over unchanged. Again, the transition matrices appearing in these relations are strongly constrained by translational symmetry. To utilize this, we first state some of the analogues of Eqs. \eqref{Tx_coherent}, \eqref{Ty_coherent}:
\begin{subequations}
   \begin{align}
	T_x \left| \psi_{c} (h_1,h_2) \right> &= 
	e^{-i\beta \kappa(h_{1y}+h_{2y})-2i\beta\delta} \left| \psi_{T(c)} (h_1+\kappa,h_2+\kappa) \right> \label{Tx1_pf} \\
	T_y \overline{\left| \psi_{c} (h_1,h_2) \right>} &= 
	e^{ i\beta \overline{\kappa}(h_{1x}+h_{2x})+2i\beta\delta} 
	\overline{\left| \psi_{T(c)} (h_1+i\overline{\kappa},h_2+i\overline{\kappa}) \right>}\,.\label{Ty1_pf}
   \end{align} 
\end{subequations}
\begin{table}[!tbp]  
   \begin{tabular}{ccc | c | c } 
	& $c$ && $T(c)$ & $F(c)$  \\
	\hline
	& 1 && 2 & $3+\eta$ \\
	& 2 && 1 & $4-\eta$ \\
	& 3 && 4 & 2  \\
	& 4 && 3 & 1 \\
   \end{tabular} 
   \caption{Transformation properties of the states shown in Table \ref{pf_sectors}. 
         Here, it is assumed
          that the sector $c$ refers to the original zero-mode basis, defined through the $L_y\rightarrow 0$ limit.
         Translating 
   	the state with $T_x$ would transition the state into sector $T(c)$. After dragging a quasihole 
	along the path $a$ in \Fig{globalpaths} the state would transition from sector $c$ into sector 
	$F(c)$, which is dependent on the particle number parity $\eta$.}  
   \label{pf_translations} 
 \end{table}
These properties again follow straightforwardly from the associated transformation properties of the dressed domain-wall states, Eqs. \eqref{Tx} and \eqref{Ty}. However, the relation of the shifted sector $T(c)$ to the original sector $c$ is different in the present case. These relations can easily be read off the patterns in Table \ref{pf_sectors} and are summarized in Table \ref{pf_translations}. 
The remaining two transformation laws 
depend more critically on the value of $\beta$, and allow us to determine its value. We focus on the action of $T_y$ on $ \left| \psi_{c} (h_1,h_2) \right>$ first. Since by duality, $ \left| \psi_{c} (h_1,h_2) \right>$  is a superposition of the states $\overline{\left| \psi_{c'} (h_1,h_2) \right>}$, \Eq{Ty1_pf} implies that
\be \label{Ty2_prelim}
	T_y \left| \psi_{c} (h_1,h_2) \right>=
	\left| \psi_{c} (h_1+i\overline{\kappa},h_2+i\overline{\kappa}) \right>\times \,\text{phase factor}\,.
\ee
Here, we have also used that $T_y$ does not change the topological sector $c$ when acting on $|a_1,a_2,c\rangle$, \Eq{Ty}. The left-hand side of the last equation is easily evaluated using \Eq{Ty} inside the coherent state expression. For $c=1$ domain-wall states, e.g., one finds $\sum n_i=\frac 12 L^2-\frac 12(a_1+a_2)$ for the sum in \Eq{Ty}. With this one finds that \Eq{Ty2_prelim} indeed holds, {\em provided that}
\be 
	\beta=1/2\,,
\ee
as anticipated earlier in the preceding section. With this, one then finds
\begin{subequations}
   \be
	T_y {\left| \psi_{c} (h_1,h_2) \right>} = 
	e^{i\pi N} 
	{\left| \psi_{c} (h_1+i\overline{\kappa},h_2+i\overline{\kappa}) \right>}\times \left\{
	\begin{array}{lr}
		1 &  \text{ for } c=1,2 \\
		-1& \text{ for }  c=3,4
	\end{array} \right. \;,
	\label{Ty2_pf}
   \ee
and similarly
   \be \begin{split}
   	T_x \overline{\left| \psi_{c} (h_1,h_2) \right>} = 
	e^{i\pi N} 
	\overline{\left| \psi_{c} (h_1+i\overline{\kappa},h_2+i\overline{\kappa}) \right>} \\
	\times \left\{
	\begin{array}{lr}
		1&  \text{ for } c=1,2\\
		-1& \text{ for } c=3,4
	\end{array} \right. \;.
	\label{Tx2_pf}
   \end{split} \ee
\end{subequations}
The relations worked out above impose strong constraints on the transition matrices $u_{cc'}(h_1, h_2)$ defined in \Eq{udef}. We apply $T_y$ to \Eq{udef} using Eqs. \eqref{Ty1_pf} (with $\beta=1/2$) and \eqref{Ty2_pf}. On the resulting equation, we use \Eq{udef} again, obtaining a relation between the coherent states $\overline{\left| \psi_{c} (h_1,h_2) \right>}$:
\be
   \begin{split}\label{urel_pf1}
	\chi(c) e^{i\pi N}  \sum_{c'} u^{}_{cc'}(h_1+i\overline{\kappa},h_2+i\overline{\kappa})
	\overline{\left| \psi_{c'}(h_1+i\overline{\kappa},h_2+i\overline{\kappa}) \right>} = \\
	\sum_{c'} u^{}_{cc'}(h_1,h_2) e^{i\overline{\kappa}(h_{1x}+h_{2x})/2+i\delta} 
	\overline{\left| \psi_{T(c')}(h_1+i\overline{\kappa},h_2+i\overline{\kappa}) \right>}\;,
   \end{split}
\ee
where $\chi(c)=1$ ($\chi(c)=-1$) for $c=1,2$ ($c=3,4$). For the local dependence of functions $u_{cc'}(h_1, h_2)$ on coordinates, \Eq{laugh_def_xi} can again be derived, using the same method as in Sec. \ref{duality}, assuming again $|h_{1x}-h_{2y}|\gg 1$, $|h_{1y}-h_{2y}|\gg 1$. When plugged into \Eq{urel_pf1}, the dependence on quasihole coordinates drops out, except for the dependence on the quasihole configurations shown in \Fig{config}, which is again denoted by $\sigma=\pm 1$. This gives the following equation for the coefficients $\xi_{cc'}^\sigma$, \Eq{laugh_def_xi},
\be\label{urel_pf2}
	\chi(c) e^{i\pi N-i\delta} \xi^{\sigma}_{cc'}=\sum_{c''} \delta_{T(c'),c''} \xi^\sigma_{cc''}\;,
\ee
where the linear independence of the kets in \Eq{urel_pf1} was used. For fixed $c$, $\sigma$, this can be looked at as an eigenvalue problem for the quantities $\xi^\sigma_{cc'}$, $c'=1\dotsc 4$. Obviously, solutions only exist if $\pm e^{i\pi N-i\delta}$ is an eigenvalue of the matrix $\delta_{T(c'),c''}$ on the right-hand side. This is only the case for
\be\label{delta}
	\exp(2i\delta)=1\;.
\ee
The coherent states are invariant, up to an unimportant phase, under $\delta\rightarrow \delta+2\pi$. Hence \Eq{delta} narrows possible values of $\delta$ down to two inequivalent possibilities. Our result for the statistics, however, will be the same for $\delta=0$ and $\delta=\pi$. We will thus keep $\delta$ as a parameter, but use \Eq{delta} wherever convenient.

Since the eigenvalues of $\delta_{T(c'),c''}$ are doubly degenerate, \Eq{urel_pf1} does not completely determine the coefficients $\xi^\sigma_{cc'}$. To this end, we must also consider the equation that is obtained by acting with $T_x$ on \Eq{udef}. In an analogous manner, this gives rise to the equation
\be\label{urel_pf3}
	\sum_{c''} \delta_{T(c),c''} \xi^\sigma_{c''c'}=   \chi(c') e^{i\pi N-i\delta} \xi^{\sigma}_{cc'}\;,
\ee
which differs from \Eq{urel_pf2} only by a replacement of the $\xi$ matrix by its transpose.

To explicitly solve the constraints \eqref{urel_pf2}, \eqref{urel_pf3}, the following transformation is useful. We define new topological sector labels $(\mu\nu)$, $\mu ,\nu=\pm 1$ via the following superposition of states carrying $c$ labels:
\be \label{pf_munu_def}
   \begin{split}
	\left| \psi_{\mu\nu} \right> &= 
	\frac{1}{\sqrt{2}} \left[ \left| \psi_{c=2-\nu} \right> + \mu e^{i\pi\eta-i\delta} \left| \psi_{c=3-\nu} \right> \right] \\
	\overline{\left| \psi_{\mu\nu} \right>} &= 
	\frac{1}{\sqrt{2}} \left[ \overline{\left| \psi_{c=2-\mu} \right>} + %
	\nu e^{i\pi\eta-i\delta} \overline{\left| \psi_{c=3-\mu} \right>} \right]\;,
   \end{split} 
\ee
where the dependence on $h_1$ and $h_2$ has been suppressed. The significance of the states $| \psi_{\mu\nu} \rangle$ is that under translations in both $T_x$ and $T_y$, they are now diagonal in the $\mu\nu$ label. Transition matrices $\tilde u_{\mu\nu,\mu'\nu'}$ and coefficients $\tilde\xi^\sigma_{\mu\nu,\mu'\nu'}$ can be defined analogous to Eqs. \eqref{udef} and \eqref{laugh_def_xi}, and are related to the quantities $u_{cc'}$ and $\xi^\sigma_{cc'}$ via the transformation \Eq{pf_munu_def}. In terms of the matrices $\tilde{\boldsymbol \xi}^\sigma$,  the constraints \eqref{urel_pf2}, \eqref{urel_pf3} read
\be\label{xiconstraint}
	\tilde{\boldsymbol \xi}^\sigma=
	{\mathbf D}\,{\tilde{\pmb \xi}^\sigma}{\mathbf D}={\mathbf D'}\,{\tilde{\pmb \xi}^\sigma}{\mathbf D'}\;,
\ee
where $\mathbf D$ and $\mathbf D'$ are the diagonal matrices $\mathbf D=\text{diag}(1,1,-1,-1)$ and $\mathbf D'=\text{diag}(1,-1,1,-1)$, respectively. It is clear from \Eq{xiconstraint} that only the diagonal elements of $\tilde{\boldsymbol \xi}^\sigma$ are unconstrained, whereas the remaining ones must vanish. The transition matrix is thus diagonal in the $\mu\nu$ basis. We write:
\be
	\tilde\xi^\sigma_{\mu\nu,\mu'\nu'}=\delta_{\mu,\mu'} \delta_{\nu,\nu'}\,\xi^\sigma_{\mu\nu}\;,
\ee
\be\label{diagonal}
\begin{split}
	\left| \psi_{\mu\nu}(h_1,h_2) \right>&=
	u_{\mu\nu}(h_1,h_2) \overline{\left| \psi_{\mu\nu}(h_1,h_2) \right>}\\
	&=\xi^\sigma_{\mu\nu}u(h_1,h_2)\overline{\left| \psi_{\mu\nu}(h_1,h_2) \right>}\;,
\end{split}
\ee
where $u(h_1,h_2)$ is as defined below \Eq{laugh_def_xi}, and no summation over indices is implied. We drop the tilde from now on, since there should be no confusion between the coefficient $\xi_{\mu\nu}^\sigma$ above and the coefficient $\xi_{cc'}^\sigma$ defined earlier. (Note again that $\mu\nu$ should be viewed as the single index of a diagonal matrix element). By unitarity of the transition matrixes, the $\xi_{\mu\nu}^\sigma$'s are pure phases.

The subscript $\mu\nu$ carries direct information about the properties of the states $\left| \psi_{\mu\nu}(h_1,h_2) \right>$, $\overline{\left| \psi_{\mu\nu}(h_1,h_2) \right>}$ under translation. From the definitions \eqref{pf_munu_def}, it is easily verified directly that
\begin{widetext}
\be \label{pf_TxTy_eig}
   \begin{split}
	\left< \psi_{\mu\nu} (h_1,h_2) \right | T_y \left| \psi_{\mu\nu} (h_1,h_2) \right>  &\approx
	\nu\,  e^{-\frac{i}{2}\overline{\kappa}(h_{1y}+h_{2y})+i\pi\eta}\approx
	\overline{\left< \psi_{\mu\nu} (h_1,h_2) \right |} T_y \overline{\left| \psi_{\mu\nu} (h_1,h_2) \right>}\;,\\
	\left< \psi_{\mu\nu} (h_1,h_2) \right | T_x \left| \psi_{\mu\nu} (h_1,h_2) \right>  
	&\approx\mu \, e^{\frac{i}{2}\kappa(h_{1x}+h_{2x})+i\pi\eta}
	 \approx \overline{\left< \psi_{\mu\nu} (h_1,h_2) \right |} T_x \overline{\left| \psi_{\mu\nu} (h_1,h_2) \right>}\;.
   \end{split} 
\ee
\end{widetext}
Since $T_x$, $T_y$ are unitary operators, an expectation value of almost unit modulus implies that the states $\left| \psi_{\mu\nu}(h_1,h_2) \right>$, $\overline{\left| \psi_{\mu\nu}(h_1,h_2) \right>}$ are, to good approximation, eigenstates of these operators, with the approximate eigenvalue given by the expectation value. Even though the $\left| \psi_{\mu\nu}(h_1,h_2) \right>$, $\overline{\left| \psi_{\mu\nu}(h_1,h_2) \right>}$ describe states of localized quasiholes, this is possible since $T_x$ and $T_y$ translate by distances $\kappa$ and $\overline\kappa$, respectively, which are small compared to the size of the quasiholes (on the order of a magnetic length). To the extent that we can regard these states as $T_x$, $T_y$ eigenstates, the different associated eigenvalues already imply that the transition functions must be diagonal in the $\mu\nu$ basis, \Eq{diagonal}.  This argument has been used in Ref. \onlinecite{seidel_pfaffian}. Naively, however, in treating the states $\left| \psi_{\mu\nu}(h_1,h_2) \right>$, $\overline{\left| \psi_{\mu\nu}(h_1,h_2) \right>}$ as $T_x$, $T_y$  eigenstates one neglects terms that scale as $1/\sqrt{L}$. The present treatment shows that no such approximation is necessary in deriving \Eq{diagonal}.

\subsection{\label{pf_paths}The transition matrix: Constraints from global paths} 
The transition functions are thus far described by eight unknown phase parameters $\xi^\sigma_{\mu\nu}$, \Eq{diagonal}. Each of these parameters describes the relation between the pair of coherent states $\left| \psi_{\mu\nu}(h_1,h_2) \right>$ and $\overline{\left| \psi_{\mu\nu}(h_1,h_2) \right>}$ within various patches of the two-hole configuration space. As already discussed in Sec. \ref{symmetries}, these patches may be connected through paths where one quasihole is dragged across a frame boundary, \Fig{globalpaths}. This then leads to relations between the $\xi$ parameters on different patches. In the case of the Laughlin state, all patches have been so connected, and there was only one independent parameter. It turns out that in the present case, the configuration space comes in two disjoint segments, which cannot be linked through paths as shown in \Fig{globalpaths}, or any paths that maintain the conditions that the two quasiholes remain well separated in both $x$ {\em and} $y$. 
 
Equation \eqref{psi_trafo} is straightforwardly generalized to the present case, following the same reasoning:
\be \label{psi_trafo_pf}
	| \psi_c(h_1, h_2) \rangle_f 
	\doteq e^{\frac{1}{2}i h_{2y}L_x+iL\delta/2 }\, |\psi_{F(c)}(h_2-L_x, h_1)\rangle\,.
\ee
Here again, $f$ denotes a frame that will allow us to extend $h_{2x}$ beyond $L_x$, which has been assumed in the above equation. Equation \eqref{psi_trafo} is just a special case of \Eq{psi_trafo_pf} for $L=2N+2$, $\delta=\pi$, as befits the $\nu=1/2$ Laughlin 2-hole state. For the $\nu=1$ Moore-Read state, however, one has $L=N+1$ in the presence of two quasiholes. Also, the function $F(c)$ assigns to $c$ the new sector that one enters when the second quasihole is dragged across the frame boundary along the path shown in \Fig{globalpaths}. The value of $F(c)$ can easily be read off the patterns that define the 2-hole sectors in Table \ref{pf_sectors}. Note however, that the patterns shown in the table correspond to the case of even particle number $N$, as the $1111$ strings are even in length. As a new feature, $F(c)$ depends on the particle number parity as shown in Table \ref{pf_translations}.

Likewise, \Eq{psibar_trafo} may be generalized to
\be \label{psibar_trafo_pf}
	\overline{|\psi_{c'}(h_1, h_2)\rangle}=  e^{-i \pi f_2(c')} \overline{|\psi_{c'}(h_1, h_2-L_x)\rangle} \,.
\ee
When analyzed in the $\mu\nu$ basis, \Eq{pf_munu_def}, both the above equations imply that the sector labeled $\mu\nu$ transitions into the sector labeled $\mu,-\nu$ when the quasihole with coordinate $h_2$ is dragged along path $a$ shown in \Fig{globalpaths}. Specifically, \Eq{psi_trafo_pf} implies
\begin{widetext}
\be \label{psi_trafo_munu}
	|\psi_{\mu\nu}(h_1, h_2)\rangle_f 
	\doteq e^{\frac{1}{2}i h_{2y}L_x+iL\delta/2 }\, |\psi_{\mu,-\nu}(h_2-L_x, h_1)\rangle
	\times \left\{
	\begin{array}{lr}
		1 & \text{for}\;  N \;\text{even}, \nu=1,\\
		\mu e^{i\delta+i\pi N} & \text{otherwise},
	\end{array} \right.
\ee
\end{widetext}
while \Eq{psibar_trafo_pf} gives
\be \label{psibar_trafo_munu}
	\overline{|\psi_{\mu\nu}(h_1, h_2)\rangle}=  e^{-i \pi f_2(2-\mu)} \overline{|\psi_{\mu,-\nu}(h_1, h_2-L_x)\rangle} \,.
\ee
Using  the same arguments given below \Eq{udef2}, we may apply  \Eq{diagonal} to an $f$-frame state $|\psi_{\mu\nu}(h_1, h_2)\rangle_f$ with two quasiholes in the $\sigma=+$ configuration:
\be
	|\psi_{\mu\nu}(h_1, h_2)\rangle_f = 
	\xi^+_{\mu\nu}  u(h_1,h_2)  \overline{|\psi_{\mu\nu}(h_1, h_2)\rangle}\,.
\ee
Here again $h_{2x}>L_x$, such that $(h_1,h_2)$ can be taken to be the final configuration of the path $a$ shown in \Fig{globalpaths}. Plugging in Eqs. \eqref{psi_trafo_munu} and \eqref{psibar_trafo_munu} gives a relation between the states $|\psi_{\mu,-\nu}(h_2-L_x, h_1)\rangle$ and $ \overline{|\psi_{\mu,-\nu}(h_1, h_2-L_x)\rangle}$. On the other hand, these equations are, by definition, related via
\begin{widetext}
\be
	|\psi_{\mu,-\nu}( h_2-L_x, h_1)\rangle_f = 
	\xi^-_{\mu,-\nu}  u(h_2-L_x,h_1)  \overline{|\psi_{\mu,-\nu}(h_1,h_2-L_x)\rangle}\,.
\ee
Comparing these two relations, recalling  $u(h_1,h_2)=e^{\frac{i}{2}(h_{1x}h_{1y}+h_{2x}h_{2y})}$, gives the following relation between $\xi_{\mu\nu}^+$ and $\xi_{\mu,-\nu}^-$. 
\be \label{pf_xi_rel1}
	\xi^-_{\mu,-\nu}=\xi^{+}_{\mu\nu}
	\,e^{-iL\delta/2 -i\pi f_2(2-\mu)}	\times \left\{
	\begin{array}{lr}
		1 & \text{for}\;  N \;\text{even}, \nu=1,\\
		\mu e^{-i\delta+i\pi N} & \text{otherwise},
	\end{array} \right.
\ee
We may also link patches of configuration space labeled by different $\mu$, $\nu$, and $\sigma$ by dragging one of the quasiholes along path $b$ shown in \Fig{globalpaths}. This is obviously a dual version of the process just considered, and by following completely analogous reasoning, we find the following relation complementing \Eq{pf_xi_rel1}:
\be \label{pf_xi_rel2}
	\xi^-_{-\mu,\nu}=\xi^{+}_{\mu\nu}
	 \,e^{-iL\delta/2 -i\pi f_2(2-\nu)}	\times \left\{
	\begin{array}{lr}
		1 & \text{for}\;  N \;\text{even}, \mu=1,\\
		\nu e^{-i\delta+i\pi N} & \text{otherwise},
	\end{array} \right.
\ee
\end{widetext}
The above two equations allow us to relate any of the parameters $\xi^\sigma_{\mu\nu}$ with the same value of $\sigma\mu\nu=\pm 1$. The transition functions have thus been reduced to two unknown phases, where only the relative phase will be of interest. 
Together with the shift parameter $s$, 
this phase will be determined in the final step by using the locality considerations of Sec. \ref{locality}.

\subsection{\label{pf_braid}Pfaffian braiding} 

Given that the transition functions are diagonal in the $\mu\nu$ basis (\Eq{diagonal}), the result of adiabatic exchange of the two quasiholes in the state $\left| {\psi}_{\mu \nu} (h_1,h_2) \right>$ is necessarily diagonal in this basis as well. Even in a non-Abelian state, it is of course possible to diagonalize any given generator of the braid group, which describes the (counter-clockwise) exchange of any two quasiholes. The phase picked up during the exchange will, however, depend on the index $\mu\nu$. Given the parameters $\xi_{\mu\nu}^\sigma$ defining the transition functions, we can calculate this phase in a manner that is completely analogous to that discussed in Sec. \ref{braid}. In particular, the expressions \eqref{berrycons} for the Berry connections carry over to the present case. The calculation is thus the same within each $\mu\nu$ sector. In particular, we recall that the statistical part
of the Berry phase could be directly read off  \Eq{laugh_statistics}. Equation \eqref{laugh_braid_simplified} can therefore be generalized to read
\be \label{pfaff_braid} 
	\left| \psi_{\mu\nu}(h_1,h_2) \right> \rightarrow 
	e^{i \Phi_{AB}}\,\frac{ \xi^+_{\mu\nu}}{\xi^-_{\mu\nu}}  \,\left| \psi_{\mu\nu}(h_1,h_2) \right>\,.
\ee
We denote by $\gamma_{\mu\nu}$ the topological part of this phase:
\be
	e^{i\gamma_{\mu\nu}}=\frac{ \xi^+_{\mu\nu}}{\xi^-_{\mu\nu}}\;.
\ee
By means of the relations \eqref{pf_xi_rel1} and \eqref{pf_xi_rel2} derived in the preceding section, it is clear that all phases $\gamma_{\mu\nu}$ can be related to $\gamma_{++}$. These relations depend both on the parameter $s$, as well as the particle number parity $\eta$. We must, therefore, distinguish the case of even ($\eta=0$) and odd ($\eta=1$) particle number $N$. In each case, using $L=N+1$ we find that only even multiples of $\delta$ enter, which are zero modulo $2\pi$. Hence the parameter $\delta$ does not enter the result, as anticipated earlier.
 For $N$ even (superscript $e$), we find:
\be\label{pf_gamma1}
	\gamma^e_{+-}=\gamma^e_{-+}=-\gamma^e_{++}+2\pi s,\quad \gamma^e_{--}=\gamma^e_{++}+\pi-4\pi s\;.
\ee
Likewise, for $N$ odd (superscript $o$), we find:
\be \label{pf_gamma2}
	\gamma^o_{+-}=\gamma^o_{-+}=-\gamma^o_{++}+2\pi s,\quad \gamma^o_{--}=\gamma^o_{++}-4\pi s\;.
\ee
There are thus three remaining parameters in the theory, which can be taken to be the phases $\gamma^e_{++}$ and $\gamma^o_{++}$, and the shift parameter $s$. It turns out that these parameters are highly constrained by locality considerations of the kind discussed in Sec. \ref{locality}.

The adiabatic transport of the quasiholes is facilitated by local potentials that pin the quasiholes to a certain location that gradually changes with time. The matrix elements of these local potentials in the dressed domain-wall basis  are subject to the general considerations for local operators made in Sec. \ref{locality}. From these considerations it follows that the patterns contributing to the coherent states before and after the quasihole exchange can only change for orbitals whose $x$ position ($\kappa n$, where $n$ is the orbital index) is within a magnetic length (plus the range of the local potentials) of the exchange path. 
Regions far to the left or right of the initial quasihole positions do not participate in the exchange process, i.e., orbitals in this region are far away from any point on the exchange path. According to the above, this implies that in this region, the pattern associated with dressed domain-wall states entering the coherent state is unaffected during the exchange process.

Let us consider the implications of this for the case where the initial state is in the sector labeled $c=3$, Table \ref{pf_sectors}. Since for a state initially in the $c=3$ sector, all patterns form one of the two possible $2020$ strings far to the left and far to the right of the quasiholes, this must also be true after the exchange process, with the $2020$ patterns unchanged.  This, however, implies that the state is still in the $c=3$ sector after the exchange. Identical observations can be made for the $c=4$ sector.

It is easy to translate these statements into the $\mu\nu$ basis. In order for the exchange process to be diagonal in the sectors $c=3$ and $c=4$, the phases $\gamma_{\mu\nu}$ must be independent of $\mu$ when $\nu=-1$. This is true for both even and odd particle number. We thus have:
\be\label{pf_loc1}
	e^{i\gamma^e_{+-}}=e^{i\gamma^e_{--}}\;,\quad  e^{i\gamma^o_{+-}}=e^{i\gamma^o_{--}}\;.
\ee
Note that in the case of even or odd particle number, the $1111$ strings linking domain walls in the sectors $c=3$, $c=4$ are even/odd in length, respectively. The locality assumptions made in Sec. \ref{locality} further imply that the matrix elements of local operators cannot depend on the length of the $1111$ string as long as the domain walls are well separated, since in this case such matrix elements do not depend on the separation of the quasiholes. In particular, this implies that the Berry connection is insensitive to particle number parity (which is solely encoded in the length of $1111$ strings) for well separated quasiholes. This is manifest in equations \eqref{berrycons} which hold independent of particle number. However, this reasoning breaks down for dressed domain-wall states whose domain walls are not well separated. Referring to the original basis $|a_1,a_2,c\rangle$, this happens when two quasiholes are {\em not} well separated in $x$. 
In this regime, it is reasonable to expect that matrix elements between dressed domain-wall states do depend on whether the (short) $1111$ strings of patterns entering the coherent states are even or odd in length. This is not manifest in our formulation, since in this regime, we always work with the dual $\overline{|a_1,a_2,c\rangle}$ basis. However, the transition functions that we calculated can be expected to ``know'' of these parity effects. 
Hence, we expect that the phases in \Eq{pf_loc1}, which describe braiding in the $c=3,4$ sectors, will depend on particle number parity. 

The situation is quite the opposite for the sectors $c=1$ and $c=2$. Here, locality requires that the string pattern to the far left and right of the dressed domain-wall states forming the coherent states remain of the $1111$ form before and after the exchange. This only forbids transitions from the sectors $c=1,2$ into the sectors $c=3,4$. This we already know from the fact that exchange processes are diagonal in the $\mu\nu$ basis, which followed from properties under translation. However, this does not forbid transitions between the sectors $c=1$ and $c=2$.

On the other hand, the $2020$ strings forming the links between domain walls in these sectors, and which become short during the exchange process, carry no information about the particle number parity. This information remains hidden in the $1111$ strings, which remain arbitrarily long during the exchange, in the limit of large $L$. We thus conclude that within the $c=1,2$ subspace, the braid matrix describing the result of the adiabatic exchange of the quasiholes is independent of particle number parity. In the $\mu\nu$ basis, this leads to the following requirements:
\be\label{pf_loc2}
	e^{i\gamma^e_{++}}=e^{i\gamma^o_{-+}}\;,\quad  e^{i\gamma^e_{-+}}=e^{i\gamma^o_{++}}\;.
\ee
Using Eqs. \eqref{pf_gamma1} and \eqref{pf_gamma2}, the latter reduce to the same equation, $\gamma_{++}^e+\gamma_{++}^o=2\pi s \mod 2\pi$. Equations \eqref{pf_loc1} give two more, $2\gamma_{++}^e=6\pi s-\pi \mod 2\pi$, and $2\gamma_{++}^o=6\pi s \mod 2\pi$. The solutions to these equations are of the form
\begin{subequations}
   \begin{align}\label{pf_solution}
	s &=\frac{3}{8}-\frac{r}{4}\\
	\gamma^e_{++}&=\gamma^o_{+-}=\gamma^o_{-+}=\gamma^o_{--}=
	\frac{5}{8}\pi-\frac{3}{4}\pi r \label{gamma_odd}\\
	\gamma^o_{++}&=\gamma^e_{+-}=\gamma^e_{-+}=\gamma^e_{--}=
	\frac{1}{8}\pi+\frac{1}{4}\pi r\,, \label{gamma_even}
   \end{align}
\end{subequations}
where $r\in\mathbb Z$. This amounts to eight inequivalent possible solutions for the statistics. To discuss the relation between these different solutions, we first generalize our result to the case of $2n$ quasiholes on the torus. This will show that up to unitary transformations (taking on the form of simple phase conventions), all solutions are related by overall Abelian phases. We will further obtain a useful pictorial representation of Pfaffian statistics, and relate it to more standard ones.

\subsection{\label{pf_braidgroup}Representation of the braid group of $2n$ particles}

\begin{figure} 
	\includegraphics[width=.8\columnwidth]{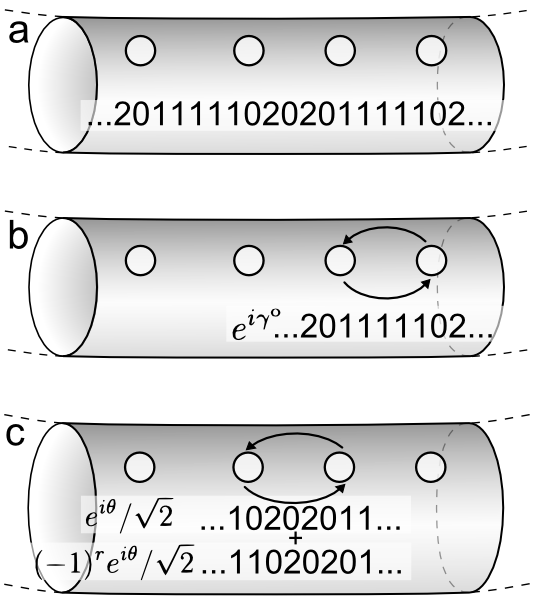}
	\caption{Graphical representation of the result of exchanging 
		two Pfaffian quasiholes for two representative pairs. a) A possible state in which four 
		quasiholes could be prepared, labeled by its associated thin torus pattern. 
		The state shown could be a four-quasihole state, in which the 20 strings at either 
		end would continue around the torus, or could be a $2n$-quasihole state for $n>2$, 
		in which the ellipses mask additional domain walls in the thin torus pattern. 
		The results of braiding any pair of quasiholes will be the same in either case. 
		In the following we show only the section of the pattern relevant to the exchange; 
		locality implies that only the segment of the pattern within a magnetic length of the 
		exchange path may be affected by the exchange and the rest remains fixed. 
		b) Upon exchange of the indicated quasiholes, the state picks up the phase 
		$\gamma^o$, given by \Eq{gamma_odd}. Had the 11 string separating the 
		quasiholes been even in length, the phase would have been $\gamma^e$, \Eq{gamma_even}. 
		In either case the thin torus pattern, and thus the topological sector of the state, 
		remains unchanged, which is shown. c) When the two indicated quasiholes are 
		exchanged the state remains in the same topological sector or transitions into a 
		sector with the linking 20 string shifted. The amplitudes for these two possibilities 
		are shown next to the thin torus patterns for the sectors, where $\theta=\pi(1/8+r/4+(-1)^r/4)$.
		$r$ is an integer labeling the eight possible values for the overall Abelian phase, where $r=0$
		 reproduces the representation given by conformal block monodromies.} 
	\label{pf_exchange_patterns}
 \end{figure}

The locality arguments used above immediately allow one to generalize the results obtained thus far for two quasiholes to the general case of $2n$ quasiholes. Consider the result of exchanging two quasiholes in a topological sector as defined by taking in the $L_y\rightarrow 0$ limit, e.g. \Fig{pf_exchange_patterns}. Such states are the analogue of the states $|\psi_c(h_1,h_2)\rangle$ defined above, generalized to $2n$ quasiholes. Locality then implies that the result of exchanging two quasiholes can at most affect the string linking the associated domain walls in the sector label. Furthermore, the presence of other quasiholes, which are assumed to be well away along the $x$ axis, cannot affect the result of the exchange. One can therefore infer the result of exchanging any two quasiholes in a state of $2n$ quasiholes from the results established above for states of two quasiholes. 

These results can be generally stated as follows:
\begin{itemize}
\item
If the two quasiholes to be exchanged are linked by a $1111$ string in the topological sector label, the state merely picks up a phase as a result of the exchange. This phase is given by \Eq{gamma_odd} when the linking $1111$ string is odd in length (\Fig{pf_exchange_patterns}b), and by \Eq{gamma_even} when the linking $1111$ string is even in length.
\item
If the two quasiholes are linked by a $2020$ string, then upon exchange, the state will remain in the same topological sector with an amplitude $e^{i\theta}/\sqrt{2}$, where $\theta=\pi(1/8+r/4+(-1)^r/4)$. It will transition with an amplitude $(-1)^r i e^{i\theta}/\sqrt{2}$ into the sector with the linking $2020$ string shifted.\footnote{Here, an additional phase factor $e^{i\delta}$ that was present in \Eq{pf_munu_def}, which would arise in the off-diagonal matrix
element with the conventions of the preceding sections, has been absorbed into a sign convention for the
adiabatically continued domain-wall state basis.} 
\end{itemize}

These rules are represented graphically in \Fig{pf_exchange_patterns}. To make connection with the standard way to represent these statistics \cite{nayakwilczek, ivanov,oppen,oshikawa}, we introduce a Majorana fermion degree of freedom $\eta_i$ associated with the $i$-th domain wall in the string patterns associated with our topological sectors. Let the pair $\eta_{2j}$, $\eta_{2j+1}$ be associated with the left and right domain wall of a $1111$ string. We then introduce fermion operators $c_j=\frac{1}{2}(\eta_{2j}+ i\eta_{2j+1})$. Each $c_j$ is now associated to a $1111$ string. The topological Hilbert space can be constructed by acting with the operators $c_j^\dagger$ on the vacuum of the $c_j$ operators, where states have the $j$-th fermion occupied if the $j$-th $1111$ string in the associated topological sector label is odd in length, and unoccupied otherwise. It is easy to check that according to the above rules,  the exchange of the $i$-th and $(i+1)$-th quasihole is then represented by the operator
\be
	e^{i\theta}\,e^{(-1)^r\frac{\pi}{4}\eta_i\eta_{i+1}}
\ee
within this fermionic space, as expected for the Pfaffian state. The sign of $\eta_i\eta_{i+1}$ in the above can be absorbed by a unitary transformation, facilitated by the operator $\prod_j \eta_{2j}$. With this, the non-Abelian part of the statistics is thus determined unambiguously by the present formalism, whereas for the overall Abelian phase $e^{i\theta}$, there are eight possible values. In the present case, these are all the values that are consistent with the $SU(2)_2$ fusion rules \cite{bonderson_thesis, kitaev06}. For $r=0$ one obtains the value that agrees \cite{nayakwilczek} with the transformation properties of the conformal blocks from which the Pfaffian many-body wave functions are constructed \cite{mooreread}. The approach discussed here is thus consistent with the CFT approach. 
For the Pfaffian case, the CFT approach was recently
 backed more rigorously through plasma analogy methods \cite{bonderson_gurarie_nayak}.
Similar results  
can also be obtained
from the $p+ip$-wave superconductor analogy \cite{ivanov, oppen, read_green}, although the present approach yields more information about the overall Abelian phase.
\section{\label{rr}$k=3$ Read-Rezayi}

We have seen that the method developed above is sufficiently general to obtain the statistics of Abelian FQH states, and, 
 with some adaptations, the non-Abelian Moore-Read state. 
Here we will show that the techniques developed in the preceding sections are indeed general enough 
to allow us, essentially without modification, to obtain the statistics of a more complicated non-Abelian state as well.
We will demonstrate this for the $k=3$ Read-Rezayi (RR) state \cite{readrezayi}.

Again, we focus on the bosonic ``root'' (highest filling factor, or $M=0$) state of the $k=3$ sequence. This state has $\nu=3/2$ and a torus degeneracy of 4. In taking the thin torus limit, the ground states are adiabatically evolved into the patterns $0303\dotsc$, $3030\dotsc$, $1212\dotsc$ and $2121\dotsc$ \cite{haldanebernevig,Ardonne2008}. Elementary excitations evolve into charge $1/2$ domain walls between the $3030$ and $2121$ ground-state patterns, or into charge $1/2$ domain walls between $2121$ and $1212$ (Table \ref{rr2_sectors}). Periodic boundary conditions require that the former type of domain wall must come in pairs, but allow the latter type to exist singly. 
Since we will need to study states with $n$ quasiholes where $n=1$, $2$, or $3$, we will begin with some
considerations for general $n$.

\subsection{\label{rr_n}States with $n$ quasiholes}

In the Moore-Read case, we introduced sector labels $(\mu, \nu)$
 that encode the properties of states under translations. The conventions
 used there
 made use of the fact that at filling factor $\nu=1$, $T_x$ and $T_y$ commute.
 For the $k=3$ RR state at $\nu=3/2$, we thus have to proceed somewhat
 differently in exploiting translational properties. 

To this end, we denote a thin torus state with $n$ domain walls by $\left| a_1,\dotsc, a_n;c,\alpha\right)$, and the adiabatically evolved state by $|a_1,\dotsc,a_n;c,\alpha\rangle$. 
We introduce two labels $c$, $\alpha$ to denote topological sectors, where $\alpha$ labels classes of sectors that are not related by translation
(see Tables \ref{rr1_sectors}, \ref{rr2_sectors}), and ${c=0, 1}$ labels the two members of each class that are related by translation. 
The meaning of $c$ is thus very much the same as in our discussion of Laughlin states. The utility of this labeling will become apparent shortly; the dependence of various quantities on the $c$ label will be constrained by translational symmetries, and 
 $c$ is conserved during braiding, in much the same way as for the Laughlin states. In contrast, the interesting non-Abelian behavior will be associated with the $\alpha$ label.

We use the same mutually 
dual coherent state expressions as before (see Eqns. \eqref{2holes} and \eqref{pf_psi_def}),
\begin{widetext}
\be \label{rr_psi} 
	\left| \psi_{c,\alpha}(\{h\}) \right> = 
	{\cal N} \sum_{a_1<\dotsc<a_n} 
	\prod_{j=1}^n \phi_{\alpha,j}(h_j,\kappa a_j) \left| a_1,\dotsc,a_n;c,\alpha \right> 
\ee
\be \label{rr_psibar} 
	\overline{\left| \psi_{c,\alpha}(\{h\}) \right>} = 
	{\cal N}' \sum_{a_1<\dotsc<a_n} \prod_{j=1}^n 
	\overline\phi_{\alpha,j}(h_j,\overline{\kappa} a_j) \overline{\left| a_1,\dotsc,a_n;c,\alpha \right>} 
\ee
where the first is defined for $n$ quasiholes that are well separated along the $x$ axis, and the second for $n$ quasiholes that are well separated along the $y$ axis. We have used $\{h\}$ for the set of quasihole positions $h_1,\dotsc,h_n$. 
For the same reasons that we discussed in Sec. \ref{pfaff} originally for the Pfaffian,
we will assume the generic Gaussian form of $\phi_{\alpha,j}(h,x)$ given in \Eq{phigeneral}: 
\be \label{rr_phi}
	\phi_{\alpha,j}(h_j,x) = \exp \left[ \frac{1}{2} i (h_{jy} + \delta(\alpha,j)/\kappa) x - \gamma (h_{jx} - x)^2 \right]  
\ee
and 
\be \label{rr_phibar}
	\overline\phi_{\alpha,j}(h_j,y) = \phi_{\alpha,j}(-i h_j, y)|_{\kappa \rightarrow \overline\kappa } 
	=  \exp\left[ -\frac12 i(h_{jx}+\delta(\alpha,j)/\overline\kappa)y - \gamma (h_{jy} - y)^2 \right] \;.
\ee
\end{widetext}
In the above, we have already set $\beta=1/2$, which follows in exactly the same way as for the Pfaffian. We have written $\phi$ as a function of the sector $\alpha$, to allow for the possibility that the momentum shift $\delta$ may take on different values for quasiholes associated with different types of domain walls. However, $\phi$ is independent of $c$ since the type of the $j$-th domain wall is invariant under translation. 

Again, the two bases \eqref{rr_psi} and \eqref{rr_psibar} are related to each other by a transition matrix. In general, the elements of this matrix depend on both $c$ and $\alpha$.
\be \label{rr_ugen}
	\left| \psi_{c,\alpha}(\{h\}) \right> = 
	\sum_{c',\alpha'} u^\sigma_{c,c',\alpha,\alpha'}(\{h\}) 
	\overline{\left| \psi_{c',\alpha'}(\{h\}) \right>} 
\ee
In complete analogy with \Eq{laugh_def_xi}, we can derive the local dependence of the transition matrix within each of the regions labeled by $\sigma$, which are components of the quasihole configuration space with quasihole coordinates well separated in both $x$ and $y$  (cf. \Fig{config} as well as \Fig{config3h} below),
\be \label{rr_u}
	u^\sigma_{c,c',\alpha,\alpha'}(\{h\}) = 
	\xi^\sigma_{c,c',\alpha,\alpha'}\, u(\{h\}),
\ee 
again with $u(\{h\})=e^{\frac i2 \sum_j h_{jx}h_{jy} }$. For $n=2$, there are 72 of these parameters $\xi^\sigma_{c,c',\alpha,\alpha'}$: we distinguish two configurations $\sigma$ (\Fig{config}), and for each there is a $6\times 6$ matrix in the sector labels. 

We first state the translational properties of the $n$--domain-wall states, which are the same as in Eqs. \eqref{Tx} and \eqref{Ty}, since $\alpha$ is a spectator under translations. We now adopt a natural definition for the $c$ labels. Recall that the action of $T_y$ is given as follows,
\be \label{rr_Ty_0}
	T_y \left| a_1,\dots,a_n;c,\alpha \right>  =  
	\exp \left[ -\frac{2 \pi i}{L}\sum_j n_j \right] \left| a_1,\dots,a_n;c,\alpha \right> 
\ee
where the $n_j$ are the orbitals occupied in the pattern labeling the state. We find that the sum over the $n_j$ takes on the following form,
\be \label{rr_cdef}
	\sum_j n_j = \frac{1}{2} L \left( \nu L - c \right) - \frac{1}{2} \sum_j a_j \mod L
\ee
where $c=0,1$, and the domain-wall positions are defined via $a_i=2n_i +f_i(c,\alpha)$ as before, with the orbital position $2n_i$ defined in relation to the domain wall as shown in Table \ref{rr2_sectors}. Equation \eqref{rr_cdef} then defines $c$ modulo 2, and $\alpha$ labels the three ``supersectors'' formed by the translational pairs of states. 

The translational properties of the $n$--domain-wall states then are
\begin{align} 
   	T_x \left| a_1,\dots,a_n;c,\alpha \right>  
	&= \left| a_1+1,\dots,a_n+1;1-c,\alpha \right> \nonumber \\
	T_x \left| a_1,\dots,a_n;c,\alpha \right>  
	&= e^{-i \pi \lambda + i \pi c +\frac{1}{2} \kappa \overline{\kappa}\sum_j a_j} \left| a_1,\dots,a_n;c,\alpha \right>
	\label{rr_Tx} \\
   	T_y \left| a_1,\dots,a_n;c,\alpha \right>  
	&= e^{i \pi \lambda - i \pi c -\frac{1}{2} \kappa \overline{\kappa}\sum_j a_j} 
	\left| a_1,\dots,a_n;c,\alpha \right> \nonumber \\
	T_y \left| a_1,\dots,a_n;c,\alpha \right>  
	&=  \left| a_1+1,\dots,a_n+1;1-c,\alpha \right>
	\label{rr_Ty}
\end{align}
where again we write $\lambda = \nu L$, this time with $\nu=3/2$. 

As in the preceding cases, the translational properties of the coherent states follow directly from Eqs. \eqref{rr_Tx} and \eqref{rr_Ty}:
\begin{align}
	T_x \left| \psi_{c,\alpha} (\{h\}) \right> &= 
	e^{-\frac{1}{2}i \kappa \sum_j h_{jy}-i \frac{1}{2}\sum_j \delta(\alpha,j) } 
	\left| \psi_{1-c,\alpha} (\{h+\kappa\}) \right>  \nonumber\\
	T_x \overline{\left| \psi_{c,\alpha} (\{h\}) \right>} &= 
	e^{i \pi \lambda - i \pi c} \overline{\left| \psi_{c,\alpha} (\{h+\kappa\}) \right>}
	 \label{rr_Tx_psi} \\
	T_y \left| \psi_{c,\alpha} (\{h\}) \right> &= 
	e^{ -i \pi \lambda+i \pi c} \left| \psi_{c,\alpha} (\{h+i\overline\kappa\}) \right> \nonumber\\
	T_y \overline{\left| \psi_{c,\alpha} (\{h\}) \right>} &= 
	e^{ \frac{1}{2}i \overline{\kappa} \sum_j h_{jx}+\frac{1}{2}i \sum_j \delta(\alpha,j) } 
	\overline{\left| \psi_{1-c,\alpha} (\{h+i\overline\kappa\}) \right>} 
	\label{rr_Ty_psi}\;,
\end{align}
where we have used the notation 
$\{h+\kappa\} = h_1+\kappa,\dotsc,h_n+\kappa$, and similarly used $\{h+i\overline\kappa\}$.  
We can use these translational properties to completely determine the dependence of the transition matrices on $c$. To make this decoupling more explicit, we introduce two-component objects, denoted by a $\Psi$:
\be \label{rr_Psi}
   \begin{split}
	\left| \Psi_\alpha (\{h\}) \right> = 
	\left( \begin{array}{c}
		\left| \psi_{0,\alpha} (\{h\}) \right> \\
		e^{\frac12i\sum_j\delta(\alpha,j) } \left| \psi_{1,\alpha} (\{h\}) \right>
	\end{array} \right) \\
	\overline{\left| \Psi_\alpha (\{h\}) \right>} = 
	\left( \begin{array}{c}
		\overline{\left| \psi_{0,\alpha} (\{h\}) \right>} \\
		e^{\frac12i\sum_j\delta(\alpha,j)} \overline{\left| \psi_{1,\alpha} (\{h\}) \right>}
	\end{array} \right) ,
   \end{split}
\ee
where the phase splitting between the $c=0$ and $c=1$ states has been introduced with foresight to keep later phases in check. Correspondingly, we may view the full transition matrix as a ``supermatrix'' $\Xi^\sigma$, i.e., an $\alpha_\text{max}\times\alpha_\text{max}$ matrix, the elements of which are each $2\times2$ matrices denoted $\Xi_{\alpha,\alpha'}^\sigma$. So we write the equation between the original and dual bases 
as 
\be \label{rr_Xi_def}
	\left| \Psi_\alpha (\{h\}) \right> = 
	\sum_{\alpha'} u(\{h\}) \Xi_{\alpha,\alpha'}^\sigma
	\overline{\left| \Psi_{\alpha'} (\{h\}) \right>} \;.
\ee
Note the similarity of \Eq{rr_Xi_def} to the Laughlin transition matrix \Eq{laugh_Xidef}, to which \Eq{rr_Xi_def} reduces for ${\alpha_{\text{max}}\!=\!1}$.

We rewrite Eqs. \eqref{rr_Tx_psi} and \eqref{rr_Ty_psi} in terms of the two-component basis.
\begin{widetext}
\be
	T_x \left| \Psi_{\alpha} (\{h\}) \right> = 
	e^{-\frac{1}{2}i \kappa \sum_j h_{jy} }
	\left( \begin{array}{cc}
		0 & e^{-i \sum_j \delta(\alpha,j)} \\
		1 & 0
	\end{array} \right)
	\left| \Psi_{\alpha} (\{h+\kappa\}) \right> ; \quad
	T_x \overline{\left| \Psi_{\alpha} (\{h\}) \right>} = 
	e^{i \pi \lambda} 
	\left( \begin{array}{cc}
		1 & 0 \\
		0 & -1
	\end{array} \right)
	\overline{\left| \Psi_{\alpha} (\{h+\kappa\}) \right>} \;,
	\label{rr_Tx_Psi} 
\ee
\be
	T_y \left| \Psi_{\alpha} (\{h\}) \right> = 
	e^{ -i \pi \lambda}
	\left( \begin{array}{cc}
		1 & 0 \\
		0 & -1
	\end{array} \right)
	\left| \Psi_{\alpha} (\{h+i\overline{\kappa} \}) \right> ; \quad
	T_y \overline{\left| \Psi_{\alpha} (\{h\}) \right>} = 
	e^{ \frac{1}{2}i \overline{\kappa} \sum_j h_{jx} } 
	\left( \begin{array}{cc}
		0 & 1 \\
		e^{i \sum_j \delta(\alpha,j)} & 0
	\end{array} \right) 
	\overline{\left| \Psi_{\alpha} (\{h+i\overline{\kappa}\} ) \right>} 
	\label{rr_Ty_Psi}\;.
\ee
\end{widetext}
As before, (cf. Eqs. \eqref{urel_pf2}, \eqref{urel_pf3}), when applied to \Eq{rr_Xi_def}, Eqs. \eqref{rr_Tx_Psi} and \eqref{rr_Ty_Psi} each give a consistency equation that must be satisfied by every $\Xi_{\alpha,\alpha'}^\sigma$:
\be \label{rr_Xi_conx}
	\Xi_{\alpha,\alpha'}^\sigma = 
	e^{i \pi \lambda }
	\left( \begin{array}{cc}
		0 & 1 \\
		e^{i \sum_j \delta(\alpha,j)} & 0
	\end{array} \right)  
	\Xi_{\alpha,\alpha'}^\sigma 
	\left( \begin{array}{cc}
		1 & 0 \\ 0 & -1
	\end{array} \right)
\ee
\be \label{rr_Xi_cony}
	\Xi_{\alpha,\alpha'}^\sigma = 
	e^{i \pi \lambda } 
	\left( \begin{array}{cc}
		1 & 0 \\ 0 & -1
	\end{array} \right) \Xi_{\alpha,\alpha'}^\sigma 
	\left( \begin{array}{cc}
		0 & 1 \\
		e^{i \sum_j \delta(\alpha,j)} & 0
	\end{array} \right)  ,
\ee
which  imply
\be \label{rr_Xi}
	\Xi_{\alpha,\alpha'}^\sigma = 
	\frac{ \xi_{\alpha,\alpha'}^\sigma }{ \sqrt2 }
	\left( \begin{array}{cc}
		1 & e^{i \pi \lambda } \\
		e^{-i \pi \lambda } & -1
	\end{array} \right),
\ee
together with the constraint
\be \label{rr_delta_constraint}
	\exp \left[ 2\pi i \lambda + i \sum_j \delta(\alpha,j)  \right] = 1 \;.
\ee
In the above,  $\xi^\sigma_{\alpha,\alpha'}$ is an overall coefficient, and $\sqrt 2$ is a normalization factor. 

The phase choice we made in \Eq{rr_Psi} has allowed us to decouple the $\alpha$ and $c$ indices within the transition function. We can write the matrix $\Xi^\sigma$ defining the transition function \Eq{rr_Xi_def}
as
\be \label{rr_xi}
	\Xi^\sigma = \xi^\sigma \otimes M \;,
\ee
where $\xi^\sigma$ is the $\alpha_\text{max} \times \alpha_\text{max}$ matrix of coefficients $\xi^\sigma_{\alpha,\alpha'}$ and $M$ is the $2\times2$ matrix
\be \label{rr_M}
	M = 
	\frac{1}{ \sqrt2 }
	\left( \begin{array}{cc}
		1 & e^{-i \pi \lambda } \\
		e^{i \pi \lambda } & -1 
	\end{array} \right)
\ee
The $\alpha$ dependence of $\Xi^\sigma$ is  completely contained in the corresponding coefficient matrix $\xi^\sigma$, and the $c$ dependence is completely contained in the $M$ matrix. 

If we consider the translational properties of the states in the case of a single quasihole, we can constrain some of the $\delta(\alpha,j)$ parameters appearing above. For a single quasihole on a torus, the only topological sectors respecting periodic boundary conditions are those with domain walls between $2121$ patterns, as shown in Table \ref{rr1_sectors}. There are two such sectors, related by translation, so for a single quasihole $\alpha_\text{max} \!=\!1$. There is then only a single $\delta(\alpha,j)$ parameter, which we call $d$. When we consider \Eq{rr_delta_constraint} and note that in this case $\lambda=\nu L=\frac 32 (\frac 13 (2N+1))$ is half-odd integral, we find $d \!=\! \pi$. 
\begin{table}[!tbp]  
   \begin{tabular}{c | c || c  }
	Sector $c,\alpha$ & Thin torus pattern & $f_1(c,\alpha)$  \\
	\hline
	0,1 & $2121212\underline{1}121212121$ & $\frac 12 $ \\
	1,1 & $1212121\underline{1}212121212$ & $-\frac 12 $
   \end{tabular} 
   \caption{Thin torus patterns for a single-quasihole k=3 Read-Rezayi state, and the 
   	offset functions of the associated domain walls. The latter are fully determined 
	by inversion symmetry of the state. The orbital positions, $2n_i$, are underlined. 
	Since the sectors are all related by translation, $\alpha$ takes on a single value.} 
   \label{rr1_sectors} 
 \end{table}

In general, the $\delta(\alpha,j)$s are each associated with a certain type of domain wall, so by fixing $d$ in the single-quasihole case we also fix any $\delta(\alpha,j)$ associated with a $2121\dw1212$-type domain wall in an $n$-quasihole state. We can constrain the other $\delta(\alpha,j)$s to be either 0 or $\pi$ by considering \Eq{rr_delta_constraint} in the case $n=2$.
For two quasiholes there are only two independent $\delta(\alpha,j)$ parameters: $\delta(3,j)$, which is associated with $2121\dw1212$-type domain walls and is thus known to be $\pi$ from the one-quasihole argument; and $\delta(1,j)$ and $\delta(2,j)$, which are associated variously with domain walls between $1212$ and $0303$ strings, and which must be equal by the argument in Ref. \onlinecite{delta_note}. For $n=2$, we have $\lambda=\nu L=\frac 32 (\frac 13 (2N+2))$, which is an integer, and \Eq{rr_delta_constraint} reduces to $\exp [ i\sum_j \delta(\alpha,j) ] =1$. This is already satisfied for $\delta(3,j)=\pi$, and can be satisfied for $\alpha=2,3$ only if $\delta(1,j)=\delta(2,j)=0,\pi$.

In the end, we want to find explicit expressions for the elements of the transition matrices $\Xi^\sigma$, which we have reduced to the problem of finding the elements of the $\xi^\sigma$ coefficient matrices. This will be our task in the following sections.
\subsection{\label{rr2}Two quasiHoles}
\begin{table}[!tbp]  
   \begin{tabular}{ccc | c || c | c || c}
	& $\alpha$ && Thin torus pattern & $f_1(\alpha)$ & $f_2(\alpha)$ & $F(\alpha)$ \\
	\hline
	& 1 && $3030303\underline{0}2121212\underline{0}30303030$ & $s$ & $-s$ & 2\\
	& 2 && $1212121\underline{2}0303030\underline{2}12121212$ & $1-s$ & $1+s$ & 1\\
	& 3 && $1212121\underline{1}2121212\underline{1}12121212$&$-\frac 12$ & $\frac 12$ & 3\\
   \end{tabular} 
   \caption{$c=0$ thin torus patterns for a two-quasihole $k=3$ Read-Rezayi state, 
   	and the offset functions of the associated domain walls. The orbital positions, $2n_i$, are underlined. 
   	Patterns for $c=1$ can be obtained by shifting each occupancy number one orbital to the right, and 
   	$c=1$ offset functions by adding or subtracting 1 to each offset function above, whichever is more convenient.} 
   \label{rr2_sectors} 
 \end{table}
The thin torus patterns for two-quasihole states with $c\!=\!0$ are given in Table \ref{rr2_sectors}. To find the statistics of these quasiholes we must constrain the transition matrices $\Xi^{+}$ and $\Xi^{-}$. Both $\Xi^\sigma$s have nine complex unknowns, the entries of the $\xi^\sigma$ matrices. To constrain these we will move the quasiholes around global paths, which we defined in Sec. \ref{symmetries}. We will then make further use of the mirror symmetry operation, which has thus far only been discussed in Sec. \ref{LLstructure} and very briefly in Sec. \ref{cons_pfaff1}. 
As in the Moore-Read case, we gain further constraints by imposing locality and unitarity. In the general solution to these equations some unknown parameters still remain.
 We will be able to constrain the latter by subsequently studying the case of three quasiholes in Sec. \ref{rr3}. 

\subsubsection{\label{rr2_globalpaths}Constraints from global paths}
As discussed in sections \ref{symmetries} and \ref{cons_pfaff1}, the transition matrices for different configurations can be connected by dragging the quasiholes through the global paths in \Fig{globalpaths}. 
We first consider two quasiholes in the $\sigma\!=\!+$ configuration, and imagine the right quasihole moving around the $x$ direction of the torus along the path $a$ in \Fig{globalpaths}. Using the reasoning of Sec. \ref{pf_paths} we find the following effects on the coherent states:
\be \begin{split} \label{rr2_xpath_1}
	| \psi_{c,\alpha}(h_1, h_2) \rangle_f 
	&\doteq e^{\frac{1}{2}i L_x h_{2y}+i \frac L2 \delta(\alpha,2)}\, |\psi_{1-c,F(\alpha)}(h_2-L_x, h_1)\rangle \\
	\overline{|\psi_{c,\alpha}(h_1, h_2)\rangle}
	&=  e^{-i \pi f_2(c,\alpha)} \overline{|\psi_{c,\alpha}(h_1, h_2-L_x)\rangle}\,.
\end{split} \ee
Moving the quasihole along this path changes the sector label $\alpha$ for the original basis
 into $F(\alpha)$, the values of which can be read off the patterns and are summarized in Table \ref{rr2_sectors}. 

To find a constraint on the $\xi^\sigma$s, we write \Eq{rr2_xpath_1} in the two-component basis.
\begin{widetext}
\be \begin{split} \label{rr2_xpath_2}
	\left| \Psi_{\alpha}(h_1,h_2) \right>_f 
	&\doteq e^{\frac{1}{2}i L_x h_{2y} + i \frac L2 \delta(\alpha,2) } 
	\left( \begin{array}{cc}
		0 & e^{-\frac12 i \sum_j \delta(\alpha,j)} \\
		e^{\frac12 i \sum_j \delta(\alpha,j)} & 0
	\end{array} \right) 
	\left| \Psi_{F(\alpha)} (h_2-L_x,h_1) \right> \\ 
	\overline{\left| \Psi_{\alpha} (h_1,h_2) \right>} 
	&= e^{ -i\pi f_2(\alpha) } 
	\left( \begin{array}{cc}
		1 & 0 \\ 0 & -1
	\end{array} \right) 
	\overline{\left| \Psi_\alpha (h_1,h_2-L_x) \right>} \;,
\end{split} \ee
where we have used that $f_j(\alpha) \equiv f_j(0,\alpha)=f_j(c,\alpha)-c \mbox{ mod } 2$.  Applying \Eq{rr2_xpath_2} to \Eq{rr_Xi_def} gives
	\be \begin{split} \label{rr2_xpath_3}
		& \left| \Psi_{F(\alpha)} (h_2-L_x,h_1) \right> = \\
		& \; \; \sum_{\alpha'} u(h_2-L_x,h_1) 
		e^{ -i \frac L2 \delta(\alpha,2) -\frac12 i \sum_j \delta(\alpha,j)} 
		\left( \begin{array}{cc}
			0 & 1 \\
			e^{i \sum_j \delta(\alpha,j)} & 0
		\end{array} \right) 
		\Xi_{\alpha,\alpha'}^{+} 
		\left( \begin{array}{cc}
			1 & 0 \\ 0 & -1
		\end{array} \right)
		e^{-i\pi f_2(\alpha')}
		\overline{\left| \Psi_\alpha (h_1,h_2-L_x) \right>} 
	\end{split} \ee
	We can simplify \Eq{rr2_xpath_3} using \Eq{rr_Xi_conx}.
	\be \label{rr2_xpath_4}
		\left| \Psi_{F(\alpha)} (h_2-L_x,h_1) \right> = 
		\sum_{\alpha'} u(h_2-L_x,h_1) 
		e^{ -i \pi \lambda -i \frac L2 \delta(\alpha,2)  -\frac12 i \sum_j \delta(\alpha,j)} 
		\Xi_{\alpha,\alpha'}^{+} e^{-i\pi f_2(\alpha')}
		\overline{\left| \Psi_\alpha (h_1,h_2-L_x) \right>} 
	\ee
\end{widetext}
We want to write this as an equation between $\xi^{-}$ and $\xi^{+}$, which we can do by noting the equivalence between \Eq{rr2_xpath_4} as written and \Eq{rr_Xi_def} evaluated at quasihole positions $(h_2-L_x,h_1)$. To make this equivalence manifest we can write the action of $F$ in matrix form as:
\be \label{rr2_Bdef}
	(B)_{\alpha,\alpha'} = \delta_{\alpha,F(\alpha')}
\ee
or
\be \label{rr2_B}
	B =
	\left( \begin{array}{ccc}
	 	0 & 1 & 0 \\
	 	1 & 0 & 0 \\
	 	0 & 0 & 1
	 \end{array} \right) \;.
\ee
Since the transition matrix in \Eq{rr_Xi_def} evaluated at positions $(h_2-L_x,h_1)$ involves $\Xi^-$, and the transition matrix in \Eq{rr2_xpath_4} is a product involving $\Xi^+$, the equivalence of these two equations implies:
\be \label{rr_xpath_5}
	\xi^{-} =
	B^{-1} \mathrm{diag} \!\left[ e^{ -i \pi \lambda -i \frac L2 \delta(\alpha,2)  -\frac12 i \sum_j \delta(\alpha,j)} \right]
	\xi^{+} \mathrm{diag} \!\left[ e^{ -i\pi f_2(\alpha) } \right] \;,
\ee
where we canceled the matrix $M$ common to both $\Xi^\sigma$s,
and the argument of $\mathrm{diag}[\dots]$ specifies the $\alpha$-th diagonal entry of a diagonal matrix.
 If we use the values of $f_2(\alpha)$ from Table \ref{rr2_sectors}, \Eq{rr_xpath_5} becomes
\be \label{rr2_xpath}
	\xi^- =
	\left( \begin{array}{ccc} 
		0 & \Delta & 0 \\ 
		\Delta & 0 & 0 \\ 
		0 & 0 & 1 
	\end{array}\right)
	\xi^+ 
	\left( \begin{array}{ccc} 
		p & 0 & 0 \\ 
		0 & p^{-1} & 0 \\ 
		0 & 0 & -1 
	\end{array}\right)
	e^{-i \frac\pi2} \,.
\ee
We have defined two phases: ${p=-\exp \left[i \pi (s+\frac{1}{2}) \right]}$ and $\Delta = \exp \left[ i(L/2+1)(\pi-\delta) \right]$. Note that for two quasiholes $L$ is even and $\Delta^2=1$.

We can perform the same process in the $y$ direction and drag the quasihole around the global path marked $b$ in \Fig{globalpaths}. After an argument similar to that above we find another equation between $\xi^-$ and $\xi^+$, which can be inverted to yield the following equation:
\be \label{rr2_ypath}
	\xi^+ =   
	\left( \begin{array}{ccc} 
		p^{-1} & 0 & 0 \\ 
		0 & p & 0 \\ 
		0 & 0 & -1 
	\end{array}\right)
	\xi^- 
	\left( \begin{array}{ccc} 
		0 & \Delta & 0 \\ 
		\Delta & 0 & 0 \\ 
		0 & 0 & 1 
	\end{array}\right) 
	e^{i \frac\pi2} 
\ee
Combining Eqs. \eqref{rr2_xpath} and \eqref{rr2_ypath} gives us a nontrivial consistency relation for $\xi^+$.
\be \label{rr2_paths}
	\xi^+ =   
	\left( \begin{array}{ccc} 
		0 & \Delta p^{-1} & 0 \\ 
		\Delta p & 0 & 0 \\ 
		0 & 0 & -1 
	\end{array}\right) 
	\xi^+ 
	\left( \begin{array}{ccc} 
		0& \Delta p & 0 \\ 
		\Delta p^{-1} & 0 & 0 \\ 
		0 & 0 & -1 
	\end{array}\right) 
\ee
Equation \eqref{rr2_paths} gives us several equations between the matrix elements of $\xi^+$, the coefficients $\xi^+_{\alpha,\alpha'}$. Equation \eqref{rr2_xpath} reduces the number of unknown $\xi^\sigma_{\alpha,\alpha'}$s from eighteen to nine.
The consistency relationship \Eq{rr2_paths} further reduces the number of unknown elements from nine down to five. A particular choice for the five independent $\xi^+_{\alpha,\alpha'}$s is the following:
\be \label{rr2_xi_constrained}
	\xi^{+} = 
	\left( \begin{array}{ccc} 
		\xi_{11} & \xi_{12} & \xi_{13} \\ 
		\xi_{12} & p^2 \xi_{11} & -\Delta p\xi_{13} \\ 
		\xi_{31} & -\Delta p\xi_{31} & \xi_{33} 
	\end{array}\right) 
\ee
Note that any time the configuration index $\sigma$ is omitted as in the above equation, we take it to be ${\sigma\!=\!+}$.
\subsubsection{\label{rr2_mirror}Constraints from mirror symmetry}
We now make use of the antilinear symmetry operator $\tau$ defined in Sec. \ref{LLstructure}, i.e., the combination of time reversal and mirror symmetry. Applying $\tau$ will exchange the $x$ positions of the quasiholes across the $y$ axis. This operation changes the configuration $\sigma$, which will allow us to derive another equation between $\xi^+$ and $\xi^-$. First, we describe how this symmetry acts on an $n$-quasihole state.

From the definition \Eq{mirror1}, the effect of $\tau$ on bare LL product states is clear: it reflects the original basis states across the $y$ axis, and it has no effect on the dual states. For bare product states with domain walls, the domain-wall positions will be similarly reflected. $\tau$ commutes with the adiabatic evolution operators (as constructed, e.g., in Ref. \onlinecite{seidel_lee}) that define the delocalized quasihole states, thus its action on those states is:
\begin{subequations} \label{rr_tau_dressed}
   \begin{align}
	\label{rr_tau_dressedx}
	\tau \left|a_1,\dots,a_n;c,\alpha \right> &=
	\left| L-a_n,\dots,L-a_1;c,F_\tau(\alpha) \right> \\
	\label{rr_tau_dressedy}
	\tau \overline{\left| a_1,\dots,a_n;c,\alpha \right>} &=
	\overline{\left| a_1,\dots,a_n;c,\alpha \right>} \,.
   \end{align}
\end{subequations}
We write that the position of the $j$-th dual-basis quasihole $a_j$ goes to $L-a_j$ in \Eq{rr_tau_dressedx} so as to stay within the default frame. Also note that in general $\tau$ might or might not change $\alpha$, and we describe this change by some function $F_\tau$, the values of which can be found from the patterns. It turns out that for the case of two quasiholes, $F_\tau(\alpha) = \alpha$. Later when we analyze the case of three quasiholes, $F_\tau$ will be a nontrivial mapping.

Equation \eqref{rr_tau_dressed} allows us to derive how $\tau$ acts on coherent states of $n$ quasiholes. In terms of two-component states:
\begin{widetext}
%
%
\begin{subequations} \label{rr_tau_Psi}
   \begin{align}
	\label{rr_tau_Psix}
	\tau \left| \Psi_\alpha (\{h\}) \right> 
	&= e^{ -\frac{1}{2}iL_x\sum_jh_{jy} - i \frac L2 \sum_j\delta(\alpha,j)}
	\left| \Psi_{F_\tau(\alpha)} (\{-h^*+L_x\}) \right> \\
	\label{rr_tau_Psiy}
	\tau \overline{\left| \Psi_\alpha (\{h\}) \right>} 
	&= e^{ i\pi \sum_j f_j(\alpha) + i\sum_j \delta(\alpha,j)f_j(\alpha)}
	\overline{\left| \Psi_\alpha (\{-h^*+L_x\}) \right>} \,.
   \end{align}
\end{subequations}
\end{widetext}

For now, we will restrict ourselves to the case of two quasiholes. In this case, \Eq{rr_tau_Psi} simplifies to:
\begin{subequations} \label{rr2_tau_Psi}
   \begin{align}
	\label{rr2_tau_Psix}
	\tau \left| \Psi_\alpha (h_1,h_2) \right> 
	&= e^{-\frac{1}{2}iL_x\sum_jh_{jy}}
	\left| \Psi_\alpha (h_1',h_2') \right> \\
	\label{rr2_tau_Psiy}
	\tau \overline{\left| \Psi_\alpha (h_1,h_2) \right>} 
	&= \overline{\left| \Psi_\alpha (h_2',h_1') \right>} \,,
   \end{align}
\end{subequations}
where for all indices $j$, $h_j'=-h_j^*+L_x$. To arrive at \Eq{rr2_tau_Psi} we have used that for two quasiholes the phase factors on \Eq{rr_tau_Psix} and \Eq{rr_tau_Psiy} are both 1---the former because $L$ is even, and the latter can be seen by inserting the values of $f_j(\alpha)$ from Table \ref{rr2_sectors}---and $F_\tau(\alpha)=\alpha$ as noted above. Equation \eqref{rr2_tau_Psi} allows us to apply $\tau$ to \Eq{rr_Xi_def}. Let us begin with the two quasiholes in the ${\sigma\!=\!+}$ configuration; when we apply $\tau$ to \Eq{rr_Xi_def} and compare the resulting equation to \Eq{rr_Xi_def} evaluated at the changed spatial coordinates, we find the simple relationship ${\xi_{\alpha,\alpha'}^- = (\xi_{\alpha,\alpha'}^+)^*}$, or
\be \label{rr2_tau_xi}
	\xi^- = (\xi^+)^* \,.
\ee
For the moment, we leave the relation \eqref{rr2_tau_xi} implicit, and use it in App. \ref{app2} to further reduce the number of independent parameters.
\subsubsection{\label{rr2_braiding}Braiding}
We can perform the adiabatic exchange of two quasiholes using again the method of Secs. \ref{braid} and \ref{pf_braid} with minor generalizations. The details formally carry over from Sec. \ref{braid} because all the Berry connections along the path segments considered above are independent of the sector (see \Eq{gam1} for example). I.e., for the exchange of two quasiholes as in \Fig{exchange}, dragging the second quasihole along the path segment $\mathcal C_1$ causes the wave functions in each sector to pick up the same phase $\exp \left[ i \gamma_1 \right]$ defined in \Eq{laugh_braid_c1}.
\be \label{rr2_C1}
	\left( \begin{array}{c}
		\left| \Psi_1 (h_1,h_2) \right> \\
		\left| \Psi_2 (h_1,h_2) \right> \\
		\left| \Psi_3 (h_1,h_2) \right>
	\end{array} \right) 
	\rightarrow e^{i \gamma_1}
		\left( \begin{array}{c}
		\left| \Psi_1 (h_1,h_a) \right> \\
		\left| \Psi_2 (h_1,h_a) \right> \\
		\left| \Psi_3 (h_1,h_a) \right>
	\end{array} \right)
\ee
Reiterating the remaining steps described in Sec. \ref{braid}, the result of the adiabatic exchange is the following:
\be \label{rr2_braid}
	\left( \begin{array}{c}
		\left| \Psi_1 (h_1,h_2) \right> \\
		\left| \Psi_2 (h_1,h_2) \right> \\
		\left| \Psi_3 (h_1,h_2) \right>
	\end{array} \right)
	\rightarrow e^{i \Phi_{AB}}\Xi^+ (\Xi^-)^{-1}
	\left( \begin{array}{c}
		\left| \Psi_1 (h_1,h_2) \right> \\
		\left| \Psi_2 (h_1,h_2) \right> \\
		\left| \Psi_3 (h_1,h_2) \right>
	\end{array} \right) \;.
\ee
Once again we see that adiabatic exchange results in a path-dependent Aharonov-Bohm phase and a topological, statistical part made of a product of the transition functions, which we call the braid matrix. 
The structure of this matrix is
\be \label{rr2_chi0}
	\Xi^+ (\Xi^-)^\dagger =
	\chi \otimes \mathbb I_{2\times2} \;,
\ee
where
the translational, $c$-dependent part of the braid matrix is the product $MM^\dagger=\mathbb{I}_{2\times2}$, and we have defined the ``reduced'' braid matrix as the $\alpha$-dependent part,
\be \label{rr2_chi_def}
	\chi=\xi^+ (\xi^-)^\dagger\;.
\ee

We can constrain the form of the matrix $\chi$ by making an argument about the locality of the exchange process, analogous to the argument made in Sec. \ref{pf_braid}. 
Recall that according to the latter,
only the string of the pattern that is between the  
domain walls taking part in the exchange 
can be changed as a result of this process. 
Any regions of the pattern far to the left or right of the initial positions must remain unchanged after the exchange. 
For one, this requires the exchange processes to be diagonal in $c$. This is already manifest by the structure
of the braid matrix derived thus far, \Eq{rr2_chi0}.
However, certain transitions of the $\alpha$ label are allowed. 
To see this, we again refer to Table \ref{rr2_sectors}. One observes that
transitions into and out of the $\alpha=1$ sector are forbidden, since this is the
only sector with $3030$-type patterns far to the left and far to the right of the domain walls.
The other two sectors have $2121$-type patterns at the left and right end.
Therefore, transitions between these sectors are allowed.

These considerations imply that the reduced braid matrix \Eq{rr2_chi_def}  must be of the form:
\be \label{rr2_local}
	\chi = \xi^+(\xi^-)^\dagger =
	\left( \begin{array}{ccc}
		\cdot & 0 & 0 \\
		0 & \cdot & \cdot \\
		0 & \cdot & \cdot	
	\end{array} \right)
\ee
where dots indicate (potentially) non-zero matrix elements. Equation \eqref{rr2_local} gives two independent constraint equations for the matrix elements $\xi_{\alpha,\alpha'}$. We will also use constraints for the $\xi_{\alpha,\alpha'}$s gained from the fact that $\xi^+$ must be unitary:
\be \label{rr2_unitary}
	\xi^+ (\xi^+)^\dagger = \mathbb{I}_{3\times3} \;.
\ee

Together, \Eq{rr2_local} and \Eq{rr2_unitary} provide enough constraint equations to fix the $\xi_{\alpha,\alpha'}$ up to the parameter $p$, introduced after \Eq{rr2_xpath}, which is in turn defined by the shift parameter $s$ defined in Table \ref{rr2_sectors}. These constraints allow us to write explicit expressions for the elements of the braid matrix in terms of only the parameter $p$:
\be  \label{rr2_chi} 
	\chi = e^{i \frac\pi2}
	\left( \begin{array}{ccc}
		p^{-1} & 0 & 0 \\ 
		0 & p(p+p^{-1}-1) & \pm\sqrt{p+p^{-1}}(1-p) \\
		0 & \pm\sqrt{p+p^{-1}}(1-p) & p+p^{-1}-1
	\end{array} \right)
\ee
The details are presented in Appendix \ref{app2}. While $p$ is still unknown at this stage, it is no longer completely unconstrained. 
To further constrain the value of $p$ and fully determine the statistics, we must study the case of three quasiholes.
\subsection{\label{rr3}Three quasiHoles}
\begin{table}[!tbp]  
   \begin{tabular}{ccc | c || c | c | c || ccc | ccc}
	& $\alpha$ && Thin torus pattern &$f_1$&$f_2$ &$f_3$&&$F$&&&$F_\tau$& \\
	\hline
	& 1 && $30303\underline{0}21212\underline{1}12121\underline{2}03030$ 
		& $s$ & $\frac12 $ & $1-s$ && 3 &&& 1 &\\
	& 2 && $12121\underline{1}21212\underline{0}30303\underline{0}21212$ 
		& $-\frac12$ & $-s$ & $s$ && 1 &&& 3 &\\
	& 3 && $12121\underline{2}03030\underline{2}12121\underline{1}21212$ 
		& $1-s$ & $-1+s$ & $-\frac12$ && 2 &&& 2 &\\
	& 4 && $12121\underline{1}21212\underline{1}12121\underline{1}21212$ 
		& $-\frac12$ & $\frac12$ & $-\frac12$ && 4 &&& 4 &
   \end{tabular} 
   \caption{$c = 0$ thin torus patterns for a three--domain-wall $k=3$ Read-Rezayi state, 
	and the offset functions of those domain walls. The orbital positions, $2n_i$, 
	are underlined. Patterns and offset functions for $c=1$ can be obtained by, 
	respectively, shifting each pattern one orbital to the right and adding 1 to each 
	offset function.} 
	\label{rr3_sectors} 
 \end{table}
We expect that we can gain new information about the statistics by braiding two quasiholes among a system of three. 
To see this, note that as long as there are only two quasiholes, boundary conditions require that both are associated
with the same ``domain-wall type''. I.e., both domain walls must either occur between a  $3030$ string and a $2121$ string,
or between two $2121$ strings. Hence, we were not yet able to study what happens when a quasihole associated
with the former type is exchanged with one associated with the latter type. To study such processes, we must 
consider systems with three quasiholes. The relevant topological sectors are displayed in Table \ref{rr3_sectors}.
It will suffice to exchange the first two quasiholes (along $x$).
The ``new'' situation described above will then occur in the sectors $\alpha=1$ and $\alpha=2$.
Using locality arguments analogous to the preceding section, we conclude that exchanging the first
two quasiholes in these sectors is a diagonal process, since it is not possible to reach a different sector
by replacing the string linking the associated domain walls.
On the other hand, by complete analogy with the preceding section,
the same exchange processes may lead to transitions between the $\alpha=3$ and $\alpha=4$
sectors. These processes are locally the same as those discussed for the $\alpha=2$ and $\alpha=3$ sectors
in the preceding section. Invoking again locality, within the  $\alpha=3,4$ subspace the (reduced)
braid matrix must be given by the same $2\times2$ block displayed in \Eq{rr2_chi}. 
We used exactly the same argument before in Sec. \ref{pf_braidgroup}, where we constructed
the $2n$-quasiparticle representation of the braid group from the two-quasiparticle braid matrix for the Moore-Read state.
These  arguments constrain the form of the reduced braid matrix associated with the first two quasiholes to be:
\be  \label{rr3_local} 
	\chi = e^{i \frac \pi2}
	\left( \begin{array}{cccc}
		\cdot & & & \\
		& \cdot & & \\ 
		& & p(p+p^{-1}-1) & \pm\sqrt{p+p^{-1}}(1-p) \\
		& & \pm\sqrt{p+p^{-1}}(1-p) & p+p^{-1}-1
	\end{array} \right)
\ee
where the dots indicate some matrix element we do not yet know, and blank spaces represent zeros. 
In the above, $p$ is the same parameter appearing in \Eq{rr2_chi}, but we leave it understood that 
the quantities $\chi$, $\xi$ and $\Xi$ in this section refer to the three-quasihole case, and are different from their
two-quasihole counterparts. In the above, we have anticipated that braiding will again be diagonal in the
$c$ label, and $\chi$ is again defined through the action of braiding on the $\alpha$ label,
 which will follow below.  

\begin{figure} 
	\includegraphics[width=\columnwidth]{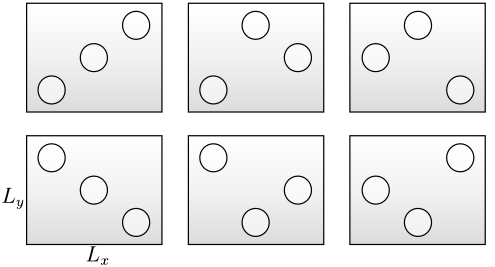}
	\caption{Configurations $\sigma=(\sigma_1,\sigma_2)$. $\sigma_1$ indicates the 
		relative position of the leftmost two quasiholes, $\sigma_2$ indicates the position of the 
		third quasihole relative to the first two. Top (left to right): $++$, $+0$, and $+-$. 
		Bottom (left to right): $--$, $-0$, and $-+$.} 
	\label{config3h}
 \end{figure}

We may again proceed by expressing $\chi$ through the transition matrix coefficients $\xi_{\alpha,\alpha'}^\sigma$
and deriving various constraints on the latter, where now additional constraints come from the $2\times2$ block in \eqref{rr3_local}.
The procedure is analogous to the preceding section, where only one aspect requires nontrivial generalization:
in the two-quasihole section there were only two transition matrices, $\Xi^{+}$ and $\Xi^{-}$, one for each configuration. For $n$ quasiholes, we must distinguish $n!$ configurations and define a transition matrix for each.
We choose the following notation to label these configurations. For an $n$-quasihole system, we let $\sigma=(\sigma_1,\dots,\sigma_{n-1})$. $\sigma_1$ takes a value $+$ or $-$, indicating the relative position of the two leftmost quasiholes, in the same manner 
as in the preceding section. 
$\sigma_2$ takes a value $+$, $-$, or $0$ and indicates the position of the third quasihole relative to the first two, as shown in \Fig{config3h} for three quasiholes. We could proceed further in the same way for $n>3$ quasiholes, but $n\leq 3$  suffices for our purposes.

With these conventions, we can find the result of exchange of two quasiholes in terms of the transition matrices. 
As pointed out, we choose to braid the two leftmost quasiholes and leave the third fixed. Further, let us say it is fixed ``above'' the other two, so ${\sigma_2 \!=\! +}$. 
\footnote{Other choices for the configurations of the quasiholes  will result in the same braiding matrices, as it should be.} 
The exchange can be broken down into segments in complete analogy 
with the two-quasihole case, yielding an equation analogous to \Eq{rr2_braid}:
\be \label{rr3_braid}
	\left( \begin{array}{c}
		\left| \Psi_1 (\{h\}) \right> \\
		\left| \Psi_2 (\{h\}) \right> \\
		\left| \Psi_3 (\{h\}) \right> \\
		\left| \Psi_4 (\{h\}) \right>
	\end{array} \right)
	\rightarrow e^{i \Phi_{AB}}\Xi^{++} (\Xi^{-+})^{-1}
	\left( \begin{array}{c}
		\left| \Psi_1 (\{h\}) \right> \\
		\left| \Psi_2 (\{h\}) \right> \\
		\left| \Psi_3 (\{h\}) \right> \\
		\left| \Psi_4 (\{h\}) \right>
	\end{array} \right).
\ee
%
%
As anticipated above, using Eqs. \eqref{rr_xi}, \eqref{rr_M}, we find
\be
	\Xi^{++} (\Xi^{-+})^\dagger
	= \chi \otimes \mathbb I_{2\times2} \;,
\ee
which again defines the reduced braid matrix $\chi$ in terms of the coefficient matrices,
\be \label{rr3_chi_def}
	\chi = \xi^{++} (\xi^{-+})^\dagger \;.
\ee
As in previous examples, we will use symmetries and global paths to constrain the $\xi^\sigma$ matrices, then use the implications of locality, \Eq{rr3_local}, to find explicit expressions for the elements of $\chi$.
\subsubsection{Constraints from mirror symmetry}
In Sec. \ref{rr2_mirror}, the action of $\tau$ on $n$ quasihole states has been discussed, \Eq{rr_tau_Psi}.
The three--domain-wall patterns are shown in Table \ref{rr3_sectors}, along with the values of $F_\tau(\alpha)$, which follow directly from these patterns. 
We can represent the map $F_\tau(\alpha)$ in matrix form:
\be
	B_\tau = 
	\left( \begin{array}{cccc}
		1 &&& \\
		& 0 & 1 & \\
		& 1 & 0 & \\
		&&& 1
	\end{array} \right) \;.
\ee
If we apply $\tau$ to \Eq{rr_Xi_def},
proceeding as in the derivation of \Eq{rr2_tau_xi} and using information from 
Table \ref{rr3_sectors}, we find
\be \label{rr3_tau}
	\xi^{g_{\tau}(\sigma)} = 
	\left( \begin{array}{cccc}
		\tilde\Delta^2 \\
		& 0 & \tilde\Delta^2 \\
		& \tilde\Delta^2 & 0 \\
		&&& 1
	\end{array} \right)
	(\xi^{\sigma})^* 
	\left( \begin{array}{cccc}
		\tilde\Delta^2 \\ &1 \\ &&1 \\ &&&1
	\end{array} \right) e^{i \pi \lambda + i \pi} \;,
\ee
where the phase $\tilde\Delta=\exp[ -i \frac L2 (\pi-\delta) ]$. Note that $L$ is odd, so $\tilde\Delta^2=\exp[i(\pi-\delta)]$ and $\tilde\Delta^4=1$.
The function $g_{\tau}(\sigma)$ gives the new configuration after 
reflection of
the quasiholes in configuration $\sigma$ across the $y$ axis, and its values are given in Table \ref{rr3_symm}.

The matrix structure of the last equation is somewhat more complicated than \Eq{rr2_tau_xi}.
Unlike the latter, \Eq{rr3_tau} is not ``self-dual'', i.e., we may obtain an analogous but different
equation by using the ``dual'' mirror symmetry operator $\bar\tau$ instead (Sec. \ref{LLstructure}).
It reads
\be \label{rr3_taubar}
	\xi^{\overline{g_\tau}(\sigma)} = 
	\left( \begin{array}{cccc}
		\tilde\Delta^2 \\ &1 \\ &&1 \\ &&&1
	\end{array} \right)
	(\xi^{\sigma})^* 
	\left( \begin{array}{cccc}
		\tilde\Delta^2 \\
		& 0 & \tilde\Delta^2 \\
		& \tilde\Delta^2 & 0 \\
		&&& 1
	\end{array} \right) e^{i \pi \lambda + i \pi} \;.
\ee
The function $\overline{g_\tau}(\sigma)$ captures the change in configuration under $\bar\tau$. Its values are given in Table \ref{rr3_symm}.

We now evaluate \Eq{rr3_tau} for $\sigma=(-,-)$, $g_\tau(-,-)\!=\!(+,+)$, and \Eq{rr3_taubar} for $\sigma=(+,+)$, $\overline{g_\tau}(+,+)\!=\!(-,-)$,  and plug one into the other.
This gives the following consistency equation for $\xi^{++}$:
\be \label{rr3_tau_xi}
	\xi^{++} = 
	\left( \begin{array}{cccc}
		1 \\
		& 0 & \tilde\Delta^2 \\
		& \tilde\Delta^2 & 0 \\
		&&& 1
	\end{array} \right)
	\xi^{++} 
	\left( \begin{array}{cccc}
		1 \\
		& 0 & \tilde\Delta^2 \\
		& \tilde\Delta^2 & 0 \\
		&&& 1
	\end{array} \right) \;,
\ee
which constrains $\xi^{++}$ to be of the form
\be \label{rr3_xi_constrained1}
	\xi^{++} =
	\left( \begin{array}{cccc}
		\xi_{11} & \tilde\Delta^2\xi_{13} & \xi_{13} & \xi_{14} \\
		\tilde\Delta^2 \xi_{31} & \xi_{22} & \xi_{23} & \tilde\Delta^2\xi_{34} \\
		\xi_{31} & \xi_{23} & \xi_{22} & \xi_{34} \\
		\xi_{41} & \tilde\Delta^2\xi_{43} & \xi_{43} & \xi_{44}
	\end{array} \right) \;.
\ee
As before, when the configuration $\sigma$ is omitted, we take it to be $++$.
\begin{table}[!tbp]  
   \begin{tabular}{c || c | c || c | c}
	$\sigma$ & $g(\sigma)$ & $\overline g(\sigma)$ & $g_\tau(\sigma)$ & $\overline{g_\tau}(\sigma)$ \\
	\hline
	$++$ & $-0$ & $+-$ & $--$ & $--$ \\
	$+-$ & $++$ & $-0$ & $-+$ & $+0$ \\
	$-0$ & $+-$ & $++$ & $+0$ & $-+$ \\
	\hline
	$--$ & $+0$ & $+0$ & $++$ & $++$ \\
	$+0$ & $-+$ & $-+$ & $-0$ & $+-$ \\
	$-+$ & $--$ & $--$ & $+-$ & $-0$
   \end{tabular} 
   \caption{The effect of various operations on the three-quasihole configuration $\sigma$.
   	$g(\sigma)$ is the resulting configuration when a quasihole is dragged around the torus along
	an $a$-type path (as shown in
          \Fig{globalpaths} for two quasiholes), and $\overline g(\sigma)$ is the same for a $b$-type
           path.  The rightmost and topmost quasihole is being dragged, respectively. $g_\tau(\sigma)$ is 
   	the resultant configuration under the mirror reflection $\tau$, $\overline{g_\tau}(\sigma)$ is similarly defined for 
   	$\overline\tau$.} 
   \label{rr3_symm} 
 \end{table}
\subsubsection{Constraints from global paths}
We continue with our program by deriving constraints from ``global paths'', as done for the two-quasihole case in
Sec. \ref{rr2_globalpaths}.
We begin by generalizing \Eq{rr2_xpath_2} (cf. \Fig{globalpaths}) to the case of three quasiholes. In the two-quasihole case we assumed the two quasiholes to be in a ${\sigma\!=\!+}$ configuration, then moved the top right quasihole around the $x$ direction of the torus to the top left. In this section we will need to derive more general behavior, allowing that the rightmost quasihole can be at the top, middle, or bottom relative to the other two quasiholes. The analogue of \Eq{rr2_xpath_1},
for a path similar to path $a$ in \Fig{globalpaths}, then becomes:
\be \begin{split} \label{rr3_xpath_1} 
	\left| \psi_{c,\alpha}(\{h\}) \right>_f 
	&\doteq e^{ \frac12 i L_x h_{3y} + i \frac L2\delta(\alpha,3) } 
	\left| \psi_{c+1,F(\alpha)} (\{h'\}) \right> \\ 
	\overline{\left| \psi_{c,\alpha} (\{h\}) \right>}
	&= e^{ -i\pi f_j(c,\alpha) } 
	\overline{\left| \psi_{c,\alpha} (\{h'\}) \right>} \,.
\end{split} \ee
Here $f_j(c,\alpha)$ can be inferred from Table \ref{rr3_sectors}, and $j$ equals 1, 2, or 3 if the quasihole encircling the torus is respectively the first, second or third when viewed from the $y$ direction. The position $\{h'\} \!=\! h_3-L_x,h_1,h_2$. As before, the change in $\alpha$ after moving the quasihole along the path is described by the function $F(\alpha)$. Its values directly follow from the associated patterns, as discussed in Sec. \ref{rr2_globalpaths}, and they are given in Table \ref{rr3_sectors}. We recast \Eq{rr3_xpath_1} in the two-component basis:
\begin{widetext}
\be \begin{split} \label{rr3_xpath_2}
	\left| \Psi_\alpha(\{h\}) \right>_f 
	&\doteq e^{ \frac{1}{2}i L_x h_{3y} + i\frac L2 \delta(\alpha,3)+\frac12i\sum_j\delta(\alpha,j)} 
	\left( \begin{array}{cc}
		0 & e^{-i\sum_j\delta(\alpha,j)} \\
		1 & 0
	\end{array} \right)
	\left| \Psi_{F(\alpha)} (\{h'\}) \right> \\ 
	\overline{\left| \Psi_\alpha (\{h\}) \right>}
	&= e^{-i\pi f_j(\alpha) } 
	\left( \begin{array}{cc}
		1 & 0 \\ 0 & -1
	\end{array} \right)
	\overline{\left| \Psi_\alpha (\{h'\}) \right>} \;.
\end{split} \ee
\end{widetext}
Just as in the two-quasihole case, \Eq{rr3_xpath_2} allows us to derive an equation between the transition matrix in the configuration $\sigma$ with the matrix in the configuration $g(\sigma)$. 
\be \label{rr3_xpath}
	\xi^{g(\sigma)} = 
	B^{-1} \textrm{diag}[ e^{-i\pi\lambda-i\frac L2 \delta(\alpha,3)-\frac12i\sum_j\delta(\alpha,j)} ]
	\xi^\sigma 
	\textrm{diag} \!\left[ e^{ -i \pi f_j(\alpha)} \right].
\ee
The pairs $\left(\sigma,g(\sigma)\right)$ are summarized in Table \ref{rr3_symm}. The matrix $B$ is defined as in \Eq{rr2_Bdef}, and for three quasiholes it has the form
\be \label{rr3_B}
	B = 
	\left( \begin{array}{cccc}
		&& 1 & \\
		1 & 0 && \\
		& 1 && \\
		&&& 1
	\end{array} \right) \;.
\ee
In any specific instance of \Eq{rr3_xpath}, one first chooses a starting configuration $\sigma$,
and identifies  the corresponding $y$-direction index $j$ of the quasihole that will encircle the torus. 
$j$ is in one-to-one correspondence with $\sigma_2$: for $\sigma_2$ is $+$, $0$, or $-$,  $j$ is respectively 3, 2, or 1. 
For instance, were we to begin in configuration $++$, then $j=3$ and after the encircling the system would be in configuration $-0$. Thus we find the relation
\be \label{rr3_ypath_++}
	\xi^{-0} = 
	\left( \begin{array}{cccc}
		& \tilde\Delta \\
		& 0 & \tilde\Delta^2 \\
		\tilde\Delta \\
		&&& 1
	\end{array} \right)
	\xi^{++}
	\left( \begin{array}{cccc}
		p \\ & p^{-1} \\ && -1 \\ &&& -1
	\end{array} \right) \;,
\ee
where $p$ is defined as before, $p = -e^{i \pi (s+\frac{1}{2})}$. 

We can go through the same derivation for a path similar to path $b$ in \Fig{globalpaths}, in which the top quasihole moves around the $y$ direction of the torus and ends at the bottom. We find
\be \label{rr3_ypath}
	\xi^{\overline g(\sigma)} = 
	\textrm{diag} \!\left[ e^{ -i \pi f_j(\alpha)} \right]
	\xi^{\sigma} 
	\textrm{diag}[ e^{-i\pi\lambda-i\frac L2 \delta(\alpha,3)-\frac12i\sum_j\delta(\alpha,j)} ] 
	B \;
\ee
It is very important to note that for this path, the meaning of the index $j$ is different from the previous path. In the previous case, the quasihole encircled the torus  in the $x$ direction, so the moving quasihole was the rightmost in horizontal ($x$) order but was the $j$-th quasihole in vertical ($y$) order; in this case, the quasihole encircles the torus in the $y$ direction, so the moving quasihole is the topmost in vertical order but is the $j$-th in horizontal order. As an example, if we begin in configuration ${-0}$, the topmost quasihole is that on the left, so ${j\!=\!1}$. Plugging in the appropriate values from the tables,
\be \label{rr3_xpath_++}
	\xi^{++} = 
	\left( \begin{array}{cccc}
		p^{-1} \\ & -1 \\ && p \\ &&& -1
	\end{array} \right)
	\xi^{-0} 
	\left( \begin{array}{cccc}
		&& \tilde\Delta & \\
		\tilde\Delta & 0 && \\
		& \tilde\Delta^2 && \\
		&&& 1
	\end{array} \right)
\ee
Combining Eqs. \eqref{rr3_xpath_++} and \eqref{rr3_ypath_++} gives us a consistency relation for $\xi^{++}$,
\be
	\xi^{++} =
	\left( \begin{array}{cccc}
		& \tilde\Delta p^{-1} && \\
		& 0 & -\tilde\Delta^2 & \\
		\tilde\Delta p &&& \\
		&&& -1
	\end{array} \right)
	\xi^{++}
	\left( \begin{array}{cccc}
		&& \tilde\Delta p & \\
		\tilde\Delta p^{-1} & 0 && \\
		& -\tilde\Delta^2 && \\
		&&& -1
	\end{array} \right) \;,
\ee
which further constrains $\xi^{++}$ in addition to \Eq{rr3_xi_constrained1}.
\be \label{rr3_xi_constrained}
	\xi^{++} =
	\left( \begin{array}{cccc}
		\xi_{11} & \tilde\Delta^2\xi_{13} & \xi_{13} & \xi_{14} \\
		\tilde\Delta^2\xi_{13} & \tilde\Delta^2p^2\xi_{11} 
			& -\tilde\Delta^{-1}p\xi_{13} & -\tilde\Delta^{-1}p\xi_{14} \\
		\xi_{13} & -\tilde\Delta^{-1}p\xi_{13} & \tilde\Delta^2p^2\xi_{11} 
			& -\tilde\Delta p\xi_{14} \\
		\xi_{41} & -\tilde\Delta^{-1}p\xi_{41} & -\tilde\Delta p\xi_{41} & \xi_{44}
	\end{array} \right)
\ee
%
\subsubsection{\label{rr3_braiding}Braid matrix}
As in the two-quasihole section, we will further determine the structure of the reduced braid matrix using constraint equations from unitarity and from locality. Enforcing locality means that we equate the matrix product for $\chi$ in \Eq{rr3_braid} with the form in \Eq{rr3_local}, which is implied by locality, as we argued above. 
The details are given in App. \ref{app3}, resulting in the following form for $\chi$:

\be  \label{rr3_chi} 
	\chi = e^{i \theta}
	\left( \begin{array}{cccc}
		p & & & \\
		& p & & \\ 
		& & p^2(1-p) & e^{i \theta_2} p^2 \sqrt{p+p^{-1}-1} \\
		& & e^{i \theta_2} p^2 \sqrt{p+p^{-1}-1} & e^{2i \theta_2} p(1-p)
	\end{array} \right)
\ee
where $\theta$ and $\theta_2$ are as yet undetermined phases. 

In deriving the above equation, only the zero matrix elements of \Eq{rr3_local} have been used.
To enforce consistency between the two- and three-quasihole braiding matrices, as dictated by locality,
we must equate the $2\times2$ block of \Eq{rr3_chi} to that of \Eq{rr3_local}. 
Equating the expressions for the element $\chi_{33}$ 
gives us a consistency relation that we can use to constrain $p$:
\be \label{rr3_consistency}
	e^{i \theta} p^2 (1-p) = e^{i \frac\pi2} p (p+p^{-1}-1) \;.
\ee
If we define ${x \!=\! p+p^{-1}}$ for convenience and take the absolute square of \Eq{rr3_consistency}, we find
\be
	2-x = (x-1)^2 \;,
\ee
which is solved when $x$ is the golden ratio, 
\be \label{rr3_gr}
	x =\varphi \equiv \frac{1+\sqrt{5}}{2} \;.
\ee
We have chosen the positive root because \Eq{app3_unit1b} implies $x\geq1$. If we define the angle $a$ by $p \!=\! \exp \!\left[ i \pi a \right]$ then $x \!=\! 2 \cos \left( \pi a \right)$ and \Eq{rr3_gr} implies 
\be \label{rr_a}
	a = \pm \frac{1}{5} \,. 
\ee
$a$ is also related to the shift parameter ${s\!=\!a+1/2}$, and so \Eq{rr_a} tells us\footnote{We leave it understood that this relation holds modulo 2.}
\be \label{rr_s}
	s = \frac12 \pm \frac15 \,. 
\ee
The phase information in \Eq{rr3_consistency} fixes the overall phase $\theta$,
\be \label{rr3_theta}
	e^{i \theta} = e^{i \pi s} \;.
\ee
There are two more consistency equations found from equating Eqs. \eqref{rr3_local} and \eqref{rr3_chi}. One yields $\exp \left[ i \theta_2 \right] = \pm1$, and the other is trivially satisfied when $x = \varphi$.
Up to some signs, the braid matrices for  two- and three-quasihole systems have thus been completely
solved for. We will discuss our solution(s) in the following section.

\section{\label{discussion}Discussion}
\begin{figure} 
	\includegraphics[width=.8\columnwidth]{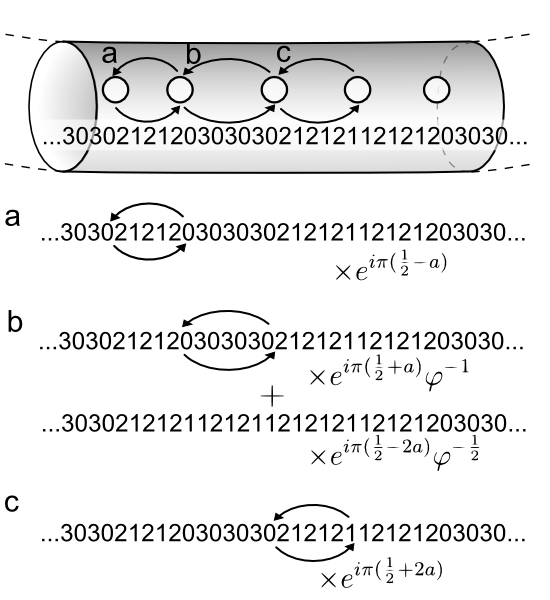}
	\caption{Graphical representation of the result of exchanging two $k=3$ 
		Read-Rezayi quasiholes for three example pairs. Top) A possible state in which 
		five quasiholes could be prepared, labeled by its associated thin torus pattern. 
		The state shown could be a five-quasihole state, in which the 30 strings at either 
		end would continue around the torus, or could be an $n$-quasihole state for $n>5$, 
		in which the ellipses mask additional domain walls in the thin torus pattern. 
		The results of braiding any pair of quasiholes shown here will be the same in either case. 
		a) Upon exchange of the indicated quasiholes, the state picks up the phase 
		$e^{i \pi (\frac12 - a)}$, where $a$ is given by \Eq{rr_a} (with the lower sign correctly describing
		the conformal block monodromies of the RR trial states). The thin torus pattern, 
		and thus the topological sector of the state, remains unchanged after the exchange, as shown. 
		b) When the two indicated quasiholes are exchanged the state remains in the same 
		topological sector or transitions into a sector with the linking 30 string changed to a 21 string. 
		The amplitudes for these two possibilities are shown beneath the thin torus patterns for the 
		sectors, where $\varphi$ is the golden ratio, \Eq{rr3_gr}. 
		c) Upon exchange of the indicated quasiholes, the state picks up the phase 
		$e^{i \pi (\frac12 + 2a)}$. The thin torus pattern, and thus the topological sector of the state, 
		remains unchanged after the exchange, as shown.} 
	\label{rr_exchange_patterns}
 \end{figure}
In Sec. \ref{rr}, we have found solutions for the braid matrices describing exchange processes
between 
two and three quasiholes that are consistent with the coherent state ansatz for the $k=3$, $\nu=3/2$ Read-Rezayi
state. In the following, we discuss how many {\em independent} solutions we have found, how they lead to general
rules for the braiding of $n$ quasiholes, and how these solutions compare to those obtained by other 
methods.

By means of \Eq{rr3_gr}, we may now
express the braid matrices for two quasiholes, \Eq{rr2_chi}, and three quasiholes, \Eq{rr3_chi}, in terms of only the golden ratio $\varphi$ and 
the parameter $a=\pm\frac15$. The two-quasihole matrix is then
\be  \label{end_chi2} 
	\chi = e^{i \frac\pi2}
	\left( \begin{array}{ccc}
		e^{-i \pi a} & & \\ 
		& e^{i \pi a} \varphi^{-1} & e^{-2i \pi a} \varphi^{-\frac12} \\
		& e^{-2i \pi a} \varphi^{-\frac12} & \varphi^{-1}
	\end{array} \right) \;,
\ee
and the three-quasihole matrix is
\be  \label{end_chi3} 
	\chi = e^{i \frac\pi2}
	\left( \begin{array}{cccc}
		e^{2i \pi a} & & & \\
		& e^{2i \pi a} & & \\ 
		& & e^{i \pi a} \varphi^{-1} & e^{-2i \pi a} \varphi^{-\frac12} \\
		& & e^{-2i \pi a} \varphi^{-\frac12} & \varphi^{-1}
	\end{array} \right) \;.
\ee
In writing these matrices,
we have removed the $\pm$ from the off-diagonal elements, choosing the $+$ sign.
Choosing the negative sign instead leads to a unitarily equivalent solution, where the
transformation is facilitated through multiplication of each state by $(-1)^{\#30}$, where $\#30$ is the number of $3030\dots$ strings in the thin torus pattern associated with that state. The arguments given below will make it obvious that this equivalence also carries over to general $n$-quasihole sectors.
We have thus obtained only two unitarily inequivalent solutions. It is clear from the above that these two solutions are closely related, namely
by complex conjugation and an overall Abelian phase $-1$. Thus, the non-Abelian content of the $k=3$ state has been determined uniquely 
by our method.

We will now use the locality arguments already made in Sec. \ref{rr2_braiding} for states of three quasiholes, 
and applied earlier in Sec. \ref{pf_braidgroup} to the Pfaffian case, to generalize these solutions to the case of $n$ quasiholes.
In essence, these arguments implied that the result of exchanging two neighboring quasiholes can only affect the
ground-state pattern linking the associated domain walls in the sector label, and only depend on the sequence of three patterns that
are separated by these two domain walls. For this, however, all possibilities have been exhausted by considering
two and three quasiholes, respectively. We can thus list the following rules, applicable to general $n$-quasihole
states, obtained directly from Eqs. \eqref{end_chi2} and \eqref{end_chi3}:

\begin{itemize}
\item
If the two quasiholes to be exchanged are associated with domain walls between ground-state strings $\dots3030\dw2121\dw121\dots$ or $\dots2121\dw1212\dw030\dots$, then after exchange the state remains in the same sector and picks up the phase $e^{i \pi (\frac12+2a)}$.
\item
If the two quasiholes are associated with the pattern  $\dots3030\dw21212\dw030\dots$, then after the exchange the state merely picks up the phase $e^{i \pi(\frac12- a)}$.
\item
If the quasiholes are associated with the pattern $\dots212\dw03030\dw21\dots$, after exchange the state will stay in same topological sector with amplitude $e^{i \pi (\frac12+a)} \varphi^{-1}$ and transition with amplitude $e^{i\pi(\frac12-2a)} \varphi^{-\frac12}$ into a sector that has the exchanged quasiholes associated with the pattern $\dots2121\dw121\dw121\dots$.
\item
If the quasiholes are associated with the pattern $\dots2121\dw121\dw121\dots$, then after exchange the state will stay in same topological sector with amplitude $e^{i \frac\pi2 } \varphi^{-1}$ and transition with amplitude $e^{i\pi(\frac12-2a)} \varphi^{-\frac12}$ into a sector that has the exchanged quasiholes associated with the pattern $\dots212\dw03030\dw21\dots$.
\end{itemize}
\begin{figure} 
	\includegraphics[width=.75\columnwidth]{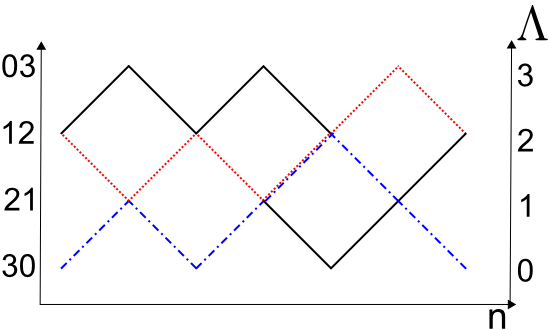}
	\caption{Bratteli diagram of the $k=3$ Read-Rezayi state with two 
	          possible paths indicated.
	 	The dashed line (red) corresponds to the sector label 
		$12\mathbf{11}2\mathbf{11}2\mathbf{11}2\mathbf{11}\mathbf{20}3\mathbf{02}12$ and 
		the dot-dashed line (blue) corresponds to the sector label 
	 	$3\mathbf{02}1\mathbf{20}3\mathbf{02}\mathbf{11}2\mathbf{11}\mathbf{20}30$. 
		} 
	\label{rr_bratteli_diagram}
 \end{figure}

These rules make it easy to visualize what is going on as a result of braiding in this non-Abelian state, as depicted in \Fig{rr_exchange_patterns}.
It remains to see which of our two solutions, if any, agrees with the representation of the braid group obtained from conformal block monodromies \cite{slingerland}. To make contact between these various representations, let us observe that our representation of topological sectors as patterns
separated by domain walls is in natural one-to-one correspondence with the representation given by paths meandering through a Bratteli diagram, \Fig{rr_bratteli_diagram}. Here, the vertices of the diagram are associated with the various ground-state patterns according to their ``height''
in the diagram, and the links represent the possible domain walls between them. A left-to-right path 
along the links of the diagram 
then represents an allowed sequence of patterns separated by domain walls, hence, a topological sector. The same diagrammatic labeling
of sectors also naturally arises through fusion rules in the CFT analysis of the RR states. With this identification, it becomes easy to see
that the above rules describing our solution are, for $a=-1/5$, in one-to-one correspondence with the ``tensor representation''
established in Ref. \onlinecite{slingerland} based on the analysis of conformal blocks.

To make this point, we briefly review the latter.
In the tensor representation given by Slingerland and Bais \cite{slingerland}, 
topological sectors, or paths through the Bratteli diagram, are represented by tensor products of vectors $v_{\Lambda_i,\Lambda_{i+1}}$
 of the ``domino'' form $v_{\Lambda_1,\Lambda_2}\otimes v_{\Lambda_2,\Lambda_3}\otimes v_{\Lambda_3,\Lambda_4}\otimes \dots \otimes v_{\Lambda_{n-1},\Lambda_n}$. 
Here, $\Lambda_i$ represents the ``height'' of the $i$-th vertex in the path (\Fig{rr_bratteli_diagram}),  $\Lambda_{i+1}=\Lambda_i\pm 1$,
and $v_{\Lambda_i,\Lambda_{i+1}}$ is a formal vector representing a link between two neighboring vertices at heights $\Lambda_i$, $\Lambda_{i+1}$,
respectively. At general level $k$, $\Lambda_i$ takes on  values $0,\dotsc,k$.
 
 In this tensor product basis, exchange of the quasiholes with indices $i$ and $i+1$ is represented by a matrix $R_{k,i}$ which acts only on the $i$-th and $(i+1)$-th factors \cite{slingerland}:
\begin{widetext}
\be \begin{split} \label{end_sling_ab1}
	R_{k,i} \, v_{\Lambda_i,\Lambda_i+1} \otimes v_{\Lambda_i+1,\Lambda_i+2}
		&= \alpha \, v_{\Lambda_i,\Lambda_i+1} \otimes v_{\Lambda_i+1,\Lambda_i+2} \\
	R_{k,i} \, v_{\Lambda_i,\Lambda_i-1} \otimes v_{\Lambda_i-1,\Lambda_i-2}
		&= \alpha \, v_{\Lambda_i,\Lambda_i-1} \otimes v_{\Lambda_i-1,\Lambda_i-2}
\end{split} \ee
\be \begin{split} \label{end_sling_ab2}
	R_{k,i} \, v_{\Lambda_i,\Lambda_i+1} \otimes v_{\Lambda_i+1,\Lambda_i} 
		&= -\alpha q^{-1} v_{\Lambda_i,\Lambda_i+1} \otimes v_{\Lambda_i+1,\Lambda_i} \;(\Lambda_i = 0)\\
	R_{k,i} \, v_{\Lambda_i,\Lambda_i-1} \otimes v_{\Lambda_i-1,\Lambda_i}	 
		&= -\alpha q^{-1} v_{\Lambda_i,\Lambda_i-1} \otimes v_{\Lambda_i-1,\Lambda_i} \;(\Lambda_i = k)
\end{split} \ee
\be \label{end_sling_nab} 
	\left( \begin{array}{c}
		R_{k,i} \, v_{\Lambda_i,\Lambda_i+1} \otimes v_{\Lambda_i+1,\Lambda_i} \\
		R_{k,i} \, v_{\Lambda_i,\Lambda_i-1} \otimes v_{\Lambda_i-1,\Lambda_i}
	\end{array} \right) = 
	\left( \begin{array}{cc}
		-\alpha q^{-\frac{\Lambda_i}2 -1} \frac{ 1 }{ \lfloor \Lambda_i +1 \rfloor_q }  &
		-\alpha q^{-\frac12} 
			\frac{ \sqrt{ \lfloor \Lambda_i +2 \rfloor_q \lfloor \Lambda_i \rfloor_q } }{ \lfloor \Lambda_i+1 \rfloor_q } \\
		-\alpha q^{-\frac12} 
			\frac{ \sqrt{ \lfloor \Lambda_i +2 \rfloor_q \lfloor \Lambda_i \rfloor_q } }{ \lfloor \Lambda_i+1 \rfloor_q } &
		\alpha q^{\frac{\Lambda_i}2}\frac{ 1 }{ \lfloor \Lambda_i+1\rfloor_q }
	\end{array} \right)
	\left( \begin{array}{c}
		v_{\Lambda_i,\Lambda_i+1} \otimes v_{\Lambda_i+1,\Lambda_i} \\
		v_{\Lambda_i,\Lambda_i-1} \otimes v_{\Lambda_i-1,\Lambda_i}
	\end{array} \right) \; (0 < \Lambda_i <k )
\ee
\end{widetext}
where $q = e^{ \frac{ 2\pi i}{k+2} }$, $\alpha = q^{ \frac{1-M}{2(kM+2)} }$, $M$ is related to the filling factor via
$\nu=3/(3M+2)$, and
``$q$-deformed integers''
$\lfloor m \rfloor_q$ are defined as,
\be
	\lfloor m \rfloor_q = \frac{ q^{\frac m2} - q^{-\frac m2} }{ q^{\frac12} - q^{-\frac 12} } \;.
\ee
In our case $k=3$ and $M=0$, so $q=e^{\frac{2 \pi i}5}$ and $\alpha = e^{\frac{i \pi}{10}}$.
In this case, it is not difficult to check that Eqs. \eqref{end_sling_ab1}-\eqref{end_sling_nab} 
reduce to the rules established in the beginning of this section, once tensor products
are reinterpreted as sequences of patterns via paths in the Bratteli diagram.


To wit, 
our first rule is equivalent to \Eq{end_sling_ab1}. To see this, observe that the two domain walls defined by the ground-state sequence $\dots3030\dw2121\dw121\dots$ could be represented on the Bratteli diagram \Fig{rr_bratteli_diagram} by $v_{0,1}\otimes v_{1,2}$ or by $v_{3,2} \otimes v_{2,1}$, both of which follow the form of \Eq{end_sling_ab1}. A similar observation can be made about $\dots2121\dw1212\dw030\dots$. The phase picked up by the states in \Eq{end_sling_ab1} is $\alpha=e^{\frac{i \pi}{10}}$, which is equivalent to the phase in our first rule $e^{i \pi (\frac12 +2a)}$, where $a=-\frac15$ here and in the following. Similarly, one can observe that the pattern in our second rule is represented by the vectors in \Eq{end_sling_ab2}. The phase in that equation is $-\alpha q^{-1}=e^{i\pi \frac7{10}}$, which is equivalent to the phase in the second rule, $e^{i \pi (\frac12-a)}$. Finally, our third and fourth rules are together equivalent to \Eq{end_sling_nab}. The patterns $\dots212\dw03030\dw21\dots$ and $\dots2121\dw121\dw121\dots$ can be written as $v_{1,0} \otimes v_{0,1}$ and $v_{1,2} \otimes v_{2,1}$ or as $v_{2,3} \otimes v_{3,2}$ and $v_{2,1} \otimes v_{1,2}$, which appear
in \Eq{end_sling_nab} for $\Lambda_i = 1$ and $\Lambda_i = 2$, respectively. Up to a change in the order of the basis states, the matrix in \Eq{end_sling_nab} for either value of $\Lambda_i$ gives the matrix elements stated in the third and fourth rules; this equivalence is shown here for $\Lambda_i=2$:
\begin{widetext}
\be
	\left( \begin{array}{cc}
		-\alpha q^{-2} \frac{ 1 }{ \lfloor 3 \rfloor_q }  &
		-\alpha q^{-\frac12} 
			\frac{ \sqrt{ \lfloor 4 \rfloor_q \lfloor 2 \rfloor_q } }{ \lfloor 3 \rfloor_q } \\
		-\alpha q^{-\frac12} 
			\frac{ \sqrt{ \lfloor 4 \rfloor_q \lfloor 2 \rfloor_q } }{ \lfloor 3 \rfloor_q } &
		\alpha q \frac{ 1 }{ \lfloor 3\rfloor_q }
	\end{array} \right) =
	\left( \begin{array}{cc}
		e^{i \pi \frac3{10}} \varphi^{-1} & e^{i \pi \frac9{10}} \varphi^{-\frac12} \\
		e^{i \pi \frac9{10}} \varphi^{-\frac12} & e^{i\pi\frac12} \varphi^{-1} 
	\end{array} \right) =
	\left( \begin{array}{cc}
		e^{i\pi(\frac12+a)} \varphi^{-1} & e^{i\pi(\frac12-2a)} \varphi^{-\frac12} \\
		e^{i\pi(\frac12-2a)} \varphi^{-\frac12} & e^{i\pi\frac12} \varphi^{-1} 
	\end{array} \right).
\ee
\end{widetext}
We hence see that one of our two solutions does indeed agree with the prediction based on conformal block monodromies, with the other one being closely related.

Furthermore, it appears that the solutions we obtained form a true subset of the solutions 
that can be derived by imposing the relevant fusion rules,
together with the axioms defining general anyon models (see, e.g., Refs. \cite{preskill_notes, bonderson_thesis, kitaev06}).
If, in addition to the pentagon and hexagon equations, one imposes unitarity and modularity,
these admit four solutions \cite{bonderson_thesis, bonderson_private_communication}.
Two of these appear to be identical to ours, with the other two related to the former by
complex conjugation.
We observe that in our approach, there is no reason to expect that solutions 
automatically come in complex conjugate pairs. This is so since the coherent state ansatz
explicitly assumes a holomorphic dependence on quasihole coordinates (see Sec. \ref{gencoherent}),
corresponding to a choice of  sign for the magnetic field that renders
trial wave functions for the RR
state holomorphic (in both electron and quasihole coordinates). Our findings thus seem to imply that for the ``missing'' two solutions,
one cannot construct holomorphic trial wave functions that can be adiabatically deformed
(through a continuous family of local Hamiltonians)
into the thin torus patterns we work with.

\section{\label{conclusion}Conclusion}

In this work, we have further developed an approach
to construct the braiding statistics associated to a given
fractional quantum Hall state through adiabatic
transport of quasiparticles. This approach
is based on the notion of adiabatic continuity
between FQH on the torus and simple product state---or ``patterns''---in the thin torus limit.
We have demonstrated that this notion, together
with a suitable coherent state ansatz for localized quasihole states,
allows one to work out the result of adiabatic transport,
using locality arguments and the information contained in the patterns.
The latter include properties under translation and the 
transformation of sectors when particles are rearranged
along certain topologically nontrivial paths on the torus.
The approach also makes heavy use of modular invariance.
We have presented a refined and unified treatment of simpler cases
studied earlier in this formalism, and then moved on to demonstrate
the applicability of the approach to the $k=3$ Read-Rezayi state.
In all cases, we found results consistent with conformal block
monodromies of trial wave functions. This is of particular interest
in the RR case, where to our knowledge, there are no rigorous results
yet that guarantee the agreement of these monodromies with
the statistics defined through adiabatic transport.
Our approach also has the benefit of giving rise to intuitive
pictures representing the transformation of topological sectors during braiding,
and allows for a logically independent, self-consistent derivation of non-Abelian statistics without
heavy mathematical machinery.

The findings presented in this work make us hopeful that the 
coherent state approach developed here, and earlier, quite generally
yields a sufficiently constraining set of equations that allows
one to construct the statistics of a given state of interest.
This should be possible, at least for the large class of states
that can be described by the recent paradigm of 
thin torus or dominance patterns. We 
have also tested this in detail on the gaffnian state \cite{gaffnian}.
This may be of particular interest since this state is associated
with a nonunitary CFT, and has been argued to have gapless excitations
 in its bulk spectrum \cite{gaffnian, read09}.
The general question of a well-defined notion of braiding statistics is therefore 
quite subtle. However, the question of whether or not our approach yields
well defined statistics in this case, and whether they are consistent with
conformal block monodromies, is well posed. We 
found the answer to be affirmative, and will present details of the
calculation elsewhere.

\section*{Acknowledgements}
We thank P. Bonderson and C. Nayak for insightful discussions.
A.S. would like to thank D.-H. Lee for collaboration on related publications.
This work was supported  by the National Science Foundation under NSF Grant No. DMR-0907793.

%
\appendix
\section{\label{app2}$k=3$ Read-Rezayi solution: Two quasiholes} 
We begin with Eqs. \eqref{rr2_local} and \eqref{rr2_unitary}, the locality and unitarity conditions, respectively, and seek to constrain the $\xi^\sigma_{\alpha,\alpha'}$ coefficients. 
As written, \Eq{rr2_unitary} 
does not provide information about the overall phase of $\xi^\sigma$, which is the overall phase relation
between the two mutually dual bases. This phase is, \emph{a priori}, arbitrary. We have, however, chosen a
phase convention by defining the action of the antilinear operator $\tau$ for both bases (in agreement
with the phase relation chosen in \Eq{dualvarphi2}). 
The symmetry under $\tau$ gave rise to 
\Eq{rr2_tau_xi},  which we can use together with 
\Eq{rr2_xpath} (from the ``global path'' along $x$) 
to replace $(\xi^+)^\dagger$ in favor of $\xi^+$, rewriting \Eq{rr2_unitary} as
\be \label{app2_unitary}
	\xi^+
	\left( \begin{array}{ccc} 
		p & 0 & 0 \\ 
		0 & p^{-1} & 0 \\ 
		0 & 0 & -1 
	\end{array}\right)
	(\xi^+)^T
	\left( \begin{array}{ccc} 
		0 & \Delta & 0 \\ 
		\Delta & 0 & 0 \\ 
		0 & 0 & 1 
	\end{array}\right)
	e^{-i \frac\pi2}  
	= \mathbb I_{3\times3} \;.
\ee
We will expand this matrix product by  plugging  
the form \Eq{rr2_xi_constrained} for $\xi^+$ derived 
from global path constraints.
%
This gives  four independent constraint equations for the $\xi_{\alpha,\alpha'}$s:
\bsub \label{app2_sol}
   \begin{align}
	\label{app2_sol3}
	\Delta p^3\xi_{11}^{\phantom{11}2} + \Delta p \xi_{12}^{\phantom{12}2}
	- \Delta p^2\xi_{13}^{\phantom{13}2} &= 0\\
	\label{app2_sol4}
	\xi_{31} \left[ - p^2 \xi_{11} + \Delta p\xi_{12} \right] + p \xi_{13}\xi_{33} &= 0 \\
	\label{app2_sol5}
	2 \Delta p \xi_{11}\xi_{12} + p \xi_{13}^{\phantom{13}2} &= e^{i \frac{\pi}{2}}\\
	\label{app2_sol6}
	2p \xi_{31}^{\phantom{31}2} - \xi_{33}^{\phantom{33}2} &= e^{i \frac{\pi}{2}}
   \end{align}
Recall ${p=-\exp \left[i \pi (s+\frac{1}{2}) \right]}$ and $\Delta^2=1$.
Similarly, we may use \Eq{rr2_tau_xi} in the definition of the reduced braid matrix 
$\chi$, \Eq{rr2_chi_def}, writing $\chi$ as $\xi^+(\xi^+)^T$. Expanding the latter again with \Eq{rr2_xi_constrained}
and comparing the result to the locality constraint \Eq{rr2_local}, we find two additional independent constraint equations:
   \begin{align}
	\label{app2_sol1}
	(1+p^2) \xi_{11}\xi_{12} - \Delta p \xi_{13}^{\phantom{13}2} &= 0 \\
	\label{app2_sol2}
	\xi_{31} \left[ \xi_{11} - \Delta p \xi_{12} \right]+\xi_{13} \xi_{33} &= 0\;.
   \end{align}
\esub
We will use the constraint equations \eqref{app2_sol} to solve for the unknown elements of $\chi$, the dots in \Eq{rr2_local}, which can be found from the expansion of the product $\xi^+(\xi^+)^T$ to be:
\bsub \label{app2_chi_el}
   \begin{align}
	\label{app2_chi_el11} 
	\xi_{11}^{\phantom{11}2} + \xi_{12}^{\phantom{12}2} + \xi_{13}^{\phantom{13}2} 
	&= \chi_{11}\\
	\label{app2_chi_el22} 
	p^4 \xi_{11}^{\phantom{11}2} + \xi_{12}^{\phantom{12}2} + p^2 \xi_{13}^{\phantom{13}2} 
	&= \chi_{22} \\
	\label{app2_chi_el33} 
	2p \xi_{31}^{\phantom{31}2} + \xi_{33}^{\phantom{33}2} 
	&= \chi_{33}\\
	\label{app2_chi_el23} 
	\Delta \xi_{31} \left[ - p^3 \xi_{11} + \Delta\xi_{12} \right] - \Delta p \xi_{13} \xi_{33}
	&= \chi_{23} = \chi_{32}
   \end{align}
\esub

We can break the solution of Eqs. \eqref{app2_sol} into two major sections, which are based on the two ways to satisfy the equation we obtain by combining Eqs. \eqref{app2_sol1} and \eqref{app2_sol3}:
\be \label{app2_sola}
	p^2 \xi_{11}^{\phantom{11}2} + \xi_{12}^{\phantom{12}2} 
	-\Delta\xi_{11} \xi_{12} - \Delta p^2 \xi_{11}\xi_{12} = 0
\ee
There are two solutions to this equation: 
%
\bsub \label{app2_xi12}
   \begin{align}
	\xi_{12} &= \Delta p^2 \xi_{11} \label{app2_xi12_1}\\
	\textrm{or } \xi_{12} &= \Delta\xi_{11} \label{app2_xi12_2}\,.
   \end{align}
\esub
We will now show that the first of the above equations never leads to 
consistent independent solutions, except in the special case $\xi_{13}=0$.
To see this, 
 we feed Eqs. \eqref{app2_xi12}
back into \Eq{app2_sol1}, and find, respectively, that 
\begin{subequations} \label{app2_xi13}
   \begin{align}
	\xi_{13}^{\phantom{13}2} &= p^2 (p+p^{-1}) \xi_{11}^{\phantom{11}2} \\
	\label{app2_xi13_2}
	\textrm{or } \xi_{13}^{\phantom{13}2} &= (p+p^{-1}) \xi_{11}^{\phantom{11}2}\,.
   \end{align}
\end{subequations}
We first utilize the above to study all cases with $\xi_{13}=0$.
This implies either $\xi_{11}=0$, or $p\in \{i,-i\}$. The former leads to 
a contradiction in \Eq{app2_sol5}.
It is then straightforward to show that for  $p\in \{i,-i\}$, the solutions of the system \eqref{app2_sol} 
 produce the braid matrix
\be \label{app2_chiwrong}
	\chi = e^{-i \frac\pi2}
	\left( \begin{array}{ccc}
		\mp p & 0 & 0 \\
		0 & \mp p & 0 \\
		0 & 0 & 1
	\end{array} \right)\,,
\ee
with the upper (lower) sign corresponding to \Eq{app2_xi12_1} (\Eq{app2_xi12_2}).
Equation \eqref{app2_chiwrong} corresponds to a consistent solution to the constraint equations \eqref{app2_sol}.
However, when \Eq{app2_chiwrong} is generalized to an $n$-quasihole system
using the locality arguments of Sec. \ref{pf_braidgroup},
it is not difficult to see that the resulting braid matrix violates the Yang-Baxter equation. 
While this might suffice to rule out this solution, we have emphasized in the beginning that 
our approach requires no \emph{a priori} assumption that any aspect of quasiparticle exchange is 
topological. We will thus show more directly in App. \ref{app3} that \Eq{app2_sol} leads to contradictions
in the present framework when three quasiholes are considered.  
Since we can rule out the special solution 
leading to the upper sign in
 \Eq{app2_chiwrong}, this case has not been mentioned in the main text.


We now proceed by exploring solutions with $\xi_{13}\neq 0$. We first show that 
\Eq{app2_xi12_1} does not lead to further independent solutions. 
To this end, we plug Eqs. \eqref{app2_xi12} first into \Eq{app2_sol2},
\bsub \label{app2_xi33_1}
   \begin{align}
	\label{app2_xi33_1_1}
	\xi_{13} \xi_{33} &= -(1- p^3 ) \xi_{31} \xi_{11} \\
	\label{app2_xi33_1_2}
	\textrm{or } \xi_{13} \xi_{33} &= -(1- p)\xi_{31} \xi_{11} \,,
	   \end{align} 
\esub
and similarly into \Eq{app2_sol4}:
\bsub \label{app2_xi33_2}
   \begin{align}
   	\label{app2_xi33_2_1}
	\xi_{13} \xi_{33} &= -(p^2 -p) \xi_{31} \xi_{11} \\
	\label{app2_xi33_2_2}
	\textrm{or } \xi_{13} \xi_{33} &= -(1- p)\xi_{31} \xi_{11} \,.
   \end{align} 
\esub
While Eqs. \eqref{app2_xi33_1_2} and \eqref{app2_xi33_2_2} are identical,
Eqs. \eqref{app2_xi33_1_1} and \eqref{app2_xi33_2_1} 
turn out to be consistent with one another only in cases where both sides vanish on both equations.
We have already discussed all cases with $\xi_{13}=0$.
To satisfy Eqs.   \eqref{app2_xi33_1_1} and \eqref{app2_xi33_2_1}, we may thus
focus  on the case $\xi_{33}=0$.
On the right-hand side, we can rule out $\xi_{31}\!=\!0$ because, with $\xi_{33}\!=\!0$, it contradicts \Eq{app2_sol6}. We can similarly rule out $\xi_{11}\!=\!0$ because, with \Eq{app2_xi12} and \Eq{app2_xi13}, it violates \Eq{app2_sol5}. 
The only other way to solve both Eqs. \eqref{app2_xi33_1_1} and \eqref{app2_xi33_2_1} is to
have $p \!=\! \pm 1$. 
In this case, however, both equations \eqref{app2_xi12} are identical.
Thus, \Eq{app2_xi12_1} does not produce independent valid solutions,
except for $p\in \{i,-i\}$, leading to the braid matrix \Eq{app2_chiwrong} (upper sign).
As mentioned, the latter leads to inconsistencies in the case of three quasiholes.

To find the solution to the constraints \eqref{app2_sol} that will be consistent with the three-quasihole case, 
we now discard \Eq{app2_xi12_1} and proceed to work from \Eq{app2_xi12_2},
and the equations \eqref{app2_xi13_2}, \eqref{app2_xi33_1_2} derived from it. 
First, we plug Eqs. \eqref{app2_xi12_2} and \eqref{app2_xi13_2} into \Eq{app2_sol5}, which will give us an explicit form for $\xi_{11}^{\phantom{11}2}$:
\be \label{app2_xi11}
	\xi_{11}^{\phantom{11}2} = \frac{e^{i \frac{\pi}{2}} }{(1+p)^2}\,.
\ee
In particular $\xi_{11}\neq 0$. From \Eq{app2_xi33_1_2} we thus obtain
\be \label{app2_xi33_3}
   \begin{split} 
	(1-p)^2{\xi_{31}^{\phantom{31}2}}
	&= \frac{\xi_{13}^{\phantom{13}2}}{\xi_{11}^{\phantom{11}2}} {\xi_{33}^{\phantom{33}2}}\\
	&= (p+p^{-1}){\xi_{33}^{\phantom{33}2}}\,,
   \end{split} 
\ee
where we have used \Eq{app2_xi13_2}.
Elimination of $\xi_{33}$ by means of \Eq{app2_sol6} then gives
\be \label{app2_xi31}
   \begin{split}
	\xi_{31}^{\phantom{31}2} &= (p+p^{-1}) \frac{e^{i \frac{\pi}{2}} }{(1+p)^2}  \\
	&= (p+p^{-1}) \xi_{11}^{\phantom{11}2} 
   \end{split}
\ee
We can now revisit the unknown elements of $\chi$. We rewrite \Eq{app2_chi_el} using the equations we have developed above.
\bsub \label{app2_chi}
   \begin{align}
	\label{app2_chi11} 
	p^{-1}(1+p)^2 \xi_{11}^{\phantom{11}2} &= \chi_{11}\\
	\label{app2_chi22} 
	p(p+p^{-1}-1) (1+p)^2 \xi_{11}^{\phantom{11}2} &= \chi_{22} \\
	\label{app2_chi33} 
	(p+p^{-1}-1) (1+p)^2 \xi_{11}^{\phantom{11}2} &= \chi_{33}\\
	\label{app2_chi23} 
	\Delta (1-p) (1+p)^2 \xi_{31} \xi_{11} &= \chi_{23} = \chi_{32}
   \end{align}
\esub
We need only plug into \Eq{app2_chi23} the square root of \Eq{app2_xi31} to write each element of $\chi$ in terms of $\xi_{11}^{\phantom{11}2}$, for which we have the expression in \Eq{app2_xi11}. We can also absorb the $\Delta$ factor in \Eq{app2_chi23} into the $\pm$ induced by taking this square root. Thus we reach the following form of the braid matrix 
\be  \label{app2_chifinal} 
	\chi = e^{i \frac\pi2}
	\left( \begin{array}{ccc}
		p^{-1} & 0 & 0 \\ 
		0 & p(p+p^{-1}-1) & \pm\sqrt{p+p^{-1}}(1-p) \\
		0 & \pm\sqrt{p+p^{-1}}(1-p) & p+p^{-1}-1
	\end{array} \right),
\ee
which was presented in the main text as \Eq{rr2_chi}.
\section{\label{app3}$k=3$ Read-Rezayi solution: Three quasiholes} 
Here we will solve a system of equations for the elements of the three-quasihole transition matrix elements
$\xi_{\alpha,\alpha'}$ and the resulting braid matrix. 
The procedure is the same as that employed for two quasiholes:
Using various constraints on $\xi^\sigma$ already derived in the main text, 
we write out the matrix elements of the unitarity equation,
$\xi^{++}(\xi^{++})^\dagger \!=\! \mathbb{I}_{4\times4}$,
and the locality constraint \Eq{rr3_local}. This gives a system 
for the remaining unknown elements of $\xi^{++}$.
However, we must recall that the form in \Eq{rr3_local} was based, in part, on the two-quasihole braid matrix \Eq{rr2_chi}. In App. \ref{app2} we found 
one other ``special'' solution for this matrix, namely  \Eq{app2_chiwrong} (upper sign),
that was not presented in the main text. Here we will consider this special solution also,
giving rise to a modified version of \Eq{rr3_local}, and show that this solution leads to inconsistencies
with three-quasihole braiding.

Just as we did in App. \ref{app2}, we first use the (antilinear) mirror symmetry
to eliminate complex conjugation from the definition of the reduced braid matrix,
 $\chi = \xi^{++} (\xi^{-+})^\dagger$.
 This is achieved by using Eqs.
\eqref{rr3_tau} and  \eqref{rr3_ypath_++}. The result is
\begin{widetext}
\be \label{app3_chi1}
	\chi = \xi^{++}
	\left( \begin{array}{cccc}
		\tilde\Delta^2p \\ & p^{-1} \\ && -1 \\ &&&-1
	\end{array} \right)
	(\xi^{++})^T
	\left( \begin{array}{cccc}
		& \tilde\Delta^{-1} \\ \tilde\Delta^{-1} & 0 \\ && 1 \\ &&& 1
	\end{array} \right)
	e^{-i \pi \lambda + i \pi}\,.
\ee
\end{widetext}
We will expand this matrix product using the constrained form of $\xi^{++}$ in \Eq{rr3_xi_constrained}, reproduced here: 
\be \label{app3_xi_constrained}
	\xi^{++} =
	\left( \begin{array}{cccc}
		\xi_{11} & \tilde\Delta^2\xi_{13} & \xi_{13} & \xi_{14} \\
		\tilde\Delta^2\xi_{13} & \tilde\Delta^2p^2\xi_{11} 
			& -\tilde\Delta^{-1}p\xi_{13} & -\tilde\Delta^{-1}p\xi_{14} \\
		\xi_{13} & -\tilde\Delta^{-1}p\xi_{13} & \tilde\Delta^2p^2\xi_{11} 
			& -\tilde\Delta p\xi_{14} \\
		\xi_{41} & -\tilde\Delta^{-1}p\xi_{41} & -\tilde\Delta p\xi_{41} & \xi_{44}
	\end{array} \right),
\ee
where we recall $\tilde\Delta^4=1$, and $p$ is defined in terms of the shift parameter
$s$ as before.
We gain a system of constraint equations for the $\xi_{\alpha,\alpha'}$s by plugging \Eq{app3_xi_constrained} into \Eq{app3_chi1} and equating the product to one of the following expressions for $\chi$ that have been
derived from locality and from consistency with the two-quasihole solution.
For generic parameter $p$, we found that the latter must be of the form \Eq{rr3_local}, which we reproduce here
as
%
\be  \label{app3_chiright} 
	\chi_{\mbox{loc}} = e^{i \frac \pi2}
	\left( \begin{array}{cccc}
		\cdot & & & \\
		& \cdot & & \\ 
		& & p(p+p^{-1}-1) & \pm\sqrt{p+p^{-1}}(1-p) \\
		& & \pm\sqrt{p+p^{-1}}(1-p) & p+p^{-1}-1
	\end{array} \right),
\ee
with blanks denoting zeros.
The 2x2 block in the above was taken directly from the two-quasihole solution, \Eq{app2_chifinal}, as explained in the
main text. For $p\in \{i,-i\}$,
 however, we found an additional solution to the two-quasihole system of equations, leading
to the form of the braid matrix \Eq{app2_chiwrong}. 
Using this form and the same reasoning that lead to \Eq{app3_chiright}, for $p\in \{i,-i\}$ the reduced braid matrix 
must be of the form
\be \label{app3_chiwrong}
	\chi_{\mbox{loc}} = e^{-i \frac\pi2} 
	\left( \begin{array}{cccc}
		\cdot\\&\cdot\\&&\mp p\\&&&1 
	\end{array} \right) \;,
\ee
where the lower sign is just a special case of \Eq{app3_chiright}, but the upper sign corresponds to the
``special'' solution.

We equate $\chi_{\mbox{loc}}$ to \Eq{app3_chi1}.
We first focus on those matrix elements
for which $\chi_{\mbox{loc}}$ is identically
zero in all cases. 
By means of \Eq{app3_xi_constrained}, this gives 
rise to the following three equations:

\bsub \label{app3_eqns}
   \begin{align}
   	\label{app3_eqn1}
	\tilde\Delta(p-p^2)\xi_{13}^{\phantom{13}2} + \tilde\Delta^{-1}p^3\xi_{11}^{\phantom{11}2}
		-\tilde\Delta p^2\xi_{14}^{\phantom{14}2} &= 0 \\
	\label{app3_eqn2}
	\tilde\Delta^{-1}p\xi_{13}^{\phantom{13}2} + (p^3-p^2)\xi_{11}\xi_{13}
		-\tilde\Delta^{-1}p^2\xi_{14}^{\phantom{14}2} &= 0 \\
	\label{app3_eqn3}
	\xi_{41} \left[ -p^2\xi_{11}+\tilde\Delta^{-1}(p-p^2)\xi_{13} \right]
		+\tilde\Delta^2p \xi_{14}\xi_{44} &= 0.
   \end{align}
We will also use equations gained from enforcing the unitarity of $\xi^{++}$. If we expand $\xi^{++}(\xi^{++})^\dagger \!=\! \mathbb{I}_{4\times4}$ using \Eq{app3_xi_constrained} we find the following independent equations:
   \begin{align}
   	\label{app3_eqn4}
	\abs{\xi_{11}}^2 + 2\abs{\xi_{13}}^2 + \abs{\xi_{14}}^2 &= 1 \\
	\label{app3_eqn7}
	3\abs{\xi_{41}}^2 + \abs{\xi_{44}}^2 &= 1 \\
	\label{app3_eqn6}
	\xi_{41} \left[ \xi_{11}^{\phantom{11}*} -2\tilde\Delta p \xi_{13}^{\phantom{13}*} \right]
		+\xi_{14}^{\phantom{14}*}\xi_{44} &= 0 \\
	\label{app3_eqn5}
	\tilde\Delta^2\xi_{13}\xi_{11}^{\phantom{11}*}+p^2\xi_{11}\xi_{13}^{\phantom{13}*}
		-\tilde\Delta^{-1}p\abs{\xi_{13}}^2 -
		&\tilde\Delta^{-1}p \abs{\xi_{14}}^2 = 0
   \end{align}
\esub
For convenience, we may also write the unitarity condition in the form
$(\xi^{++})^\dagger\xi^{++} \!=\! \mathbb{I}_{4\times4}$, yielding a similar
(and equivalent) set of equations, one of them being $3\abs{\xi_{14}}^2 + \abs{\xi_{44}}^2 = 1$.
By comparison with \Eq{app3_eqn7}, this implies
\be\label{app3_1441}
  \abs{\xi_{14}}^2=\abs{\xi_{41}}^2\;.
\ee
Once the $\xi_{\alpha,\alpha'}$ are known, 
\Eq{app3_chi1} allows us to obtain the following expressions for the
unknown elements of $\chi$:
\bsub \label{app3_chi_el}
   \begin{align} 
   	\label{app3_chi_el11}
	\chi_{11} &= \chi_{22} =  \\
	&\quad \left( 2\tilde\Delta^{-1}p\xi_{11}\xi_{13} + \tilde\Delta^2 p \xi_{13}^{\phantom{13}2} + 
		\tilde\Delta^2 p\xi_{14}^{\phantom{14}2} \right)e^{-i\pi\lambda + i \pi} \nonumber \\
	\label{app3_chi_el33} 
	\chi_{33} &=
		\left( -p^4\xi_{11}^{\phantom{11}2} + 2\tilde\Delta^2 p \xi_{13}^{\phantom{13}2} - 
		\tilde\Delta^2 p^2 \xi_{14}^{\phantom{13}2} \right)e^{-i\pi\lambda + i \pi} \\
	\label{app3_chi_el44} 
	\chi_{44} &=
		\left( \tilde\Delta^2(2p-p^2)\xi_{41}^{\phantom{41}2} - \xi_{44}^{\phantom{44}2} \right)
		e^{-i\pi\lambda + i \pi} \\
	\label{app3_chi_el34} 
	\chi_{34} &= \chi_{43} = \\
	&\quad \left( \xi_{41} \left[ \tilde\Delta^{-1} p^3 \xi_{11} + 2\tilde\Delta^2 p \xi_{13} \right] + 
		\tilde\Delta p\xi_{14} \xi_{44} \right)e^{-i\pi\lambda + i \pi}. \nonumber
   \end{align}
\esub
If we subtract \Eq{app3_eqn2} (times $\tilde\Delta^2$) from \Eq{app3_eqn1}, the resultant equation can be solved two ways.
\bsub \label{app3_loc}
   \begin{align}
	\label{app3_loca}
	\xi_{13} &= -\tilde\Delta p \xi_{11} \\
	\label{app3_locb}
	\textrm{or } \xi_{13} &= \tilde\Delta \xi_{11}
   \end{align}
\esub
We can quickly eliminate one of these possibilities by comparing to the equations from unitarity. When Eqs. \eqref{app3_loc} are put into \Eq{app3_eqn5}, they respectively produce the equations
\bsub \label{app3_unit1}
   \begin{align}
	\label{app3_unit1a}
	\abs{\xi_{14}}^2 &= -3\abs{\xi_{11}}^2 \\
	\label{app3_unit1b}
	\textrm{or } \abs{\xi_{14}}^2 &= \abs{\xi_{11}}^2  (p+p^{-1}-1) \;.
   \end{align}
\esub
Whereas we can put Eqs. \eqref{app3_loc} into \Eq{app3_eqn4} and get the same equation for both cases:
\be \label{app3_unit2}
	\abs{\xi_{14}}^2 = -3\abs{\xi_{11}}^2 +1 \;,
\ee
which clearly contradicts \Eq{app3_unit1a}. Thus Eqs. \eqref{app3_loca} and \eqref{app3_unit1a} are not true. Eliminating $\abs{\xi_{14}}^2$ from Eqs. \eqref{app3_unit1b} and \eqref{app3_unit2} gives us an expression for $\abs{\xi_{11}}^2$, which we can turn into an expression for $\xi_{11}^{\phantom{11}2}$ with the inclusion of some phase $\theta_1$.
\be \label{app3_xi11}
	\xi_{11}^{\phantom{11}2} = \frac{e^{i \theta_1}p}{(1+p)^2}\,.
\ee
Furthermore,  putting \Eq{app3_locb} into either \Eq{app3_eqn1} or \Eq{app3_eqn2} gives:
\be \label{app3_xi14}
	\xi_{14}^{\phantom{14}2} = \tilde \Delta^2(p+p^{-1}-1)\,\xi_{11}^{\phantom{11}2}\,
\ee
which fixes the phase between $\xi_{14}^{\phantom{14}2}$ and $\xi_{11}^{\phantom{11}2}$.
Together with \Eq{app3_1441}, this also implies
\be \label{app3_xi41}
	\xi_{41}^{\phantom{41}2} = e^{2i \theta_2} \tilde \Delta^2 (p+p^{-1}-1)\, \xi_{11}^{\phantom{11}2} \,,
\ee
where we introduced another phase $\theta_2$. Assuming first that $\xi_{14}\neq 0$, 
we define $e^{i\theta_2}=\xi_{41}/\xi_{14}$ (cf. \Eq{app3_1441}),
we may solve
\Eq{app3_eqn3} for $\xi_{44}$:
\be\label{app3_xi44}
   \xi_{44}=(2p-1)\tilde\Delta^2 e^{i\theta_2} \xi_{11}\,.
\ee
It is easy to see that the last equation also holds in cases where
$\xi_{14}=\xi_{41}=0$.\footnote{
In this case, Eqs. \eqref{app3_eqn4}, \eqref{app3_locb} imply $\abs{\xi_{11}}=1/\sqrt{3}$,
and $\abs{\xi_{44}}=1$ from \Eq{app3_eqn7}.
From \Eq{app3_xi14}, we must then have $p+p^{-1}-1=0$, hence $(2p-1)^3=-3$. 
The absolute values in \Eq{app3_xi44} therefore work out, and \Eq{app3_xi44}
must thus hold for some phase $\theta_2$, which is then defined through this equation.
}
With Eqs. \eqref{app3_locb}, \eqref{app3_xi14}, \eqref{app3_xi41}, and \eqref{app3_xi44}  we can rewrite the unknown elements of $\chi$ in terms of $\xi_{11}^{\phantom{11}2}$, and Eqs. \eqref{app3_chi_el} become
\bsub
   \begin{align}
	\label{app3_chi11} 
	\chi_{11} &= \chi_{22} =
		(1+p)^2 \xi_{11}^{\phantom{11}2} e^{-i\pi\lambda + i \pi}  \\
	\label{app3_chi33} 
	\chi_{33} &= 
		p(1-p)(1+p)^2 \xi_{11}^{\phantom{11}2} e^{-i\pi\lambda + i \pi} \\
	\label{app3_chi44} 
	\chi_{44} &=
		e^{2i \theta_2} (1-p) (1+p)^2 \xi_{11}^{\phantom{11}2} e^{-i\pi\lambda + i \pi} \\
	\label{app3_chi34} 
	\chi_{34} &= \chi_{43} =
		e^{i \theta_2} \sqrt{p+p^{-1}-1} 
		(1+p)^2 \xi_{11}^{\phantom{11}2} e^{-i\pi\lambda + i \pi}
   \end{align}
\esub
Or, using \Eq{app3_xi11},
\be  \label{app3_chi} 
	\chi = e^{i \theta}
	\left( \begin{array}{cccc}
		p & & & \\
		& p & & \\ 
		& & p^2(1-p) & e^{i \theta_2} p^2 \sqrt{p+p^{-1}-1} \\
		& & e^{i \theta_2} p^2 \sqrt{p+p^{-1}-1} & e^{2i \theta_2} p(1-p)
	\end{array} \right)
\ee
where we have defined $e^{i \theta} \!= \!e^{ i \theta_1-i \pi \lambda +i \pi}$. 
This is the result quoted in the main text as \Eq{rr3_chi}. 
It is worth noting that once again, the $\delta$ parameters have dropped out.
The derivation of \Eq{app3_xi11}
is valid irrespective of the value of $p$, since we did thus far not use the
diagonal matrix elements of $\chi_{\mbox{loc}}$, which may take on special
values for $p \in \{i,-i\}$. We are now able to rule out $p \in \{i,-i\}$, and thus the
``special'' solution obtained in App. \ref{app2}. For in this case,
\Eq{app3_chi} has non-zero off-diagonal matrix elements, whereas 
\Eq{app3_chiwrong} does not.
This justifies
Eqs. \eqref{rr2_chi} and \eqref{rr3_local} in the main text, which ignore the ``special'' solution.
Requiring consistency between the non-zero matrix elements of Eqs. \eqref{rr3_local}
and \eqref{rr3_chi}, which we have not done in this Appendix, finally provides information about the phase $p$,
relating it to the golden mean.
This short argument is presented in the main text.


\begin{thebibliography}{74}
\bibitem [{del()}]{delta_note}%
  \BibitemOpen
  \href@noop {} {}\bibinfo {note} {{One may consider a generalized version of
  the coherent state \eqref{pf_psi_def}, with $\delta$ replaced by $\delta_1$
  in the first $\phi$-factor, and by $\delta_2$ in the second. Consistent
  behavior of this expression under $\bar\tau$ requires $\delta_1=\delta_2 \mod
  \pi/\beta$. Consistent behavior under $I$ requires $\delta_1=-\delta_2 \mod
  \pi/\beta$. This yields that either $\delta_{1,2}=0 \mod \pi/\beta$, or
  $\delta_{1,2}=\pi/(2\beta) \mod \pi/\beta$. We can then take
  $\delta_1=\delta_2\equiv \delta$ without loss of generality, since shifting
  $\delta_i$ by $\pi/\beta$ only results in an overall change of phase. Cf.
  Sec. \ref{cons_pfaff1}, where furthermore $\beta=1/2$ is
  derived.}}\BibitemShut {Stop}%
\end{thebibliography}
%

\end{document}